\documentclass[prd,nofootinbib,showkeys,showpacs]{revtex4}
\usepackage{graphicx}%
\usepackage{amsmath}

\usepackage{amsmath}
\usepackage{amsfonts}
\usepackage{amssymb}
\usepackage{amsthm}
\usepackage{amscd}

\usepackage{dsfont}

\usepackage{epsfig}

\renewcommand{\slash}[1]{\ensuremath{#1 \!\!\! / }}
\newcommand{\slashp}{\ensuremath{p \!\!\! / }}

\newcommand{\sups}[1]{\ensuremath{{}^{#1}}}
\newcommand{\subs}[1]{\ensuremath{{}_{#1}}}

\newcommand{\spinor}{\ensuremath{\Psi} }

\renewcommand{\Im}{\ensuremath{\textit{Im }}}
\renewcommand{\Re}{\ensuremath{\textit{Re }}}

\renewcommand{\vec}[1]{\ensuremath{\mathbf{#1}}} 

\newcommand{\Pdreidrei}{\ensuremath{P_{33}\,}}
\newcommand{\Deinsdrei}{\ensuremath{D_{13}\,}}

\newcommand{\rhomeson}{\ensuremath{\rho}-meson }

\begin{document}

\title{Complete Relativistic Description of the $\mathbf{N^*(1520)}$}
\author{Lukas Jahnke and Stefan Leupold}
\affiliation{Institut f\"ur Theoretische Physik, Universit\"at Giessen, Germany}

\begin{abstract}
A relativistic description of spin 3/2 resonances and their decay channels is presented 
by calculating their selfenergies and spectral functions. The full vector-spinor structure
is taken into account. Special emphasis is put on the $N^*(1520)$ and its decay channels
$\pi N$, $\rho N$ and $\pi \Delta$. All interactions are formulated such that only the 
correct number of degrees of freedom of a spin 3/2 state is propagated. 
The obtained results are compared with several 
approximations frequently used to avoid the complicated vector-spinor structure.
Since this structure is taken fully into account here, the quality of the approximations
can be judged.
\end{abstract}


\maketitle

\section{Introduction}

In the last years there has been a large interest in the description of hadronic properties in a strongly interacting medium. This interest is induced by the fact that various experiments give indications for modifications of the mass and the width of hadrons when put into the medium. The changes of hadron masses in a medium are probably related to the restoration of chiral symmetry as one of the fundamental symmetries of QCD which is spontaneously broken in the vacuum (for further discussion see \cite{Rapp:1999ej}).

The \rhomeson is an ideal probe for in-medium studies because it decays into dileptons. They can travel nearly undisturbed through the medium making it possible to measure the in-medium properties. In-medium effects are particularly strong for $\rho$-mesons at rest as could be shown e.g. by Post et al.~\cite{Post:2000qi} or Gale and Kapusta \cite{Gale:1990pn}. Especially in a nucleon dominated medium (nuclear matter) scattering of such low energetic $\rho$-mesons with nucleons will create the $N^*(1520)$ resonance, making it especially important for the in-medium properties of the $\rho$-meson. A coupled channel in-medium calculation involving the $N^*(1520)$ and several other baryon resonances and in the meson sector $\rho$, $\pi$ and $\eta$ has been performed by Post et al.~\cite{Post:2003}. Typically due to the complexity of the problem a number of approximations are involved in such calculations. Especially the selfenergies of the considered particles are either calculated non-rel\-a\-tiv\-is\-tic\-ally  \cite{Post:2003} or at least a simplified spin structure is assumed \cite{Post:2000qi}. As a prelude to a more complete treatment it is important to understand first the vacuum case in its full relativistic structure which can serve as a starting point for more involved in-medium computations. The aim of the present work is to provide a framework for a complete relativistic description of spin 3/2 resonances in the vacuum. In a previous relativistic calculation of Post et al.~\cite{Post:2000qi} on the in-medium properties of the \rhomeson it could be shown that for a resonance with negative parity an often used approximation for the spin structure leads to a wrong sign for the imaginary part of the self\-energy of the resonance. This in turn leads to a negative cross section. By changing the propagator by hand it was possible to overcome this short coming. This shows that the relativistic effects can be non-trivial. A correct and fully relativistic description of resonances and in our case especially of the $N^*(1520)$ in vacuum is therefore desirable.

A study of the correct description of higher spin particles is also of interest due to the non-trivial character of their interactions. Though field theoretical descriptions for higher spin particles were already introduced and discussed in the 40s of the last century by Fierz and Pauli \cite{Fierz:1939ix} and Rarita and Schwinger  \cite{Rarita:1941mf}, the question how to introduce couplings in a consistent way is still discussed. When electromagnetic interactions are introduced via minimal coupling acausal propagations \cite{Velo:1969bt} and non positive definite anticommutation relations \cite{Johnson:1960vt} arise. In \cite{Cox:1989hp} it was shown that both inconsistencies appear because the interaction violates the proper number of degrees of freedom (DOF) of the free theory.

A first analysis for hadronic interactions was performed in the 70s. A still widely used coupling for $N \Delta \pi$ was proposed in \cite{Nath:1971wp} is inconsistent because it leads to acausal propagation \cite{Singh:1973gq} and non positive definite anti-commutation relations \cite{Hagen:1972ea}. These inconsistencies arise because the interaction violates the number of DOF of the free theory \cite{Cox:1989hp}. Many general forms of interactions were ruled out on this ground. But still fully relativistic calculations were performed with these "inconsistent" couplings leading to reasonable description of  experimental data \cite{Korpa:1997fk}. At the same time an idea was proposed how to introduce "consistent" couplings \cite{Pascalutsa:1999zz}. It is based on the finding that interactions which have the same symmetries as the massless free theory will not violate the DOF of the theory. The mass term is introduced to break the symmetry and rise the number of DOF to the correct value. Because the interaction does not introduce further DOF the mass term which breaks the symmetry correctly in the free case will do it also in the interacting case. Such an approach always leads to a consistent interaction and the spin 1/2 and spin 3/2 parts of the interaction are clearly separated.

So far such calculations were performed only for the $\Delta(1232)$ isobar. The corresponding calculation for the $N^*(1520)$ is a non trivial extension for two reasons. First, it is a particle with negative parity leading to the above mentioned complications. Second, the $N^*(1520)$ decays also into unstable particles leading to a more complicated structure of the selfenergy.

In this work we will investigate the $\Delta(1232)$ and $N^*(1520)$ by calculating the full relativistic structure of the propagator. This will be done in a framework proposed by Pascalutsa and Timmermans \cite{Pascalutsa:1999zz}. The aim is to find out how feasible or how complicated it is to calculate fully relativistic propagators in the Pascalutsa framework. One interesting finding will be that the selfenergies are actually simpler compared to the conventional approach. While our main emphasis is on the $N^*(1520)$ we will also discuss the $\Delta(1232)$ in some detail since one of the decay channels of the $N^*(1520)$ is just $\pi \Delta$ \cite{PDBook}. Therefore we need a correct description of the $\Delta$ anyhow for a proper treatment of the $N^*(1520)$.

For a further understanding of higher spin particles comparison with the experiment will be important. For a description of baryonic resonances formed e.g.~in $\pi N$ collisions a coupled-channel approach is needed including also background terms from t-channels etc. Typically such calculations do not consider the full selfenergy structure of the baryon resonances. To give an example, in K-matrix calculations \cite{Penner:2002ma} only the on-shell part of the involved two-particle propagators is taken into account. In this way analyticity is violated to some extent. The present work aims at a fully relativistic calculation of baryon resonance properties respecting all constraints of a local field theory like analyticity and unitarity. (Note that also a K-matrix calculation is unitary by construction.) On the other hand, as compared to coupled-channel calculations, the present work is more modest by concentrating mainly on the $N^*(1520)$ and the $\Delta(1232)$. Therefore a detailed description of scattering data is beyond the scope of the present work.

The work is structured in the following way: In section \ref{sec:bosonspec} the spectral function for a scalar particle is presented. It is the simplest case and serves as an example how to treat spectral functions and read off mass and width of the resonance. The spin 3/2 case is discussed in section \ref{sec:fermprop}. We derive the free propagator and include interactions. From the analytical structure of the full propagator the spectral function is determined. In section \ref{D13channels} the couplings for the three main decay channels of the $N^*(1520)$ are introduced and their selfenergies calculated. This is presented in a somewhat more general way by including not only the $N^*(1520)$ resonance but also particles with positive parity and different isospin. To give a better understanding of the results the non-relativistic limit of the widths is calculated and compared to non-relativistic calculations. The specific results for the $\Delta(1232)$ and $N^*(1520)$ are presented in section \ref{sec:results}. In the beginning the parameters used for the calculations are determined. Next the selfenergies are calculated and discussed for different cut-off parameters. The widths are discussed next followed by the spectral functions. Different parameters are discussed and they are compared to simplified versions. In section \ref{sec:summary} a summary of the main findings is given.

\section{Pedagogical Introduction to Spectral Functions}
\label{sec:bosonspec}

The propagator, or more precisely the Feynman propagator of an arbitrary field $\phi$ is defined as the two-point correlation function or two-point Green's function \cite{Peskin:1995ev}
\begin{equation}
 \mathcal{D}_0(x,y) = - i \langle \Omega | T \phi(x) \phi(y) |\Omega \rangle \label{eq:FeynProp}.
\end{equation}
Here $\Omega$ is understood as the vacuum of the theory which in general will differ for interacting and non-interacting theories.

For a non-interacting scalar field of mass $m$ there are analytic solutions of the fields $\phi(x)$. Then it is possible to calculate the two-point function in (\ref{eq:FeynProp}) and its Fourier transform: 
\begin{equation}
\mathcal{D}^0_F(p^2)  = -i  \int d^4 x e^{i px } \langle \Omega| T \phi(x)  \phi(0) |\Omega \rangle 
		= \frac{1}{p^2 - m^2 + i \epsilon} \label{eq:D} \quad.
\end{equation}

To calculate a propagator for an interacting theory a resummation is appropriate. This can be done consistently using the Schwinger-Dyson equation:
\begin{equation}
\mathcal{D}(q^2) = \mathcal{D}_0(q^2) + \mathcal{D}_0(q^2) \Sigma(q^2) \mathcal{D}(q^2). \label{eq:Dyson}
\end{equation}

$\Sigma(q^2)$ is the selfenergy of the particle including all possible decay channels. In the case of a scalar field the geometric series of (\ref{eq:Dyson}) can be summed up. To highlight the character of the mass as a bare mass we rewrite $m \to m_0$. Then the result of the resummation reads:  
\begin{equation}
\mathcal{D}(q^2) = \frac{\mathcal{D}_0(q^2)}{1 - \mathcal{D}_0(q^2) \Sigma(q^2)} 
	= \frac{1}{q^2 - m_0^2 - \Sigma(q^2)} \label{eq:Ddressed} \quad.
\end{equation}
In the K\"allen-Lehmann representation \cite{BjorkenDrell:1967} the propagator is written in the form:
\begin{equation}
\mathcal{D}(q^2) = \int_0^{\infty} d \sigma^2 \rho(\sigma^2) \frac{1}{q^2 - \sigma^2 + i\epsilon} \quad. \label{eq:bosonProp}
\end{equation}
The quantity $\rho(q^2)$ is called the spectral function of the boson. It has some properties
\begin{align}
\rho(q^2) \text{ is real} &,\notag \\ 
\rho(q^2) \geq 0 &, \notag \\
\int_{0}^{\infty} dq^2 \rho(q^2) = 1 &. \notag
\end{align}
For stable bosons one sees immediately from equation (\ref{eq:bosonProp}) that the spectral function is given as
\begin{equation*}
\rho(q^2) = \delta(q^2 - m^2) .
\end{equation*}
For unstable particles the spectral function can be calculated using the property 
\begin{equation}
\Im \frac{1}{q^2 - \sigma^2 + i\epsilon} = - \pi \delta(q^2 - \sigma^2). \label{eq:deltasigma}
\end{equation}
Taking the imaginary part of the Feynman propagator leads to
\begin{equation*}
\Im \mathcal{D}(q^2) = \int_0^{\infty} ds \rho(s) \Im \frac{1}{q^2 - s + i\epsilon}
\end{equation*}
since $\rho(q^2)$ is a real quantity. Using equation (\ref{eq:deltasigma}) one gets
\begin{equation*}
\rho(q^2) = - \frac{1}{\pi} \Im \mathcal{D}(q^2) .
\end{equation*}
With the representation (\ref{eq:Ddressed}) of the dressed propagator the bosonic spectral function can be obtained as:
\begin{equation}
\rho(q^2) = - \frac{1}{\pi} \Im \mathcal{D}(q) = - \frac{1}{\pi} \frac{ \Im \Sigma}{(q^2 - m_0^2 - \Re \Sigma)^2 + \Im \Sigma^2} \quad. \label{eq:Aboson}
\end{equation}

The selfenergy or the spectral function fully describes the resonance. From them all measurable quantities can be calculated, e.g.~phase shifts. On the other hand, selfenergies and spectral functions are not directly measurable quantities. As pointed out in the introduction a detailed description of phase shifts is not our intention here. On the other hand, we have to fix our parameters introduced below, namely masses and coupling constants. Therefore we introduce simpler quantities to characterize a resonance which are closer to theory than phase shifts and closer to experiment than spectral functions. These quantities, the mass $m_R$ of the resonance and its (on-shell) width $\Gamma$, are only two numbers instead of a full spectral shape. The price to pay for such an oversimplification is an ambiguity how to define mass and width. In the following we will define these quantities by comparing our spectral function (\ref{eq:Aboson}) to a relativistic Breit-Wigner form  \cite{PDBook} 
\begin{equation}
S = \frac{1}{\pi} \frac{\sqrt{s} \, \Gamma}{(s- m_R^2)^2 + s \, \Gamma^2}  \label{eq:BWFormRho}
\end{equation}
where $\Gamma$ is the width  and $m_{R}$ the physical mass of the resonance. Note that this is not the only possibility. We could have taken as well e.g.~the peak position of $\rho$ to define the mass. In turn this means that for complicated selfenergies it should not be too surprising if the mass as we defined it deviates to some extent from the peak position.

A direct comparison of the Breit-Wigner form and the spectral function in equation (\ref{eq:Aboson}) is not possible due to the more complicated structure of the spectral function. On the other hand, it is convenient to define the particle's mass and width at the point where the real part of the inverse propagator, i.e. $s-m_0^2- \Re \Sigma(s)$ vanishes. Then we can expand the denominator of $\rho$ around this point, which is the physical mass $m_R$:
\begin{equation}
s - m_0^2 - \Re \Sigma \sim \frac{1}{c} (s - m_R^2 ) + O((s-m_R^2)^2) \label{eq:taylorexrho}.
\end{equation}
The coefficient $c$ can be extracted as the first derivative of $s-m_0^2 - \Re \Sigma$ on the mass shell:
\begin{equation}
\frac{1}{c} = \left. \frac{d}{ds} (s-m_0^2 - \Re \Sigma) \right|_{s=m^2_R} = 1 - \left. \frac{d}{ds} Re \Sigma \right|_{s=m^2_R}. \label{eq:defc}
\end{equation}
Then the spectral function has approximately the form
\begin{equation}
\label{eq:Arhoentwickelt}
 \rho(s) = - \frac{1}{\pi} \frac{ \Im \Sigma}{(s - m_0^2 - \Re \Sigma)^2 + \Im \Sigma^2}
			 \sim  - \frac{1}{\pi} \frac{ c^2 \Im \Sigma}{(s - m_R^2)^2 + c^2 \Im \Sigma^2} 
			 =  - \frac{c}{\pi} \frac{ c \Im \Sigma}{(s - m_R^2)^2 + (c \Im \Sigma)^2} 
\quad .
\end{equation}
Now the physical mass $m_R$ and the width $\Gamma$ can be extracted. This is achieved for the mass by  putting  (\ref{eq:taylorexrho})  on the mass shell
\begin{equation*}
m_R^2 - m_0^2 - \Re \Sigma(m_R^2) = 0.
\end{equation*}
One can interpret this relation such that the bare mass of the scalar particle is shifted by the real part of the selfenergy leading to the physical mass $m_R$. 

Also the width can be read off by comparing (\ref{eq:Arhoentwickelt}) to the Breit-Wigner form (\ref{eq:BWFormRho}).  One can read off the width of the  \rhomeson from the denominator or numerator leading to the same quantity 
\begin{equation}
\Gamma = - \left. \frac{c(s)}{\sqrt{s}} Im \Sigma(s) \right|_{s=m_R^2}. \label{eq:GammaRho}
\end{equation}
Note that without the additional $c$ in front of the Breit-Wigner type form in (\ref{eq:Arhoentwickelt}) the spectral function would be normalized incorrectly. The appearance of $c$ in (\ref{eq:GammaRho}) leads to a squeezed or an enlarged width, depending whether $c$ is larger or smaller than one. Or looking at the definition (\ref{eq:defc}) of $c$ the spectral function will be squeezed (enlarged) if the derivative of the real part of the selfenergy is larger (smaller) than zero. This is clear because then the real part of the selfenergy will increase (decrease).

Below we will meet much more complicated spectral functions caused by the intrinsic spin structure of non-scalar particles. In such a case one is tempted to replace the complicated structure by a simpler one, e.g. by something like (\ref{eq:BWFormRho}), but with an energy dependent width deduced from phase space considerations. It is one purpose of the present work to compare the full relativistic structure with such simplifications to judge the validity of the latter. For scalar particles by taking the structure of the spectral function and the width one can introduce a "pseudo-relativistic" spectral function in a Breit-Wigner form
\begin{equation}
\mathcal{A}_{pseudo} = \frac{1}{N} \frac{\sqrt{s} \Gamma_{pseudo}}{(s- m^2_R)^2 + s \Gamma^2_{pseudo}}  \label{eq:BWFormRhoPseudo}
\end{equation}
with $N$ as a normalization factor requiring
\begin{equation*}
1 = \int_{0}^{\infty} ds \mathcal{A}_{pseudo}(s) .
\end{equation*}
The width is given by the appropriate phase space behavior:
\begin{equation}
\Gamma_{pseudo} = a \vec{q}^{2 l+1} \label{eq:guessedwidht}
\end{equation}
and $a$ is determined by the requirement that on the mass shell this width is equal to the respective value given in \cite{PDBook} for the particle under consideration. 

\section{The Spin 3/2 Propagator}
\label{sec:fermprop}

From the Lagrangian of the free Rarita-Schwinger field derived in \cite{Pascalutsa:1998pw} we can extract the propagator as the Green's function of the equation \cite{deJong:1992wm}
\begin{equation}
\label{eq:green}
\left\{ \sigma_{\mu \nu}, ( \slashp - M ) \right \} \mathcal{G}_{0}^{ \nu \tau} = g{_{\mu}}{^{\tau}} .
\end{equation}
The free Rarita-Schwinger propagator can be expanded in the basis of projection operators of the spin states. The definition used here is taken from \cite{deJong:1992wm} apart from a misprint there. The projection operators are given by
\begin{align}
P^{3/2} =& g^{\mu \nu} - \frac{1}{3} \gamma^{\mu} \gamma^{\nu} - \frac{1}{3 p^2} (\slashp \gamma^{\mu} p^{\nu} + p^{\mu} \gamma^{\nu}  \slashp),\label{eq:projop}\\
P^{1/2}_{11} =& \frac{1}{3} \gamma^{\mu} \gamma^{\nu} - \frac{p^{\mu} p^{\nu}}{p^2} + \frac{1}{3 p^2} (\slashp \gamma^{\mu} p^{\nu} + p^{\mu} \gamma^{\nu}  \slashp), \notag \\
P^{1/2}_{22} =& \frac{p^{\mu} p^{\nu} } {p^2}. \notag
\end{align}
For a complete system one also needs
\begin{align}
P^{1/2}_{12} =& \frac{1}{\sqrt{3} p^2} (p^{\mu} p^{\nu} - \slashp \gamma^{\mu} p^{\nu}), \notag \\
P^{1/2}_{21} =& \frac{1}{\sqrt{3} p^2} (\slashp p^{\mu} \gamma^{\nu} - p^{\mu} p^{\nu}). \label{eq:P1221}
\end{align}
This set of projection operators satisfies the orthonormality and completeness conditions \cite{VanNieuwenhuizen:1981ae}
\begin{align}
(P^I_{ij})_{\mu \nu} (P^J_{kl})^{\nu \delta} =& \delta^{IJ} \delta_{jk} (P^J_{il})_{\mu} {}^{\delta}, \label{eq:orthonormality} \\
P^{3 / 2} + P^{1/2}_{11} + P^{1/2}_{22} =& g^{\mu \nu}. \label{eq:vollstaendig}
\end{align}
Expanding the propagator in these projection operators reads
\begin{equation*}
\mathcal{G}^{\mu \nu}_0 = A \, P^{3/2} + B \, P_{11}^{1/2} + C \, P_{22}^{1/2} + D \, P_{12}^{1/2} + E \, P_{21}^{1/2} 
\end{equation*}
where the Lorentz indices of the projection operators are omitted for simplicity.

Also the operator in equation (\ref{eq:green}) can be written in this basis leading to an equation for the propagator of the form
\begin{equation*}
\left [ (\slashp - M) P^{3/2} - 2 (\slashp - M) P_{11}^{1/2} + \sqrt{3} M P_{12}^{1/2} + \sqrt{3} M P_{21}^{1/2} \right] \mathcal{G}^{\mu \nu}_0 \\= P^{3/2} + P_{11}^{1/2} + P_{22}^{1/2} .
\end{equation*}
On the right hand side we used equation (\ref{eq:vollstaendig}), the completeness of these projection operators. Using also the orthonormality conditions (\ref{eq:orthonormality}) it is possible to extract the solutions
\begin{align*}
A &= \frac{\slashp + M}{p^2 - M^2}, \\
B &= 0, \\
C &= - \frac{2}{3 M^2} ( \slashp + M), \\
D &=  \frac{1}{\sqrt{3} M}, \\
E &= D, \\
\end{align*}
leading to the free propagator of the form
\begin{equation}
\mathcal{G}_0^{\mu \nu} = \frac{\slashp + M}{p^2 - M^2} P^{3/2} -  \frac{2}{3 M^2} ( \slashp + M) P_{22}^{1/2} + \frac{1}{\sqrt{3} M} (P_{12}^{1/2} + P_{12}^{1/2}) . \label{eq:RSfreeProp}
\end{equation}
When inserting the projection operators the propagator can be written in a form often found in the literature \cite{Korpa:1997fk}\cite{Pascalutsa:1999zz}
\begin{equation*}
\mathcal{G}_0^{\mu \nu} = \frac{\slashp + M}{p^2 - M^2} \left[ g^{\mu \nu} - \frac{1}{3} \gamma^{\mu} \gamma^{\nu \nu} - \frac{2}{3 M^2} p^{\mu} p^{\mu} + \frac{1}{3M} p^{\mu} \gamma^{\nu} - p^{\nu} \gamma^{\mu} \right].
\end{equation*}

An interesting observation is that only the part of the propagator proportional to the spin 3/2 state has a pole structure (first term on the right hand side of equation (\ref{eq:RSfreeProp})). Thus, on-shell the propagator only propagates spin 3/2 fields.
This propagator can be simplified in the Pascalutsa formalism. Let for a given two-body scattering Lagrangian the vertices be $\Gamma^{\mu}(p,p-q,q)$ with $p$ and $q$ as the momenta of the resonance and one of the scattered particles, respectively. In the Pascalutsa formalism the vertices will satisfy the relation \cite{Pascalutsa:1999zz}
\begin{equation}
p_{\mu} \Gamma^{\mu}(p,p-q,q) = 0.
\end{equation}
Scattering amplitudes will have the form
\begin{equation}
\Gamma^{\mu}(p,p-q^{\prime},q^{\prime}) G_{\mu \nu} \Gamma^{\nu}(p,p-q,q).
\end{equation}
Examining equations (\ref{eq:projop}) and (\ref{eq:P1221}) on sees that the whole spin 1/2 sector drops out. The effective propagator of the spin 3/2 field in the Pascalutsa formalism is then given as
\begin{equation}
\mathcal{G}^{\mu \nu}_{\rm eff} = \frac{1}{\slashp - M + i \epsilon} P^{3/2}  = \mathcal{G} P^{3/2} \label{eq:dressedspindreihalbeprop}
\end{equation}
where in the last step $\mathcal{G}$ is the spin 1/2 propagator. This means that the effective propagator of the spin 3/2 fields can be written as the propagator of the spin 1/2 field multiplied by the projection operator $P^{3/2}$.

\begin{figure}
	\epsfig{file=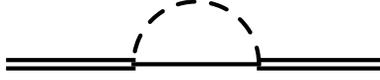,height=1cm}
	\caption{Feynman graph for a selfenergy. The double line denotes the resonance, the other lines the decay products. \label{fig:sigmanpi}}
\end{figure}

The full propagator in the presence of interactions is calculated as in equation (\ref{eq:Dyson}) by the Schwinger-Dyson equation:
\begin{equation}
\mathcal{G}^{\mu \nu}(p^2) = \mathcal{G}^{\mu \nu}_0(p^2) + \mathcal{G}^{\mu \alpha}_0(p^2) \Sigma_{\alpha \beta}(p^2) \mathcal{G}^{\beta \nu} (p^2). \label{eq:DysonSpin32}
\end{equation}
The selfenergy $\Sigma^{\mu \nu}$ is as the propagator a Dirac quantity with a Lorentz structure. 
Due to the completeness of the spin 3/2 projection operators, the selfenergy can be written as
\begin{equation*}
\Sigma_{\mu \nu} = P^{3/2} a + P^{1/2}_{11} b  + P^{1/2}_{22} c + P^{1/2}_{12} d +P^{1/2}_{21} e
\end{equation*}
with the Dirac matrices
\begin{equation*}
a = a_s(p^2) + \slashp a_v(p^2)
\end{equation*}
and the corresponding decomposition for b-e.
The gauge invariant structure of the Pascalutsa interaction \cite{Pascalutsa:1999zz} leads to a transverse selfenergy
\begin{equation*}
p_{\mu} \Sigma^{\mu \nu}(p) = p_{\nu} \Sigma^{\mu \nu} (p) = 0.
\end{equation*}
This can be written as
\begin{equation*}
(P^{1/2}_{22})_{\alpha \mu} \Sigma^{\mu \nu} =0= \Sigma^{\mu \nu} (P^{1/2}_{22})_{\nu \beta}.
\end{equation*}
Then some of the terms given above drop out. Since 
\begin{equation*}
(P^{1/2}_{22}) (P^{1/2}_{21}) = (P_{21}^{1/2})
\end{equation*}
 and 
\begin{equation*}
(P^{1/2}_{12}) (P^{1/2}_{22}) = (P_{12}^{1/2})
\end{equation*}
we find
\begin{align*}
P^{1/2}_{22} c + P^{1/2}_{21} e &= 0, \\
P^{1/2}_{22} c + P^{1/2}_{12} d &= 0.
\end{align*}
Contracting the upper (lower) equation with $P^{1/2}_{11}$ from right (left) leads to $e=0$ and $d=0$ 
since  
\begin{equation*}
(P^{1/2}_{21}) (P^{1/2}_{11}) = (P_{21}^{1/2})
\end{equation*}
and 
\begin{equation*}
(P^{1/2}_{11}) (P^{1/2}_{12}) = (P_{12}^{1/2}).
\end{equation*}
Then also $c=0$ and the number of coefficients reduces to two
\begin{equation}
\Sigma_{\mu \nu} = P^{3/2} (a_1  + \slashp a_2)  + P^{1/2}_{11} (a_3  + \slashp a_4). \label{eq:sigmamunu}
\end{equation}
We define for later purposes
\newcommand{\Tr}{\operatorname{Tr}}
\begin{equation} 
\Sigma = \frac{1}{2} P^{3/2}_{\mu \nu} \Sigma^{\mu \nu}  = a =  \Sigma_s + \slashp \Sigma_v . \label{eq:sigma}
\end{equation}
If $\Sigma_{\mu \nu}$  is known (e.g. calculated from the Feynman graph \ref{fig:sigmanpi}) the coefficients $a_i$ can be determined via traces over $\Sigma_{\mu \nu}$:
\begin{align}
  a_1 &= \frac{1}{8} \Tr\left( \Sigma \sups{\mu} \subs{\mu} - \frac{1}{3} \gamma_{\nu} \gamma_{\mu} \Sigma^{\mu \nu} \right), \notag \\
  a_2 &= \frac{1}{8 p^2} \Tr\left( \slash{p} \Sigma \sups{\mu} \subs{\mu} - \frac{1}{3} \slash{p} \gamma_{\nu} \gamma_{\mu} \Sigma^{\mu \nu} \right), \notag \\
  a_3 &= \frac{1}{12} \Tr\Bigl(\gamma_{\nu} \gamma_{\mu} \Sigma^{\mu \nu} \Bigr), \label{eq:ai} \\
  a_4 &= \frac{1}{12} \Tr\Bigl(\slash{p} \gamma_{\nu} \gamma_\mu \Sigma^{\mu \nu} \Bigr) \notag .
\end{align}
This is a striking result because in the conventional approach all ten coefficients are needed \cite{Korpa:1997fk}. We can conclude already here that it is not only feasible to work in the Pascalutsa framework but much easier at least concerning the spin structure of the selfenergy.

After resummation by the Schwinger-Dyson equation the full propagator reads
\begin{equation}
\mathcal{G}^{ \mu \nu} (p) = \frac{1}{\slashp - M_0 -\Sigma} P^{3/2} + \frac{1}{3 M_0^2} (-2(\slashp + M_0)-\tilde{b}) P^{1 / 2}_{22} + \frac{1}{\sqrt{3} M_0} (P^{1/2}_{12} + P^{1/2}_{21}) \label{eq:fullPropSpin32}
\end{equation}
with $\Sigma$ and $\tilde{b} = a_3 - \slashp a_4$ taken from (\ref{eq:sigma}) and (\ref{eq:sigmamunu}).
Also in the interacting case on the level of amplitudes only the first term in equation (\ref{eq:fullPropSpin32}) contributes. It is the full propagator of a spin 1/2 particle
\begin{equation}
\mathcal{G}(p) = \frac{1}{\slashp - M_0  - \Sigma(p)} \label{eq:Gdressed},
\end{equation}
multiplied by the spin 3/2 projection operator. The structure of this propagator has the form \cite{Henning:1994qz}
\begin{equation}
\mathcal{G}(p) = \frac{F_s + \slashp F_v}{(\tilde{s} - \tilde{M}^2 )^2 + Z^2} \label{defG}
\end{equation}
with the quantities ($s=p^2$):
\begin{align}
\tilde{s}&= s \left[ \left(1-\Re \Sigma_v\right)^2 - \left(\Im \Sigma_v \right)^2 \right], \label{eq:stilde} \\
\tilde{M}^2&= \left(M_0+\Re \Sigma_s\right)^2- \left(\Im \Sigma_s\right)^2, \label{eq:M2tilde}\\
Z &=  2 \left[s \left(1-\Re \Sigma_v \right) \Im \Sigma_v + \left(M_0+\Re \Sigma_s \right) \Im \Sigma_s \right] \label{eq:Zdef}, \\
Re F_s &= \left(M_0 + \Re \Sigma_s\right)\left(\tilde{s} - \tilde{M}^2\right) - \Im \Sigma_s Z, \\
Im F_s &=  \Im \Sigma_s \left(\tilde{s} - \tilde{M}^2 \right)+ \left(M_0+\Re \Sigma_s \right) Z, \\
Re F_v &= \left(1 - \Re \Sigma_v \right) \left(\tilde{s} - \tilde{M}^2 \right) + \Im \Sigma_v Z, \\
Im F_v &=  -\Im \Sigma_v \left(\tilde{s} - \tilde{M}^2 \right)+ \left(1-\Re \Sigma_v \right) Z. \label{eq:ImFvdef}
\end{align}
The spectral representation of the Feynman propagator of spin 1/2 particles can be written as \cite{BjorkenDrell:1967}
\begin{equation}
\mathcal{G}(p^2) = \int_0^{\infty} ds [\slashp \rho_v(s) + \rho_s(s)] \frac{1}{p^2 - s + i \epsilon} \quad. \label{eq:defrhos}
\end{equation}
$\rho_v$ and $\rho_s$ are scalar functions and have some fundamental properties:
\begin{align}
\rho_v(p^2) \text{ and } \rho_s(p^2) \text{ are both real}&, \label{eq:rho1rho2} \\ 
\rho_v(p^2) \geq 0&, \label{eq:rhovgeq0} \\
\sqrt{p^2} \rho_v(p^2) - \rho_s(p^2) \geq 0&. \label{eq:rho1rho2ende}
\end{align}
A normalization condition can be derived from the quantization condition of the fields \cite{Post:2003Phd}
\begin{equation}
\int_{0}^{\infty} ds \rho_v(s) = 1. \label{eq:normferm}
\end{equation}
Equation (\ref{eq:rhovgeq0}) and (\ref{eq:normferm}) suggest that $\rho_v$ is the fermionic quantity which is closest to the spectral function of the simpler and therefore more intuitive bosonic case. Therefore when discussing results we will mainly concentrate on $\rho_v$.

For the case of spin 3/2 fields it is a delicate task to perform a general spectral representation \cite{Korpa:1997fk}\cite{deJong:1992wm}. But in the Pascalutsa framework the effective propagator of the spin 3/2 fields has a similar structure as the spin 1/2 fields. The effective spectral representation of spin 3/2 fields in the Pascalutsa formalism can then be written as
\begin{equation*}
\mathcal{G}^{\mu \nu}_{\rm eff} (p^2) = \int_0^{\infty} ds [\slashp \rho_v(s) + \rho_s(s)] \frac{1}{p^2 - s + i \epsilon} P^{3/2}(p^2).
\end{equation*}
This means that in the Pascalutsa framework the spectral functions for spin 1/2 and spin 3/2 particles basically have the same structure which is a great simplification.

For stable spin 1/2 particles the spectral functions can be read off immediately from equation (\ref{eq:defrhos}) as
\begin{align}
\rho_v(p^2) &= \delta(p^2 - M^2), \label{eq:rhovtodelta}\\
\rho_s(p^2) &= \sqrt{p^2} \delta(p^2-M^2). \label{eq:rhostodelta}
\end{align}
Generally the spectral functions can be calculated using equation (\ref{eq:deltasigma}). Because the spectral function is defined as the imaginary part of the propagator one needs to clarify what the imaginary part of a Dirac quantity is. The imaginary part is defined via the hermitian rather than the complex conjugate \cite{BjorkenDrell:1967}\cite{Post:2003}:
\begin{equation*}
\Re \mathcal{G}(p) = \frac{1}{2} ( \mathcal{G}(p) + \gamma_0 \mathcal{G}(p)^{\dagger} \gamma_0 ) \text{ , } \Im \mathcal{G}(p) = \frac{1}{2i} ( \mathcal{G}(p) - \gamma_0 \mathcal{G}(p)^{\dagger} \gamma_0 ).
\end{equation*}
This definition treats $\slashp$ as a real quantity. Now it is possible to calculate the imaginary part of the propagator from the K\"allen-Lehmann representation  
\begin{equation}
\Im \mathcal{G}(p^2) = -\pi \, \left[\slashp \rho_v(p^2) + \rho_s(p^2) \right].
\end{equation}
For spin 3/2 particles the spectral functions can be extracted as specific traces over the imaginary part of the propagator (\ref{eq:fullPropSpin32}):
\begin{align*}
\rho_v(p^2) &= - \frac{1}{8 \pi p^2 } \Tr[\slash{p} \, \Im \mathcal{G}_{\rm eff}^{\mu \nu}(p^2) P^{3/2}_{\mu \nu}], \\
\rho_s(p^2) &= - \frac{1}{8 \pi } \Tr[\Im \mathcal{G}^{\mu \nu}_{\rm eff}(p^2) P^{3/2}_{\mu \nu}].
\end{align*}

Having calculated the dressed propagator the analytic structure of the spectral functions $\rho$ can be easily derived using the abbreviations (\ref{eq:stilde})-(\ref{eq:ImFvdef}).
\begin{align}
\rho_v(s) &= - \frac{1}{\pi} \frac{ \Im F_v }{(\tilde{s} - \tilde{M}^2 )^2 + Z^2}, \label{eq:rhov}\\ 
\rho_s(s) &= - \frac{1}{\pi} \frac{ \Im F_s }{(\tilde{s} - \tilde{M}^2 )^2 + Z^2}. \label{eq:rhos}
\end{align}
Because $\rho_v$ ($\rho_s$) and $\Im F_v$ ($\Im F_s$) are proportional to each other the relations (\ref{eq:rho1rho2})-(\ref{eq:rho1rho2ende}) also hold for $-\Im F_s$ and $-\Im F_v$:
\begin{align}
\Im F_v(s) \text{ and } \Im F_s(s) \text{ are both real} &, \notag \\ 
-\Im F_v(s) \geq 0 &, \label{eq:ImFVgeq0}\\
\Im F_s(s) - \sqrt{s} \Im F_v(s)  \geq 0 \notag &.
\end{align}

For later use we define the Bjorken-Drell function as:
\begin{equation}
BD(s) = \Im F_s(s) - \sqrt{s} Im F_v(s). \label{eq:BDfunc}
\end{equation}
Note that this function must not get negative.

As discussed already for the bosonic case an extraction of the mass and the width of the fermionic resonance is desirable to determine our input parameters. We recall the discussion in section \ref{sec:bosonspec} that there are several possibilities to define mass and width. Because the comparison with the Breit-Wigner form (\ref{eq:BWFormRho}) worked well for the bosonic case we also want to  use it for the fermionic case. This will be more complicated because the denominator of the spectral function is much more involved and not only one spectral function exists but two.

To put the spectral functions in a more convenient form for comparing to the Breit-Wigner form we expand the first term of the denominator in  (\ref{eq:rhov}) and (\ref{eq:rhos}) around the physical mass $M_R$ of the resonance:
\begin{equation}
\tilde{s} - \tilde{M}^2 \sim \frac{1}{c} (s - M_R^2) + O((s-M_R^2)^2).\label{eq:taylorexpsmtilde}
\end{equation}
The physical mass $M_R$ is defined such that for $\sqrt{s} = M_R$:
\begin{equation}
\tilde{s}(M_R^2) - \tilde{M}^2(M_R^2) = 0. \label{eq:defmass}
\end{equation}
We recall that using this definition the physical mass $M_R$ in general cannot be read off as the peak of a spectral function. The bare mass can be extracted when inserting the definitions for $\tilde{M}^2$ from equation (\ref{eq:M2tilde}):
\begin{equation}
\label{eq:M0}
M_0 = \sqrt{\tilde{s}(M_R^2) + Im \Sigma_s(M_R^2)^2} - Re \Sigma_s(M_R^2).
\end{equation}
The coefficient $c$ of the Taylor expansion (\ref{eq:taylorexpsmtilde}) is given by the first derivative of $\tilde{s} - \tilde{M}^2$:
\begin{equation*}
c = \left[ \frac{d}{ds} \left(\tilde{s}(s, \Im \Sigma_v(s), \Re \Sigma_v(s)) - \tilde{M}^2(s, \Im \Sigma_s(s), \Re \Sigma_s(s)) \right) \right]_{s = M_R^2}^{-1} \quad .
\end{equation*}
The spectral functions $\rho_v$ (\ref{eq:rhov}) and $\rho_s$ (\ref{eq:rhos}) can be approximated around $\sqrt{s} \approx M_R$ by
\begin{align*}
\rho_v &\approx - \frac{c}{\pi} \frac{c \, \Im F_v}{(s - M^2_R)^2 + c^2 Z^2} \quad, \\
\rho_s &\approx - \frac{c}{\pi} \frac{c \, \Im F_s}{(s - M^2_R)^2 + c^2 Z^2} \quad .
\end{align*}
As in the bosonic case the width can be read off by comparison with the Breit-Wigner form defined in equation (\ref{eq:BWFormRho}). Due to the complicated structure of the spectral functions not only one possible definition of a width exists but three:
\begin{align*}
\Gamma_Z(s) &= - \frac{c}{\sqrt{s}} Z(s), \\
\Gamma_V(s) &= - \frac{c}{\sqrt{s}} \Im F_v(s), \\
\Gamma_S(s) &= - \frac{c}{\sqrt{s} M_0} \Im F_s(s).
\end{align*}
$\Gamma_Z$ is read off from the denominator, whereas $\Gamma_V$ and $\Gamma_S$ are read off from the numerator of $\rho_v$ and $\rho_s$, respectively. In the bosonic case discussed in section \ref{sec:bosonspec} such an ambiguity was not present. The factor $\frac{1}{M_0}$ in the case of $\Gamma_S$ is motivated by the fact that a width has energy as the proper unit. 

When neglecting the real parts of the selfenergy and quadratic terms one can write these widths as:
\begin{align*}
\Gamma_Z &= - \frac{c}{\sqrt{s}} 2 \left[s \left(1-\Re \Sigma_v \right) \Im \Sigma_v + \left(M_0+\Re \Sigma_s \right) \Im \Sigma_s \right] \\
	& \approx - \frac{2c}{\sqrt{s}} \left[ s \Im \Sigma_v + M_0 \Im \Sigma_s \right], \\
\Gamma_V &= - \frac{c}{\sqrt{s}} \left[ -\Im \Sigma_v (\tilde{s} - \tilde{M}^2 )+ \left(1-\Re \Sigma_v \right)  Z \right] \\
	& \approx - \frac{c}{\sqrt{s}} \left[ - \Im \Sigma_v (\tilde{s} - \tilde{M}^2 ) + Z \right] \\
	& = \frac{c}{\sqrt{s}} \Im \Sigma_v (\tilde{s} - \tilde{M}^2) + \Gamma_Z, \\
\Gamma_S &= - \frac{c}{\sqrt{s} \, M_0} \left[ \Im \Sigma_s (\tilde{s} - \tilde{M}^2 ) + \left(M_0+\Re \Sigma_s \right)  Z \right] \\
	& \approx - \frac{c}{\sqrt{s} \, M_0} \left[ \Im \Sigma_s (\tilde{s} - \tilde{M}^2) + M_0 Z \right] \\
	& =  - \frac{c}{\sqrt{s} \, M_0}  \Im \Sigma_s (\tilde{s} - \tilde{M}^2) + \Gamma_Z .
\end{align*}
For $\Re \Sigma \to 0$ and $\sqrt{s} = M_R$ 
\begin{equation*}
\Gamma_Z = \Gamma_V = \Gamma_S
\end{equation*}
because on the mass shell $\tilde{s} - \tilde{M}^2=0$. But for large real parts of the selfenergy and away from the mass shell deviations occur. Examples for all three widths will be given in section \ref{sec:results}.

For the plots and fits of this work $\Gamma_V$ will been chosen as a reasonable definition for the width. This choice is motivated from the fact that it is a positive definite quantity which follows from the fact that $-\Im F_v$ is positive definite according to (\ref{eq:ImFVgeq0}).

Further difficulties arise when considering partial widths of a particle. When a particle has more than one decay channel the total selfenergy will be the sum of the selfenergies of each channel. Then the total width of the particle can be extracted by calculating the propagator with the total selfenergy. Note however that the selfenergies enter the width non-linearly as can be seen e.g. by inserting (\ref{eq:stilde})-(\ref{eq:Zdef}) in (\ref{eq:ImFvdef}). Terms up to cubic in the selfenergies appear. Therefore a partial width cannot be defined as the width where one channel is calculated independently neglecting all the other channels. Such a definition leads to a different bare mass for each channel and the sum of the partial widths would differ from the total width calculated using the full selfenergy. Therefore the partial widths must be calculated using the total selfenergy for the real parts of the selfenergy and for nonlinear terms in $Im F_v$ and $Z$. The partial selfenergies are used for the linear terms only. The sum of all partial widths will be equal to the total width. $M_0$ is calculated using the total selfenergy.

\section[Couplings and Selfenergies of a Spin 3/2 Resonance to $N \pi$, $N \rho$ and $\Delta \pi$]{Couplings and Selfenergies of a Spin 3/2 Resonance to $\bf N \pi$, $\bf N \rho$ and $\bf \Delta \pi$}
\label{D13channels}

In this section interaction Lagrangians for a spin 3/2 resonance coupled to $N \pi$, $N \rho$ and $\Delta \pi$ are introduced. According to \cite{PDBook} these are the main decay channels of the $N^*(1520)$. All three interactions are constructed in the Pascalutsa framework. For the $N \rho$ and $\Delta \pi$ channel the selfenergies are calculated also for the case where $\rho$ and $\Delta$ are stable. 

This full calculation is compared to commonly used simplifications where the widths of the particles are approximated by phase space considerations. For the case of the $N \pi$ coupling the width is also compared to a calculation using conventional coupling.

\subsection{Form Factor}

As should have become obvious by now, we spend great efforts in a fully relativistic
treatment of spin 3/2 resonances by exploring the full spin structure of propagators
and selfenergies. We use the Pascalutsa formalism to make sure that also in an
interacting theory only the correct number of degrees of freedom is propagated. We
also insist on (and will check) the fundamental properties of spectral functions
given in (\ref{eq:rho1rho2})-(\ref{eq:normferm}). As we will see below
we will also respect analyticity of the selfenergies by calculating the real part
from the imaginary part via a dispersion relation. 

Concerning the regularization/renormalization of the selfenergies, however, we admit that 
we are somewhat less ambitious. In the traditional sense the interactions which we
will use below are not renormalizable. One could imagine to formulate an effective
field theory which is renormalizable order by order in a given counting scheme.
Unfortunately, such approaches like chiral perturbation theory including 
{\em relativistic} baryons are developed only now \cite{Becher:1999he}\cite{Hacker:2005fh}\cite{Lutz:2001yb} and there is no consensus reached
yet what is the best way of doing this.

Therefore we follow in the present work the well-trotted path of hadronic model builders
and use form factors to tame the infinities. The form factor chosen throughout this work
is \cite{Korpa:1997fk}:
\begin{equation}
FF(s) = exp \left [ - \frac{s_{threshold}-s}{\Lambda^2} \right ] \quad . \label{eq:lambda}
\end{equation}
Below we will study the sensitivity of our results to the chosen value for the cut-off
$\Lambda$.

\subsection[Selfenergy of a Spin 3/2 Baryon $N \pi$ System]{Selfenergy of a Spin 3/2 Baryon $\bf N \pi$ System}
\label{sec:selfenergyNPi}

We present here the results for a spin 3/2 baryon decaying into $N \pi$. The Lagrangian for such a state is given by the following formula, the top (bottom) line for particles with positive (negative) parity:
\begin{equation*}
\mathcal{L}_{R N \pi} = g_{R N \pi} \tilde{\bar{\psi}}^{\mu \nu}_R \gamma_{\mu} \left\{ \begin{matrix} i \\ \gamma_5 \end{matrix} \right\} \spinor_N \partial_{\nu} \pi + h.c.
\end{equation*}
where h.c. denotes hermitian conjugate. $\psi^{\mu \nu}$ is the field strength tensor of the spin 3/2 baryon field and is defined in analogy to the electromagnetic case as
\begin{equation*}
\psi^{\mu \nu} = \partial^{\mu} \psi^{\nu} - \partial^{\nu} \psi^{\mu}
\end{equation*}
and its dual
\begin{equation*}
\tilde{\psi}^{\mu \nu} = \frac{1}{2} \epsilon^{\mu \nu \varrho \sigma} \psi_{\varrho \sigma}.
\end{equation*}
The imaginary part of the selfenergy can be determined by four coefficients $a_i$. They were defined in (\ref{eq:ai}) and are only nonzero for $s > (M_N + m_{\pi})^2$. All have the following form:
\begin{equation}
\Im a_i(s) = \frac{N_I}{8 \pi}  g_{R N \pi}^2 FF(s)^2 \frac{q_*}{\sqrt{s}} b_i(s) . \label{eq:Imai}
\end{equation}
$N_I$ is  $1$ ($3$) for a resonance of isospin $3/2$ ($1/2$). Furthermore we introduce the notation:
\begin{align}
q_*^2 &= \frac{1}{4 s} \left[ (M_N^2 - m_{\pi}^2 -s)^2 - 4 s m_{\pi}^2 \right], \label{eq:impulspi}\\
y &= s + M_N^2 - m_{\pi}^2. \label{eq:energynucleon}
\end{align}
For the functions $b_i(s)$ appearing in (\ref{eq:Imai}) we obtain
\begin{align*}
b_1(s) &= -  \frac{1}{3} P \, q_*^2 \, s \,M_N, \\
b_2(s) &= - \frac{1}{6} q_*^2 y, \\
b_3(s) &= 4 \,  b_1(s), \\
b_4(s) &= 4 \, b_2(s).
\end{align*}
$P$ is $+1$ ($-1$) for particles with positive (negative) parity.
The real part is calculated numerically using the dispersion relation
\begin{equation}
\Re a_i(s) = \frac{1}{\pi} \, \mathcal{P} \int_{(M_N + m_{\pi})^2}^{\infty} d\sigma \frac{\Im a_i (\sigma)}{\sigma - s} \label{eq:reaipi}
\end{equation}
where $\mathcal{P}$ denotes the principal value of the integral.
\begin{figure}
\begin{center}
\epsfig{file=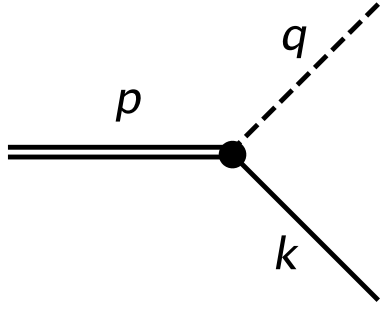,width=5cm}
\end{center}
\caption{Two-body decay.}
\label{fig:decay}
\end{figure}

The physical meaning of equations (\ref{eq:impulspi}) and (\ref{eq:energynucleon}) can be seen when exploring the kinematics of the decay process depicted in figure \ref{fig:decay}. $p$, $k$ and $q$ are the four-momenta of resonance, nucleon and pion, respectively. In the rest frame of the resonance these quantities can be written as
\begin{align}
p = (\sqrt{s}, 0), & \quad p^2 =s, \notag\\
q = (q_0, \vec{q}), & \quad q_0^2 = \vec{q}^2 + m_{\pi}^2, \label{eq:Epion}\\
k = (k_0, -\vec{q}), & \quad k_0^2 = \vec{q}^2 + M_N^2 .\label{eq:Enucleon}
\end{align}
Using momentum conservation it is possible to calculate $\vec{q}^2$ because
\begin{align}
\begin{split}
s &= p^2 = (q + k)^2 = (q_0 + k_0)^2 = q_0^2 + k_0^2 + 2q_0 k_0 = 2 \vec{q}^2 + m^2_{\pi} +M_N^2 + 2 \sqrt{\vec{q}^2 + m_{\pi}^2} \sqrt{\vec{q}^2 + M_N^2} \\
\end{split} \notag\\
\Rightarrow  \vec{q}^2 &= \frac{1}{4s} \left[ (s-m_{\pi}^2 - M_N^2)^2 - 4 m_{\pi}^2 M_N^2 \right] =  \frac{1}{4s} \left[ (s+m_{\pi}^2 - M_N^2)^2 - 4 s m_{\pi}^2 \right] =  \frac{1}{4s} \left[ (s - m_{\pi}^2 + M_N^2)^2 - 4 s M_N^2 \right] . \label{eq:pimom}
\end{align}
Comparing equation (\ref{eq:impulspi}) with (\ref{eq:pimom}) shows that $q_*=|\vec{q}|$ is the momentum of the pion in the rest frame of the resonance. Equation (\ref{eq:energynucleon}) becomes clear when calculating the energy of pion and nucleon using equation (\ref{eq:Epion}), (\ref{eq:Enucleon}) and (\ref{eq:pimom}):
\begin{align*}
k^2_0 =  \vec{q}^2 + M_N^2 = \frac{1}{4s} \left[ (s - m_{\pi}^2 + M_N^2)^2  \right], \\
q^2_0 =  \vec{q}^2 + m_{\pi}^2 =  \frac{1}{4s} \left[ (s + m_{\pi}^2 - M_N^2)^2 \right] .
\end{align*}
The quantity in (\ref{eq:energynucleon}) is proportional to the energy of the nucleon in the rest frame of the resonance.

On the mass shell the conventional and the Pascalutsa interaction are the same \cite{Pascalutsa:2000kd}. The relation between the Pascalutsa coupling $g_{R N \pi}$ and conventional coupling $f_{R N \pi}$ defined in \cite{Nath:1971wp} is given by \cite{Pascalutsa:1999zz}
\begin{equation}
g_{R N \pi} = \frac{f_{R N \pi}}{m_{\pi} M_R} \quad. \label{eq:gf}
\end{equation}
Next we analyse the phase space of the obtained width. As shown in section \ref{sec:fermprop}, when neglecting the real parts and quadratic terms the width near the on-shell point is:
\begin{equation*}
\Gamma = - \frac{2c}{\sqrt{s}} \left( s \Im \Sigma_v + M_0 \Im \Sigma_s \right).
\end{equation*}
When neglecting the real parts the bare mass $M_0$ is equal to the physical mass $M_R$. Inserting the selfenergy calculated in this section the width can be written as
\begin{align*}
\Gamma &= - \frac{2c}{\sqrt{s}} \left( s \Im a_2 + M_0 \, \Im a_1 \right) = - \frac{2c}{\sqrt{s}} \frac{N_I}{8 \pi} g_{R N \pi}^2 \frac{q_*}{\sqrt{s}} \left( s b_2 + M_R \, b_1 \right) 
	 = c \frac{N_I }{12 \pi} \, g_{R N \pi}^2 \, q_*^3 \frac{s}{\sqrt{s}} \left( \frac{1}{2 \sqrt{s}} y + P \, M_N \frac{M_R}{\sqrt{s}} \right).
\end{align*}
As shown above $k_0 = \frac{1}{2 \sqrt{s}} y$ is the energy of the nucleon and $q_*$ is the momentum of the pion. Inserting the conventional coupling (\ref{eq:gf}) the width reads
\begin{equation}
	\Gamma = c \frac{N_I }{12 \pi m_{\pi}^2} f_{R N \pi}^2 \, q_*^3 \frac{s}{M_R^2} \frac{ k_0 + P \, M_N \frac{M_R}{\sqrt{s}}}{\sqrt{s}} . \label{eq:Penner}
\end{equation}
For c=1 and on the mass shell this expression agrees with the width calculated in \cite{Penner:2002ma}.

Now we can compare this result with the fact \cite{Post:2003} that around threshold the energy dependence of the width is determined by the orbital angular momentum $l$:
\begin{equation}
\Gamma(\sqrt{s} \approx \sqrt{s_{thr}}) \sim |\vec{q}|^{2l +1} \label{eq:angmomPenner}
\end{equation}
with $\vec{q}$ as the center of mass momentum of the decay products. For small kinetic energies we can expand the nucleon energy in the non-relativistic limit as 
\begin{equation*}
k_0 = \sqrt{\vec{q}^2 + M_N^2} \approx M_N + \frac{\vec{q}^2}{2M_N} \quad.
\end{equation*}
The width (\ref{eq:Penner}) is proportional to
\begin{align*}
\Gamma \sim |\vec{q}|^3 (k_0 + P M_N) &\approx |\vec{q}|^3 \left(M_N (1+P) + \frac{\vec{q}^2}{2M_N} \right) \\
&\sim \begin{cases} |\vec{q}|^3 & \text{ for positive parity $P=+1$} \\ |\vec{q}|^5 & \text{ for negative parity $P=-1$} \end{cases} \quad .
\end{align*}
The angular momentum can be read off using equation (\ref{eq:angmomPenner}): $l=1$ for a particle with positive parity and $l=2$ for negative parity. As it should be, the Lagrangian in this section  describes for positive parity a P-wave and for negative parity a D-wave resonance decaying into $N \pi$.

\subsection[Selfenergy of a Spin 3/2 Baryon $N \rho$ System]{Selfenergy of a Spin 3/2 Baryon $\bf N \rho$ System}
\label{sec:rho}

Because the $\rho$-meson is not a stable particle the spectral function is not trivially given as a $\delta$-function. A lot of work have been devoted to the $\rho$-meson and its properties (e.g.~\cite{Klingl:1996by} \cite{Herrmann:1993za}). In this work we took the selfenergies calculated in \cite{Herrmann:1993za}. They come in a fully analytical form and preserve unitarity. Using this selfenergies the spectral function $\rho(q^2)$ of the $\rho$-meson can be derived directly. 

\begin{figure}
\begin{center}
\epsfig{file=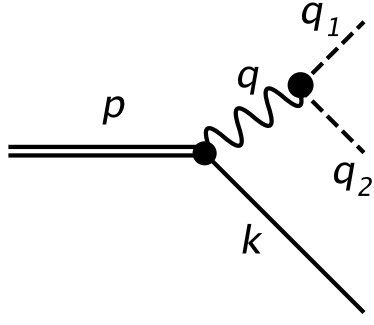,width=5cm}
\caption{Three-body decay of a resonance into $ N \pi \pi$ via a virtual $\rho$-meson. \label{fig:decayD13Nrho}}
\end{center}
\end{figure}

We construct a relativistic gauge invariant Lagrangian of the form
\begin{equation*}
\mathcal{L} = \frac{g_{R N \rho}}{2} \bar{\psi}^{\mu \nu} \left\{ \begin{matrix} i  \gamma_5 \\ 1 \end{matrix} \right\} \Psi \rho_{\mu \nu} + h.c.
\end{equation*}
with top (bottom) line for particles with positive (negative) parity. $\psi^{\mu \nu}$ and $\rho_{\mu \nu}$ are the field strength tensors of the spin 3/2 baryon and $\rho$-meson, respectively.
The imaginary parts of the coefficients $a_i$ are only nonzero for $s > (M_N + 2 \, m_{\pi})^2$  and all have the form: 
\begin{equation}
\Im a_i(s) = - \frac{N_I }{4 \pi^2} g_{R N \rho}^2 FF(s)^2  \int_{M_N}^{z_{*}(s)} dk_0 \, b_i(s,k_0)  \, \sqrt{k_0^2 -M_N^2} \, \rho \left(s - 2 \sqrt{s} k_0 + M_N^2 \right). \label{eq:rhoai}
\end{equation}
The lower limit of the integral represents the case when the nucleon is at rest. This is possible because the \rhomeson is an unstable particle and no constraint for its mass exists. The upper limit comes because the spectral function of the \rhomeson is zero ($\rho(q^2) = 0$) for $q^2 < 4 m_{\pi}^2$. The quantity
\begin{equation*}
z_{*}^2  = \frac{1}{ 4 s} (s + M_N^2 - 4 m_{\pi}^2)^2  \label{eq:energynucleonmax}
\end{equation*}
can be understood when exploring the three-body kinematics of this process in figure \ref{fig:decayD13Nrho}. In the rest frame of the resonance the two pions have the same momentum when the nucleon has maximum energy. The kinematical quantities are then:
\begin{align*}
p = (\sqrt{s}, 0), & \quad p^2=s, \\
q_1 = q_2 = (q_0,-\frac{\vec{k}}{2}), & \quad q_0^2 = \frac{\vec{k}^2}{4} + m^2_{\pi},\\
k = (k_0,\vec{k}), & \quad k_0^2 = \vec{k}^2 + M_N^2.
\end{align*}
Using momentum conservation the energy of the nucleon can be calculated
\begin{align*}
s &= p^2 = (q_1 + q_2 + k)^2 =(2 q_0 + k_0)^2 = 2 k_0^2 - M_N^2 + 4m_{\pi}^2 + 4 q_0 k_0 \\
& \Rightarrow k_0^2=\frac{1}{4s} (s + M_N^2 - 4 m^2_{\pi})^2 = z_*^2.
\end{align*}
Now the limits of the integration (\ref{eq:rhoai}) become clear. The lower limit is the energy where the nucleon is at rest and the upper limit is the energy where the nucleon has maximum energy and the \rhomeson is at rest. Because the \rhomeson is not a stable particle $k_0$ can take all possible values between these two limits. Each kinematical situation is weighted by the spectral function $\rho(q^2)$ of the $\rho$-meson.

The integral in (\ref{eq:rhoai}) is evaluated numerically.
The functions $b_i(s,k_0)$ are obtained as:
\begin{align*}
b_1(s,k_0) &= - \frac{s}{3 M_N} P \left(M_N^4 + 2(s - 3 k_0 \sqrt{s} + k_0^2) M_N^2 + s k_0^2 \right), \\
b_2(s,k_0) &= - \frac{\sqrt{s} k_0}{3 M_N^2} \left(M_N^4 + 2(s - 3 k_0 \sqrt{s} + k_0^2) M_N^2 + s k_0^2 \right), \\
b_3(s,k_0) &= b_1(s,k_0), \\
b_4(s,k_0) &= b_2(s,k_0). 
\end{align*}
The real part is given by the numerical evaluation of
\begin{equation}
\Re a_i(s) = \frac{1}{\pi} \, \mathcal{P}\int_{(M_N + 2 \, m_{\pi})^2}^{\infty} d\sigma \frac{\Im a_i (\sigma)}{\sigma - s} \quad. \label{eq:reairho}
\end{equation}

Note that the determination of $\Re a_i$ requires to calculate a double integral on account of (\ref{eq:reairho}) and (\ref{eq:rhoai}). The higher effort as compared to the $\pi N$ decay channel is caused by the unstable nature of the $\rho$-meson. It is interesting to study the simplification which occurs if the $\rho$-meson is taken as a stable particle with mass $m_{\rho}$. Now the integration in equation (\ref{eq:rhoai}) can be carried out. The coefficients $a_i$ are nonzero for $s > (M_N + m_{\rho})^2$ and one obtains
\begin{equation*}
\Im a_i(s) = - \frac{N_I}{8 \pi^2} \, g^2_{R N \rho} FF(s)^2 \, \frac{k_*}{\sqrt{s}} \, b_i(s,k_{0*}) .
\end{equation*}
With the notations
\begin{align*}
k^2_{0*} &= \frac{1}{4s}(s+M_N^2-m_{\rho}^2)^2, \\
k_*^2  &= k_{0*}^2 - M_N^2 .
\end{align*}
Equivalent to the last section $k_0=k_{0*}$ is the energy and $\sqrt{\vec{k}^2} = k_*$ the momentum of the nucleon. Inserting this notation the coefficients $b_i(s,k_{0*})$ can be calculated further leading to
\begin{align*}
b_1(s) &= - P \frac{s}{3 M_N} \, B(s), \\
b_2(s) &= \frac{\sqrt{s} k_{0*}}{3 M_N^2}  \, B(s), \\
B(s) &= \left((2 M_N^2  + s) k_*^2 +  3 m^2_{\rho} M_N^2 \right).
\end{align*}

The nonrelativistic limit of the width is calculated only for stable $\rho$-mesons. The approach is equal to the previous case of $\pi N$:
\begin{equation*}
\Gamma_{N \rho} \sim \sqrt{\vec{k}^2} (s \, b_2 + M_R \, b_1) \sim \sqrt{\vec{k}^2} B(s) (k_0 - P \, M_N).
\end{equation*}
Expanding $k_0$ and $B(s)$ in the non-relativistic limit give in leading order:
\begin{align*}
k_0 &= \sqrt{\vec{k}^2 + M_N^2} \approx M_N + \frac{\vec{k}^2}{2M_N}, \\
B(s) & \sim 3 m_{\rho}^2 M_N^2.
\end{align*}
The width is proportional to
\begin{equation*}
\Gamma_{N \rho} \sim \sqrt{\vec{k}^2} (\vec{k}^2 + M_N - P M_N)
\end{equation*}
leading to a P-wave for a resonance with positive parity and an S-wave for a resonance with negative parity.

\subsection[Selfenergy of a Spin 3/2 Baryon $\Delta \pi$ System]{Selfenergy of a Spin 3/2 Baryon $\bf \Delta \pi$ System}
\label{sec:delta}

Due to the unstable character of the $\Delta$ its spectral functions $\rho_s$ and $\rho_v$ are not trivial anymore. The $\Delta$ on its own is a spin 3/2 particle, so the Lagrangian must be invariant under a simultaneous gauge transformation of both spin 3/2 particles. A Lagrangian satisfying all symmetries has the form
\begin{equation*}
\mathcal{L} =  \frac{g_{R \Delta \pi}}{2} \bar{\psi}^{\mu \nu} \gamma_{\alpha} \left\{ \begin{matrix} i \gamma_5 \\ 1 \end{matrix} \right\} \Delta_{\mu \nu} \partial^{\alpha} \pi + h.c.
\end{equation*}
The imaginary parts of the coefficients $a_i$ are only non-zero for $s > (M_N + 2 \, m_{\pi})^2$ and have the form:
\begin{equation}
\Im a_i(s) = N_I \frac{3 }{2 \pi^3} \, g_{R \Delta \pi}^2 FF^2 \int_{ka_0(s)}^{\sqrt{s}-m_{\pi}} dk_0 q_{vec} \, \rho \left(m_{\pi}^2 + 2 \sqrt{s} k_0 -s \right)  \, b_i(s,k_0) 
\label{eq:Deltaai}
\end{equation}
with $\rho=\rho_s$ for $i=1,3$ and $\rho=\rho_v$ for $i=2,4$. $\rho_s$ and $\rho_v$ are the scalar and vector parts of the $\Delta$ spectral function.

The upper limit can be understood by the fact that if the $\Delta$ would get more energy there would be no energy left to create a pion. The lower limit is introduced when demanding that $\rho(k^2) = 0$ for $k^2 < (M_N + m_{\pi})^2$. This is the minimal energy needed to create an unstable $\Delta$. The integral is evaluated numerically. 

We introduce the notation
\begin{align*}
ka_0(s) &= \frac{1}{2 \sqrt{s}} (M_N^2 +2 M_N m_{\pi} + s), \\
q_{vec}^2(s,k_0) &= (k_0 - \sqrt{s})^2 - m_{\pi}^2,
\end{align*}
where $ka_0(s)$ is the least energy needed to create a $\Delta$ in the rest frame of the decaying resonance. $q_{vec}$ is the momentum of the pion in the same frame.

For the functions $b_i$ in (\ref{eq:Deltaai}) we obtain:
\begin{align*}
b_1(s,k_0) &= - \frac{1}{9} P \, s \left[ 2 (k_0^2 - q_{vec}^2)^2 + (7 k_0^2 - 8 \sqrt{s} k_0 + 2 s) (k_0^2 - q_{vec}^2) + k_0^2 (7 s - 10 k_0 \sqrt{s}) \right] , \\
b_2(s,k_0) &= \frac{\sqrt{s}}{9} \left[4 (k_0 - \sqrt{s}) (k_0^2 - q_{vec}^2)^2 + k_0 (5 k_0^2 - 14 \sqrt{s} k_0 + 4 s) (k_0^2 - q_{vec}^2) + 5 k_0^3 s \right], \\
b_3(s,k_0) &= - \frac{2}{9} P \, s \left[ (k_0^2 - q_{vec}^2)^2 + (k_0^2 + 2 \sqrt{s} k_0 + s) (k_0^2 - q_{vec}^2) - k_0^2  (2 \sqrt{s} k_0 +s) \right], \\
b_4(s,k_0) &= - \frac{2 \sqrt{s}}{9} \left[(k_0 + 2 \sqrt{s}) (k_0^2 - q_{vec}^2)^2 + 2 k_0 (k_0^2 + 2 \sqrt{s} k_0 - s) (k_0^2 - q_{vec}^2) + k_0^3 s \right].
\end{align*}
The real part is given by the numerical integration of
\begin{equation}
\Re a_i(s) = \frac{1}{\pi} \, \mathcal{P} \int_{(M_N + 2 \, m_{\pi})^2}^{\infty} d\sigma \frac{\Im a_i (\sigma)}{\sigma - s} \quad . \label{eq:Deltareai}
\end{equation}

Let us discuss briefly the numerical effort necessary for the calculation of $\Re a_i$ for an unstable $\Delta$: Already the spectral functions $\rho_s$ and $\rho_v$ of the $\Delta$ which appear in (\ref{eq:Deltaai}) must be determined numerically since the real parts of the selfenergies which enter (\ref{eq:rhov}),(\ref{eq:rhos}) via (\ref{eq:stilde})-(\ref{eq:ImFvdef}) are given only numerically according to (\ref{eq:reaipi}). By an additional integral (\ref{eq:Deltaai}) $\Im a_i$ is obtained. Finally a third integration (\ref{eq:Deltareai}) yields $\Re a_i$ for the decay channel $\Delta \pi$.
Taking instead the $\Delta$-resonance as a stable particle with mass $M_{\Delta}$ one can carry out the integration in equation (\ref{eq:Deltaai}) leading to 
\begin{equation*}
\Im a_i(s) = - N_I \frac{1}{4 \pi} \, g_{R \Delta \pi}^2 FF(s)^2 \, k_{vec} \, b_i(s)
\end{equation*}
with $k_{vec}$ as the momentum of the $\Delta$ which is, in the rest frame of the resonance, given by
\begin{align*}
k^{2}_{vec}(s) &= \frac{1}{4 s}\left((s - m_{\pi}^2 + M_{\Delta}^2)^2 - 4s M_{\Delta}^2 \right).
\end{align*}
The functions $b_i$ are similar to the unstable case. First $b_1$ has to be multiplied by $M_{\Delta}$ (see (\ref{eq:rhostodelta}). Second $q_{vec} \to k_{vec}$ and $k_0$ is not a free parameter anymore because the $\Delta$ can only be on the mass shell with the condition $k_0^2 = k_{vec}^2 + M_{\Delta}^2$. In the rest frame of the resonance it is given by
\begin{align*}
k^{2}_{0}(s) &= \frac{1}{4 s}(s - m_{\pi}^2 + M_{\Delta}^2)^2.
\end{align*}
Before we calculate the non-relativistic limit for the width as was done in the previous sections we want to rearrange $b_1$ and $b_2$ in a more convenient form using the on-shell condition leading to:
\begin{align*}
b_1 &= - \frac{s}{9} M_{\Delta} \, P \, \left[ (M_{\Delta}^2 + s) ( 9 M_{\Delta}^2 + 7 k_{vec}^2) - 2 \sqrt{s} k_0 (9 M_{\Delta}^2 - 5 k_{vec}^2) \right], \\
b_2 &= \frac{1}{9} \left[ \sqrt{s} k_0 (M_{\Delta}^2 + s)(9 M_{\Delta}^2 + 5 k_{vec}^2) - 2 s M_{\Delta}^2 (9 M_{\Delta}^2 + 7 k_{vec}^2) \right].
\end{align*}
The on-shell width, in first order of $k_{vec}$, is proportional to
\begin{align*}
\Gamma_{\Delta \pi} &\sim k_{vec} ( M_R b_2 + b_1) \\
	&= k_{vec} \left( \frac{M_R^2}{9} \left[ (M_{\Delta}^2 + M_R^2) (k_0 - P \, M_{\Delta}) 9 M_{\Delta}^2 - 2 M_R M_{\Delta} (M_{\Delta} - P \, k_0) 9 M_{\Delta}^2 \right] \right)\\
	&= k_{vec} \left( M_R^2 M_{\Delta}^2 (k_0 - P \,  M_{\Delta}) (M_{\Delta} + P M_R)^2 \right).
\end{align*}
In the non-relativistic limit $k_0 \to M_{\Delta} + \frac{k^2_{vec}}{2 M_{\Delta}}$ the width is proportional to
\begin{equation*}
 \Gamma_{\Delta \pi} \sim k_{vec} \left( M_{\Delta} (1-P) + k_{vec}^2) \right).
\end{equation*}
This channel is an S-wave (P-wave) for a resonance with negative (positive) parity which is the correct description.

\section[Results for $\Pdreidrei(1232)$ and $\Deinsdrei(1520)$]{Results for $\bf \Pdreidrei(1232)$ and $\bf \Deinsdrei(1520)$}
\label{sec:results}

In this section the results for the \Pdreidrei(1232) and \Deinsdrei(1520) resonances will be presented. We will determine the input parameters and discuss the off-shell width of the particles, their selfenergies and spectral functions.

In general the calculations were performed using the following parameters: bare mass $M_0$ and for each decay channel $i$ a coupling $g_i$ and a cut-off $\Lambda_i$. For simplicity all $\Lambda_i$ are chosen to be the same. We will study, however, how the results change when $\Lambda$ is varied. The best way to obtain the values of the parameters would be a fit to experimental data (e.g.~phase shifts). But this approach would go far beyond the scope of this work. The emphasis of this work is to find out whether it is feasible to calculate propagators for spin 3/2 particles in a fully relativistic framework for different channels. The priority is set to the implementation of the full relativistic structure of the propagators. Calculating experimental data, as for example cross sections and phase shifts, from the derived selfenergies and spectral functions is not a trivial task involving also background terms etc.~(cf.~the corresponding discussion in the introduction). Therefore the input parameters were fitted only to the partial widths and the mass of the resonance taken from \cite{PDBook}. This leads to some complications because mass and width are not directly measurable observables leaving an ambiguity how to define them as discussed in section \ref{sec:fermprop}.

Further problems arise because the parameters $M_0$ and $g$ are coupled to each other. To calculate $M_0$ from equation (\ref{eq:M0}) the full knowledge of the selfenergy is needed which will only be possible when the couplings are known. They can be extracted by demanding that the respective partial width of the resonance on the mass shell is equal to the partial width published in \cite{PDBook}. But for the calculation of the widths $M_0$ is needed. In the case where only one channel exists this problem can be solved by inserting equation (\ref{eq:M0}) into the propagator. In the case of different channels $M_0$ needs to be calculated with the sum of all selfenergies as discussed in section \ref{sec:fermprop}. Variation of the coupling in each channel will influence every other channel which must be solved by fine tuning.

Obviously it is impossible to fit the three parameters $M_R$, $g$ and $\Lambda$, using only two input parameters (mass and width). This leaves one parameter, the cut-off $\Lambda$, open which leads to an ambiguity of the results because the selfenergy depends largely on $\Lambda$. Although it is not possible to pin down $\Lambda$ exactly it will be possible to give arguments for a reasonable range of values which resolves the ambiguity. The reasonable range of $\Lambda$ is around $\Lambda=1$ GeV. When not stated differently all plots depicted in this section are calculated using such a value for $\Lambda$.

The form factor (\ref{eq:lambda}) used in this work is 1 at threshold and decreases exponentially afterwards. This means that the value of the coupling $g$ is given at threshold energy and not on the mass shell. In the literature (e.g.~\cite{Post:2003}) the form factors are often chosen in such a way to be 1 on the mass shell of the particles. The given values of the coupling constants are hence the values on the mass shell. To make it easier to compare the values of this work with other works also the effective coupling constant on the mass shell
\begin{equation*}
g^{\rm eff} = g * FF(M_R^2).
\end{equation*}
will be given below.

\subsection[Parameters of the $ \Delta$ $ \Pdreidrei(1232)$]{Parameters of the $\bf \Delta$ $\bf \Pdreidrei(1232)$}
\label{sec:Delta}
Much work has already been devoted to the complete relativistic structure of the $\Delta$ in conventional \cite{Korpa:1997fk} and Pascalutsa \cite{Almaliev:2002tg} coupling. We present these results for completeness and as a good example for the structure of a relativistic spin 3/2 propagator. In addition, the spectral functions of the $\Delta$ are needed as an input for the decay channel $N^*(1520) \to \Delta \pi$.

The calculations are done using three parameters, bare mass $M_0$, coupling $g_{\Delta N \pi}$ and cut-off $\Lambda$. For given $\Lambda$ the parameters $M_0$ and $g_{\Delta N \pi}$ are chosen such that $M_R$ in equation (\ref{eq:M0}) takes the physical value $1.232$ GeV and the width on the mass shell is
\begin{equation*}
\Gamma_V(M_R) = \Gamma_{exp} = 120 \, {\rm MeV}.
\end{equation*}

The $\Delta$ is a particle with isospin 3/2, so $N_I=1$, and positive parity $P=+1$. The remaining parameters for different cut-offs and the effective coupling are listed in table \ref{tab:P33parameter}. 

In view of these parameters a motivation for the above mentioned value of $\Lambda=1$~GeV can be given. For small $\Lambda$ the effective coupling deviates largely from the coupling at threshold energy. This is not desirable because in the physical meaningful region from threshold energy to approximately 1 GeV above threshold the coupling should not deviate too much. Consequently, small values of $\Lambda$ can be excluded. For large values of $\Lambda$ the bare mass is shifted to a very high value. It is not reasonable that the selfenergy of the \Pdreidrei changes its own mass so dramatically. There will be a further argument why large values of $\Lambda$ lead to undesired features when discussing the normalization function of the \Pdreidrei. All this arguments raise strong indications for a reasonable cut-off parameter at around 1~GeV.
\begin{table}
\begin{center}
\begin{tabular}{c|c|c|c}
$\Lambda $ [GeV]   & $M_0$ [GeV] & $g_{\Delta N \pi}$  [GeV$^{-2}$] & $g^{\rm eff}_{\Delta N \pi}$ [GeV$^{-2}$]  \\ \hline
0.6 	& 1.237 & 34.8	& 12.80 \\
1.0	& 1.334 & 22.5	& 13.05 \\
2.0	& 2.303 & 17.0	& 15.08 \\
\end{tabular}
\caption{Parameters for the \Pdreidrei resonance \label{tab:P33parameter}.}
\end{center}
\end{table}

\subsection[Parameters of the $ N^*$ $ \Deinsdrei(1520)$]{Parameters of the $\bf N^*$ $\bf \Deinsdrei(1520)$}
\label{sec:paramD13}

There are three major decay channels for the $\Deinsdrei(1520)$, which are $N \pi$, $N \rho$ and $\Delta \pi$. We calculate the selfenergy of each channel separately. The total selfenergy is then given as the sum of each individual channel.

The $\Deinsdrei(1520)$ has negative parity and isospin 1/2. This fixes $P = -1$. $N_I=3$ for the channels $N \pi$ and $N \rho$ and $N_I=12$ for the channel $\Delta \pi$. 

The coupling constants are fitted via the partial width of each channel in such a way that the contribution of the $N \pi$ channel to the total width on the mass shell is $55 \%$ as proposed in \cite{PDBook}. The contribution of the $N \rho$ channel is chosen to be 10~MeV or 26~MeV.  The results in this work are calculated with these two values because it is hard to extract the coupling of the \rhomeson to the \Deinsdrei resonance. Information on the coupling of baryon resonances to the $N \rho$ channel originates mainly from an analysis of the reaction $\pi N \to \pi \pi N$. This is a formidable task because the $N^+(1520)$ is nominally subthreshold to a $\rho N$ final state. In other words, only the low-energy tail of the $\rho$-meson contributes at the $N^*$ mass shell. Analysis of the experimental data by Manley et al. \cite{Manley:1984jz}\cite{Manley:1992yb} leads to the value of $\Gamma_{N \rho}=26$ MeV. This value is also similar to the partial width given by the particle data group (PDG) \cite{PDBook}. A somewhat different approach was taken by Leupold and Post \cite{Leupold:2004gh} where the coupling was extracted in a QCD sum rule analysis of in-medium modifications of the $\rho$-meson. This analysis leads to the smaller value of $\Gamma_{N \rho}=10$ MeV. A similar value $\Gamma_{N \rho}=12$ MeV was deduced by Vrana \cite{Vrana:1999nt} in an analysis of the experimental data similar to Manley. In the present work both partial widths will be used to see how much the spectral function of the \Deinsdrei depends on the value of the coupling to $N \rho$. For further discussion of this issue cf. \cite{Post:2003}. The $\Delta \pi$ channel is adjusted such that one always gets a total width of $\Gamma = 120$ MeV.

Because the definition for the partial widths discussed in section \ref{sec:fermprop} leads to coupled equations the couplings are fitted such that the above requirements are fulfilled as good as possible. 

In table \ref{tab:D13parameter} the parameters for different $\Lambda$'s and $\Gamma_{N \rho}=10$ MeV can be found. In table \ref{tab:D13paraRho26} the couplings are fitted to give a contribution for the $N \rho$ channel of 26 MeV.

\begin{table}
\begin{center}
\begin{tabular}{c|c|c|c|c|c|c|c}
$\Lambda$ [GeV] &$M_0$ [GeV] 	&$g^{ }_{N^* N \pi} $ &$g^{\rm eff}_{N^* N \pi} $& $g^{ }_{N^* N \rho} $ &$g^{\rm eff}_{N^* N \rho} $ & $g^{ }_{N^* \Delta \pi} $ & $g^{\rm eff}_{N^* \Delta \pi} $ \\ \hline
0.6 	& 1.36	& 154.6	& 6.29 	& 40.3	& 3.92	&  6.4	& 0.62			\\
1.0	& 1.57	& 22.3	& 7.04	& 9.6	& 4.15	&  1.51	& 0.65			\\
2.0	& 3.42	& 9.8	& 7.35	& 5.2	& 4.21	&  0.82	& 0.66			\\
\end{tabular}
\caption{Parameters for the \Deinsdrei resonance with a partial width of $\Gamma_{\rho N} =10$ MeV. The sum of all partial widths is 120 MeV and the cut-off parameter for the $\Delta$ resonance is set to 1 GeV. The units of the couplings are [GeV$^{-2}$] for the $N \pi$ and $N \rho$ channels and [GeV$^{-3}$] for the  $\Delta \pi$ channel. \label{tab:D13parameter}}
\end{center}
\end{table}

\begin{table}
\begin{center}
\begin{tabular}{c|c|c|c|c|c|c|c}
$\Lambda$ [GeV] &$M_0$ [GeV] 	&$g^{ }_{N^* N \pi} $ &$g^{\rm eff}_{N^* N \pi}$ & $g^{ }_{N^* N \rho}$ &$g^{\rm eff}_{N^* N \rho} $& $g^{ }_{N^* \Delta \pi} $& $g^{\rm eff}_{N^* \Delta \pi} $\\ \hline
0.6 	& 1.36	& 156.1	& 6.29 	& 65.6	& 6.38	&  6.4	& 0.50			\\
1.0	& 1.66	& 22.4	& 7.04	& 15.5	& 6.68	&  1.51	& 0.52			\\
2.0	& 5.97	& 9.8	& 7.35	& 8.4	& 6.81	&  0.82	& 0.53			\\
\end{tabular}
\caption{As figure \ref{tab:D13parameter} but using $\Gamma_{N \rho} = 26$ MeV. \label{tab:D13paraRho26}}
\end{center}
\end{table}

\subsection{Selfenergies}
\label{sec:discselfenergies}

\begin{figure}
\epsfig{file=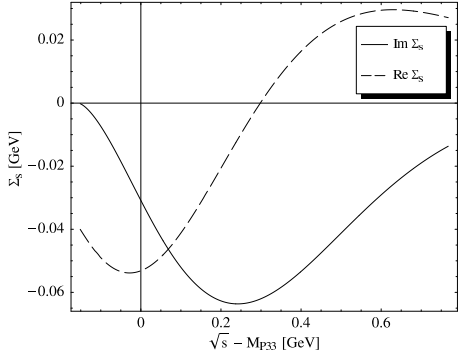,width=7.4cm} \hfill
\epsfig{file=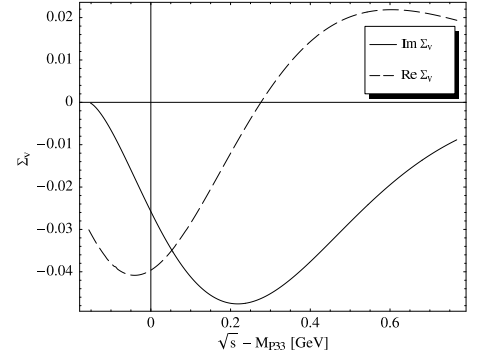,width=7.4cm}
\caption{The selfenergy of the $P_{33}(1232)$ resonance as defined in equation (\ref{eq:sigma}). Left: Scalar part. Right: Vector part.}
\label{fig:betaP33}
\end{figure}

The selfenergies of the \Pdreidrei are depicted in figure \ref{fig:betaP33}. The imaginary parts do not change sign and go to zero for large $\sqrt{s}$ due to the form factor applied to them. For the \Pdreidrei both imaginary parts of the selfenergy are negative which is characteristic for a particle with positive parity in this channel.

\begin{figure}
\epsfig{file=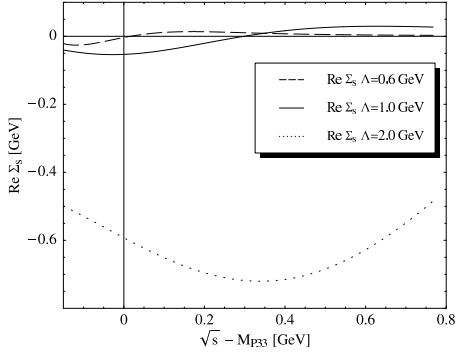,width=7.4cm} \hfill
\epsfig{file=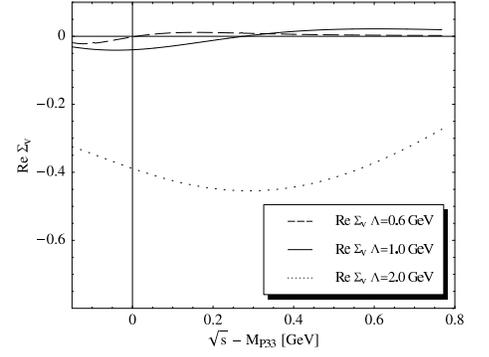,width=7.4cm} 
\caption{Real part of the selfenergies of the $P_{33}(1232)$ resonance using different form factors as defined in equation (\ref{eq:lambda}). Left: Scalar part. Right: Vector part.}
\label{fig:lambdaP33}
\end{figure}

The real parts are small compared to the mass of the \Pdreidrei but compared to the imaginary parts equally large on the mass shell. This is only the case for the cut-off parameter chosen for the plots of figure \ref{fig:betaP33} which is $\Lambda=1$ GeV. Choosing different values for the cut-off leads to a large variation of the real part, depicted in figure \ref{fig:lambdaP33}. One sees that for large values of the cut-off $\Lambda$ the real parts get very large and do not change sign anymore. As discussed above such high values of $\Lambda$ are not reasonable. When taking smaller values of $\Lambda$ the real parts become small on the mass shell. But for small values of $\Lambda$ the variation of the coupling is strong and also the suppression in the physically meaningful energy region is high.

\begin{figure}
\epsfig{file=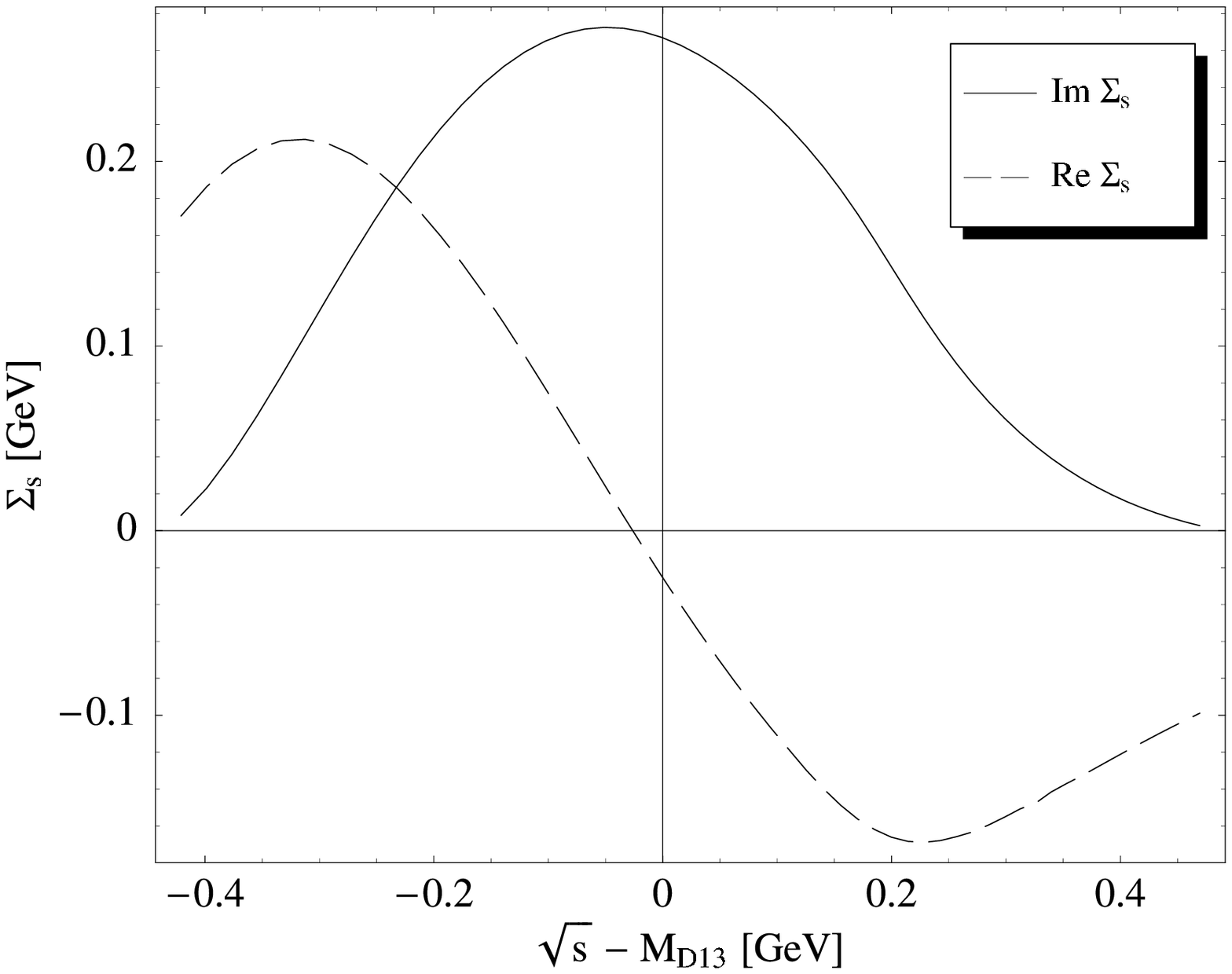,width=7.4cm}\hfill
\epsfig{file=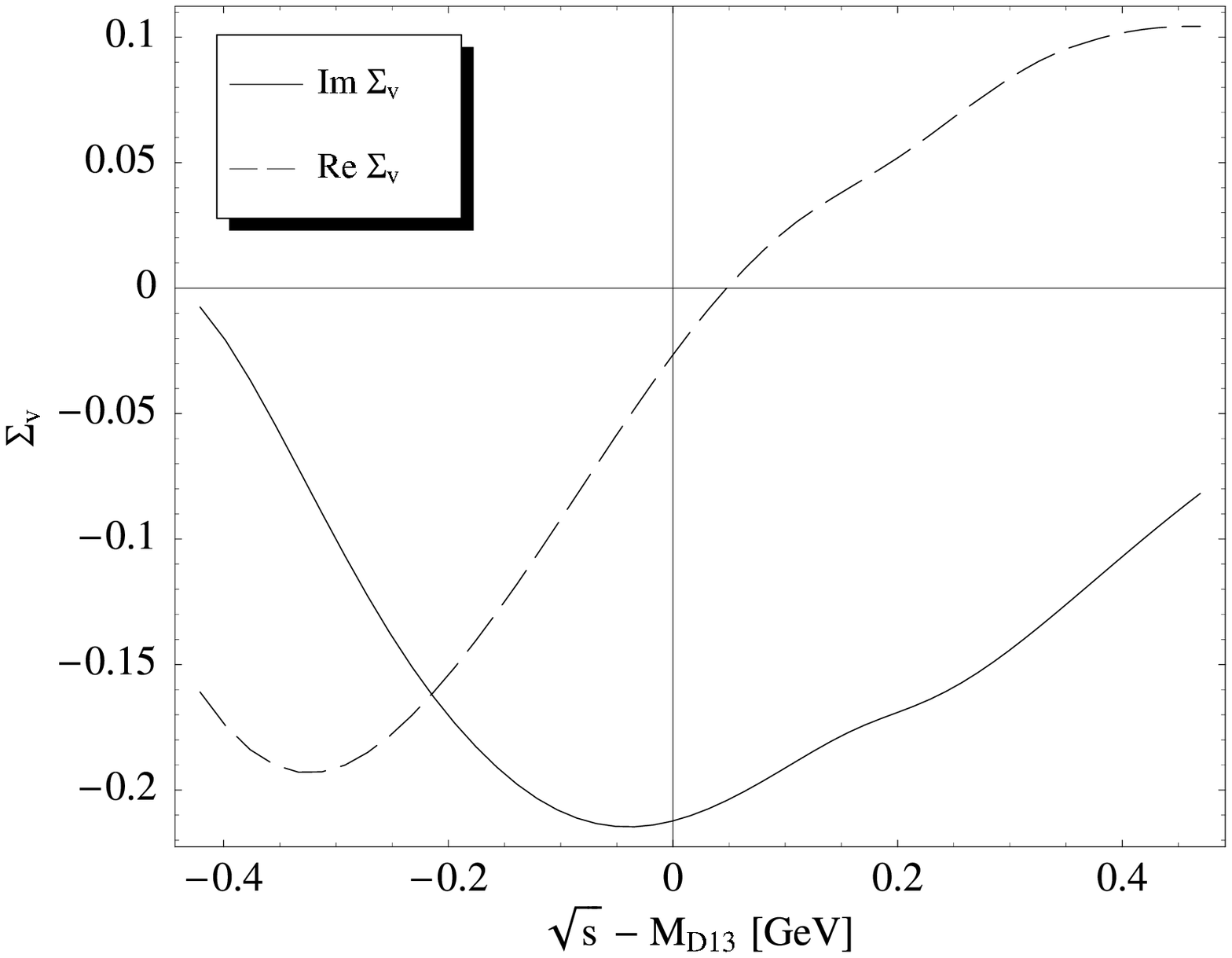,width=7.4cm}
\caption{The selfenergy of the $D_{13}(1520)$ resonance as defined in equation (\ref{eq:sigma}). Left: Scalar part. Right: Vector Part.}
\label{fig:betaD13}
\end{figure}
\begin{figure}
\epsfig{file=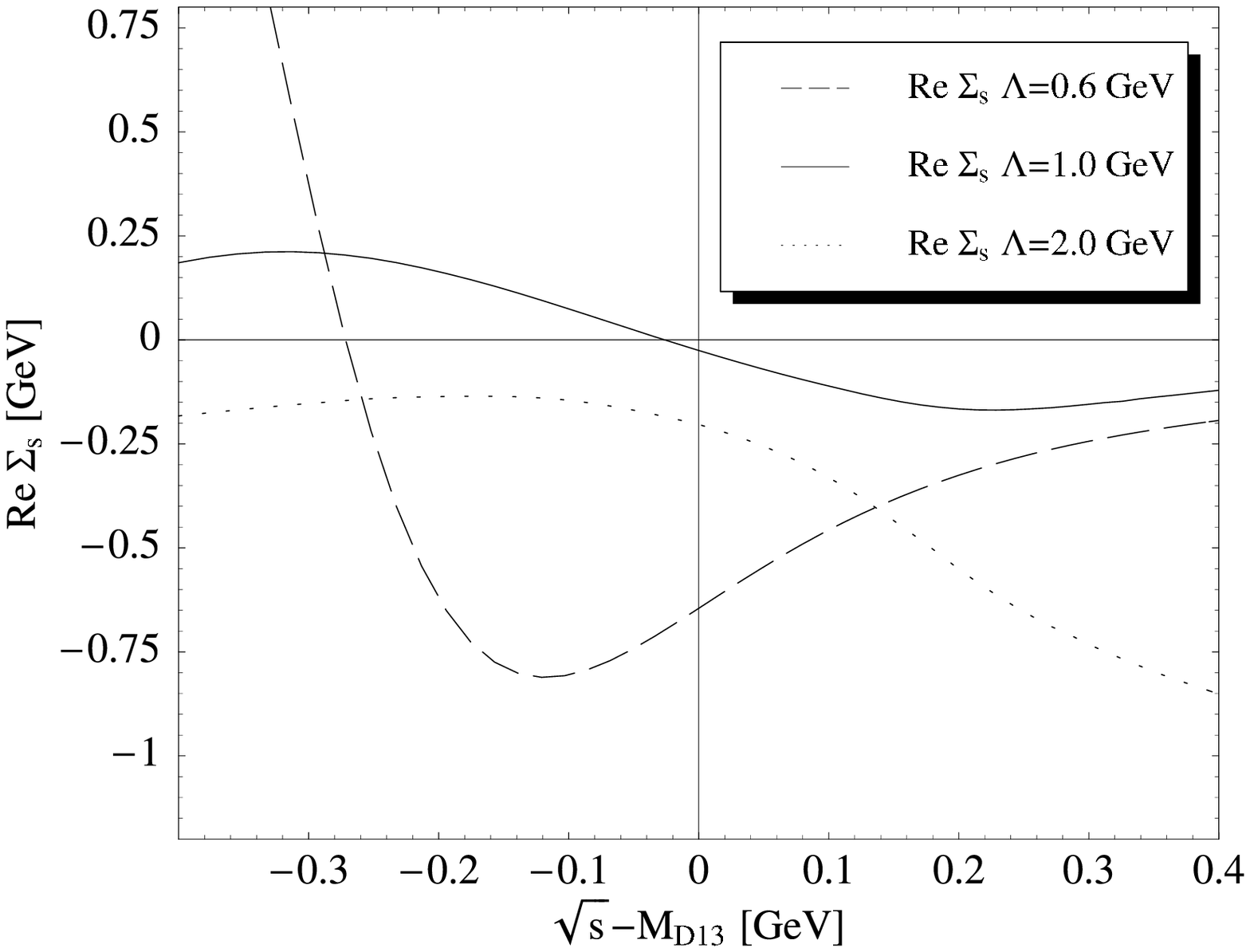,width=7.4cm}\hfill
\epsfig{file=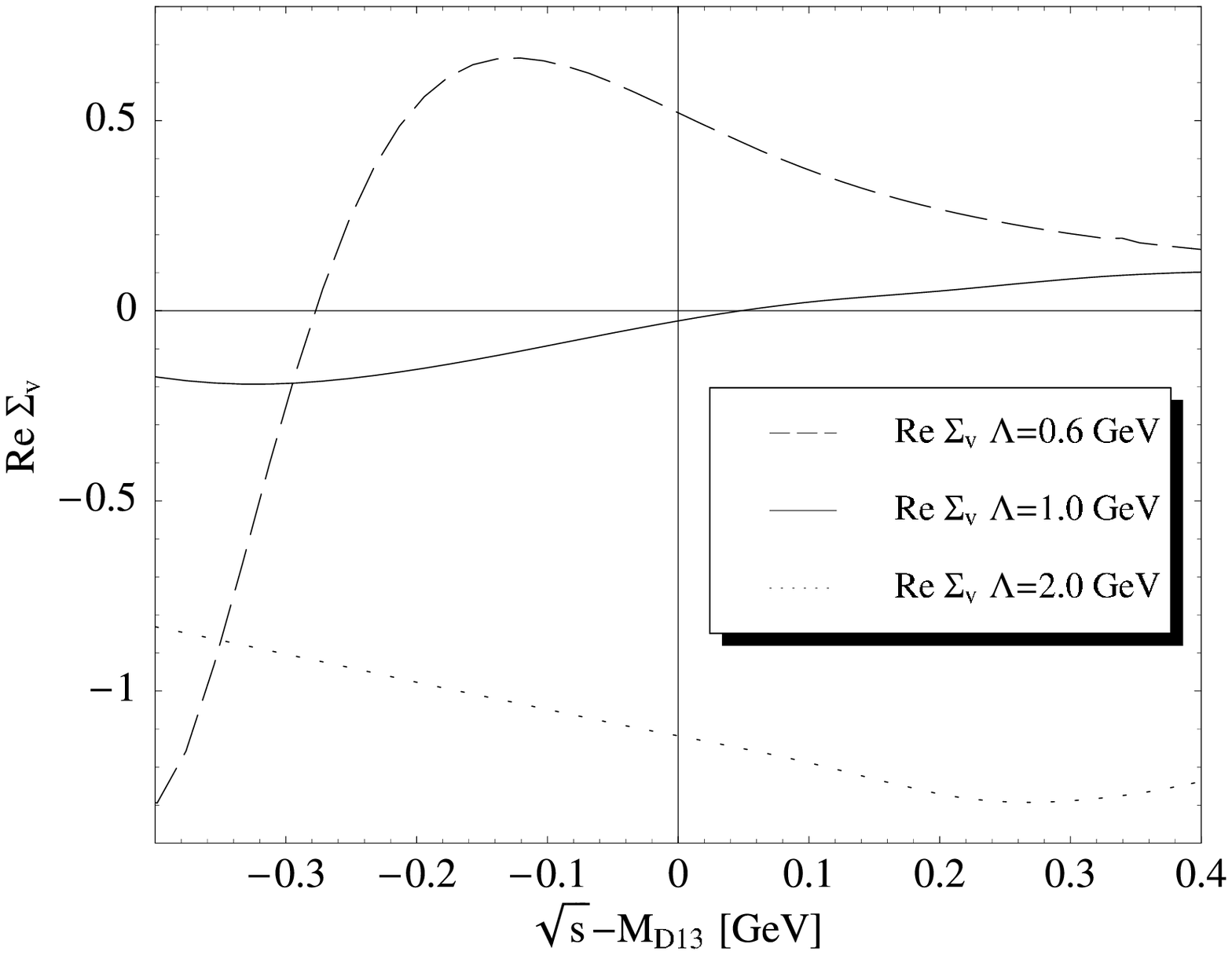,width=7.4cm}
\caption{Real parts of the selfenergies of the \Deinsdrei resonance plotted for different cut-off parameters $\Lambda$. The corresponding form factor is defined in (\ref{eq:lambda}). Left: Scalar part. Right: Vector part.}
\label{fig:lambdaD13}
\end{figure}

For the case of the \Deinsdrei the selfenergies are depicted in figure \ref{fig:betaD13}. The imaginary parts do not change sign, as for the case of the \Pdreidrei, but the sign is opposite for $\Sigma_v$ and $\Sigma_s$. The energy dependence of the selfenergy is much stronger for the case of the \Deinsdrei (note the different scales of the y-axis in figures \ref{fig:betaP33} and \ref{fig:betaD13}) leading to a larger shoulder in the spectral function which will be discussed later.

The real parts of the \Deinsdrei are small around the mass shell of the resonance and change its sign. This is, as for the case of the \Pdreidrei, only true for the cut-off parameter around $\Lambda=1$ GeV. For lower $\Lambda$ the on-shell value of the real parts can become large as depicted in figure \ref{fig:lambdaD13}. A difference to the \Pdreidrei case can be seen because there for lower $\Lambda$ the real parts became smaller which means that this is not generally true but differ from case to case. When $\Lambda$ is large the real parts do not change sign anymore and become large on the mass shell.

\begin{figure}
\epsfig{file=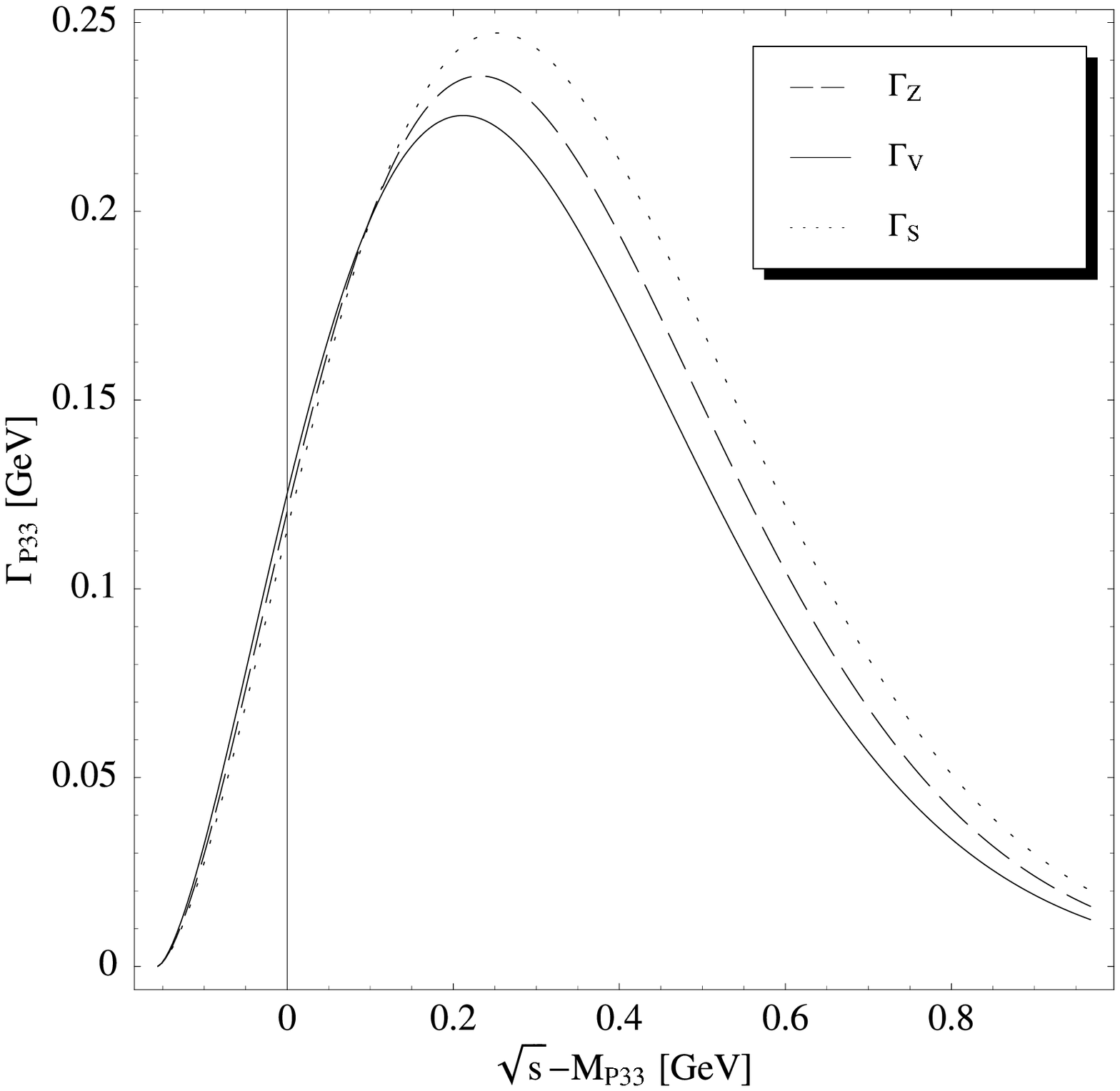,width=7.4cm} \hfill
\epsfig{file=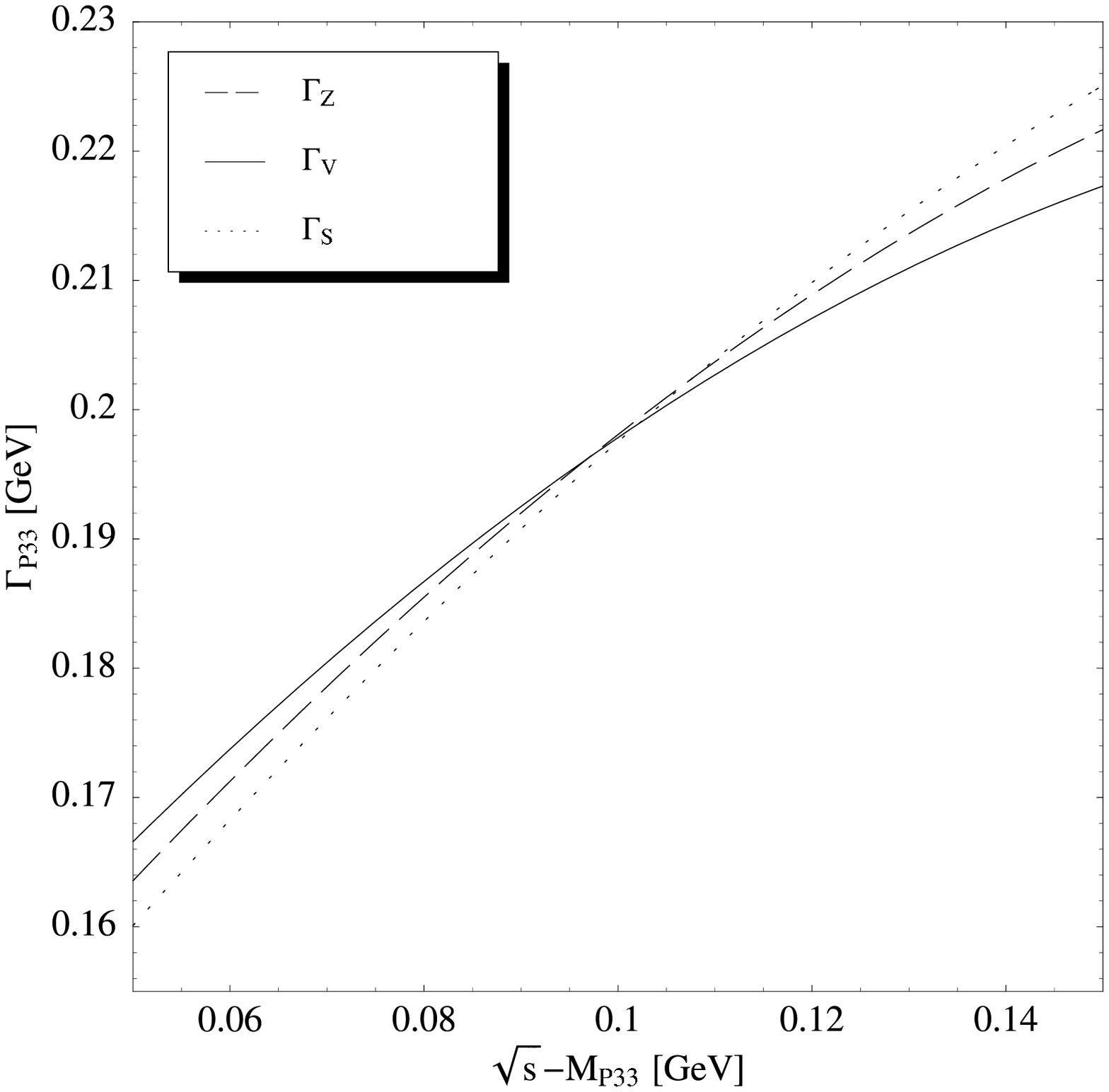,width=7.4cm}
\caption{Width of the $\Delta$. Left: $\Gamma_Z$, $\Gamma_V$ and $\Gamma_S$. Right: $\Gamma_Z$, $\Gamma_V$ and $\Gamma_S$ zoomed. \label{fig:P33altwidht}}
\end{figure}

\subsection{The Width}


As discussed in section \ref{sec:fermprop} there is an ambiguity concerning which function is to call the width when going off-shell. In figure \ref{fig:P33altwidht} all three candidates are depicted for the case of the \Pdreidrei. They have large energy dependences with a maximum above the mass shell energy. This means that spectral strength will be concentrated above mass shell energies.

\begin{figure}
\begin{center}
\epsfig{file=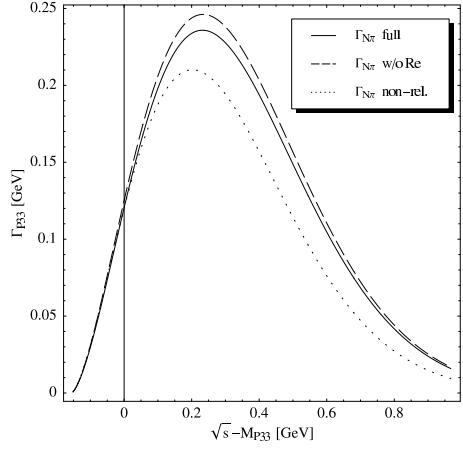,width=7.4cm}
\end{center}
\caption{Width of the $P_{33}(1232)$. The solid line is the full calculation for $\Gamma_V$. The dashed line is the width calculated in \cite{Penner:2002ma}. The dotted line is a simple non-relativistic width. \label{pl:width}}
\end{figure}

In the case of the \Pdreidrei all three widths are similar over the whole energy range. All three widths become equal and cross each other slightly above the mass shell energy. This is the case because the real parts of the selfenergies become zero and change sign also above the mass shell energy as can be seen in figure \ref{fig:betaP33}. Because the values of the real parts are small, compared to the mass, the deviations are small. $\Gamma_Z$, $\Gamma_V$ and $\Gamma_S$ are positive definite which is not in general the case because only for $\Gamma_V$ such a constraint exists (cf.~the discussion in section \ref{sec:fermprop}). But it is not surprising that all widths are positive definite in the case of the \Pdreidrei resonance because the width is approximately proportional to 
\begin{equation*}
\Gamma \sim - (s Im \Sigma_v + M_0 \Im \Sigma_s)
\end{equation*}
as was shown in section \ref{sec:fermprop}. The imaginary parts of the selfenergy are both negative when the resonance has positive parity as was shown in figure \ref{fig:betaP33}. Then $\Gamma$ will not become negative. This is in general not the case. It was shown in figure \ref{fig:betaD13} for the \Deinsdrei that the imaginary parts of the selfenergy have opposite signs. As we sill see below $\Gamma_Z$ and $\Gamma_S$ will become negative leaving $\Gamma_V$ as the only good candidate for a definition of the width. In the case of the \Pdreidrei all three candidates are good choices.

It is possible to compare the width $\Gamma_V$ of the \Pdreidrei with a calculation where the real parts of the selfenergy are neglected and conventional coupling is used \cite{Penner:2002ma}.  We have discussed this width in section \ref{sec:fermprop}, equation (\ref{eq:Penner}). Both widths agree well from threshold to slightly above the mass shell region as depicted in figure \ref{pl:width}. The similarity near the threshold region is induced by the fact that both widths have to reach the same non-relativistic limit. This can be seen when comparing both widths with a simple non-relativistic width of the form
\begin{equation}
\label{eq:non-rel}
\Gamma_{\rm non-rel.}(s) = f \vec{q}^3 FF(s)^2 = f \, \left[\frac{1}{4s} \left[ (s-m_{\pi}^2 - M_N^2)^2 - 4 m_{\pi}^2 M_N^2 \right] \right]^{3/2} FF(s)^2
\end{equation}
which is discussed in subsection \ref{sec:selfenergyNPi} and also depicted in figure \ref{pl:width}.  The quantity $\vec{q}$ is the center of mass momentum of the decay products. The constant $f$ is fitted such that on the mass shell the width is equal to 120 MeV.

Near the mass shell the full width and the approximation by Penner and Mosel \cite{Penner:2002ma} are equal if the real parts of the selfenergy are negligible as was shown in subsection  \ref{sec:selfenergyNPi}. Because the real parts are small for the cut-off parameter chosen both widths agree well on the mass shell. These two constraints, at threshold energy and at the mass shell region, keeps both widths close together. Above this energy region no further constraint occurs and both functions start to differ. The difference is small, leading to the conclusion that the approximations are reasonable.

The non-relativistic width gives a good description of the width for the \Pdreidrei up to the region of its on-shell mass. Similar as above this can be understood because the full width will give the correct non-relativistic limit for small $\sqrt{s}$ and on the mass shell both widths are fitted to be equal. Above the mass shell region the description is not reasonable indicating that the more involved character of the width plays a role in this region where relativistic effects cannot be neglected.

\begin{figure}
\epsfig{file=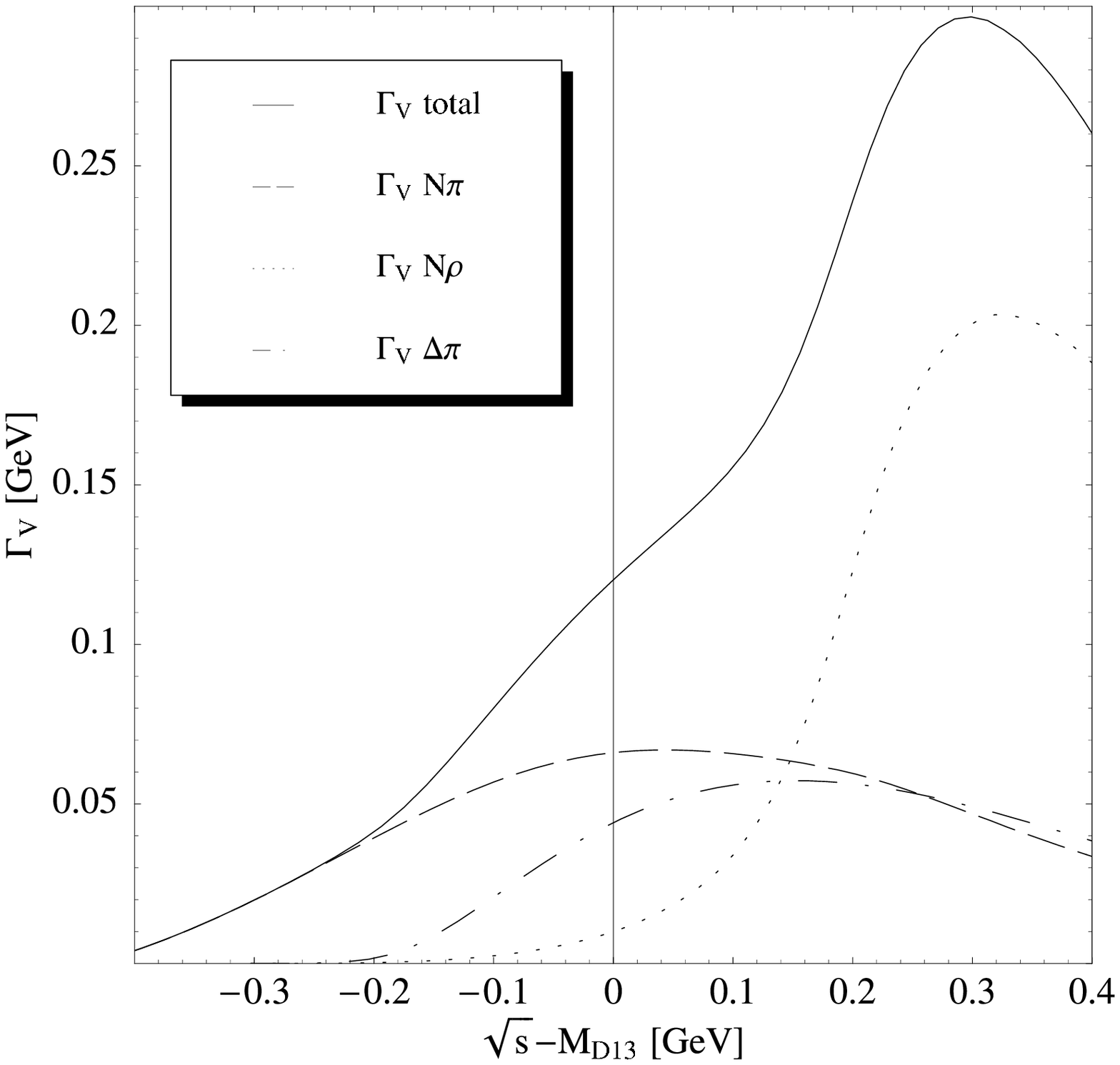,width=7.4cm} \hfill
\epsfig{file=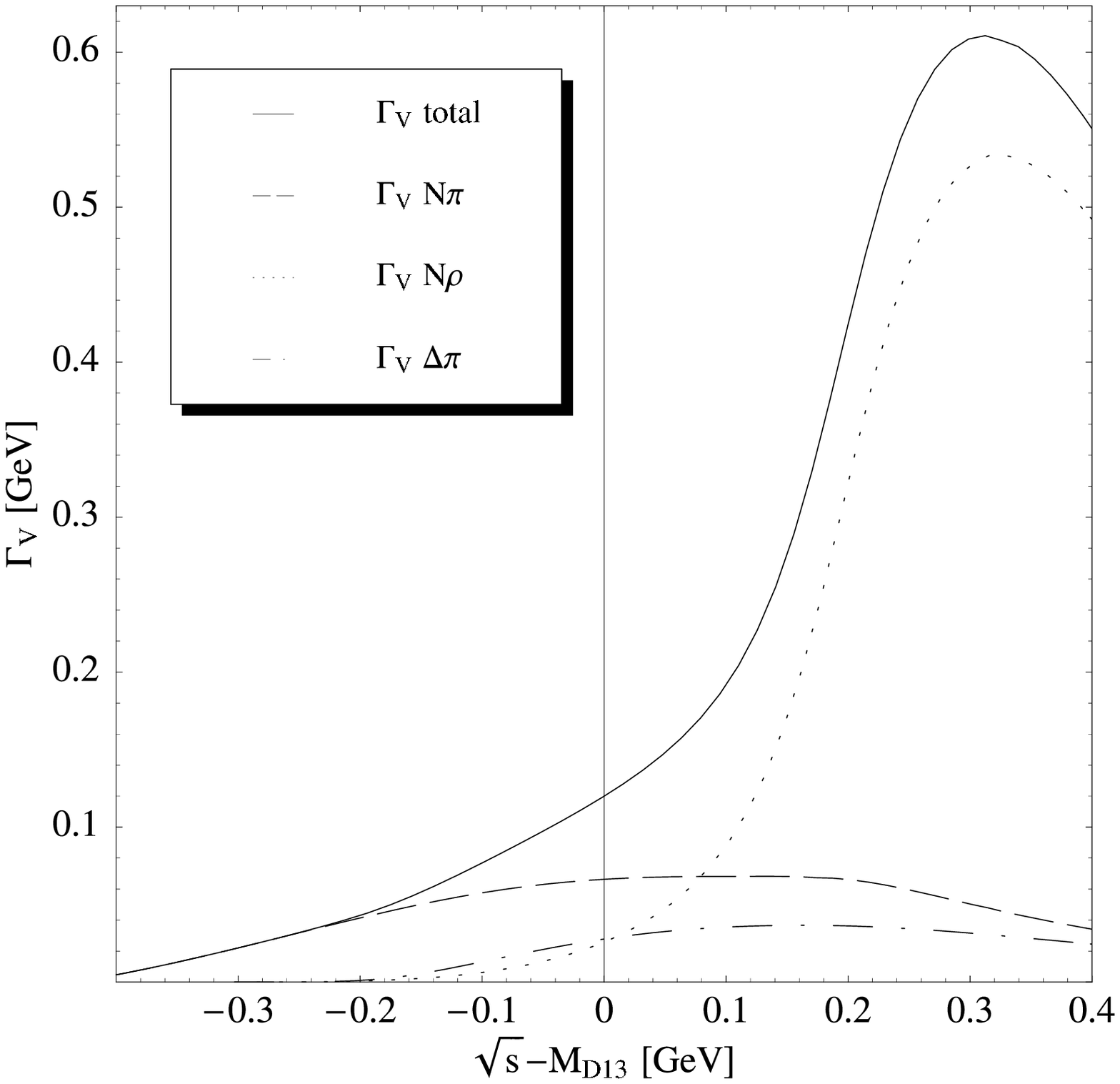,width=7.4cm}
\caption{Different widths of the $D_{13}(1520)$. The solid line is the total width $\Gamma_V$. The dashed lines are the partial widths of the various channels. Left: $\Gamma_{N \rho}=10$ MeV. Right: $\Gamma_{N \rho} = 26$ MeV. \label{fig:widthD13}}
\end{figure}

\begin{figure}
\epsfig{file=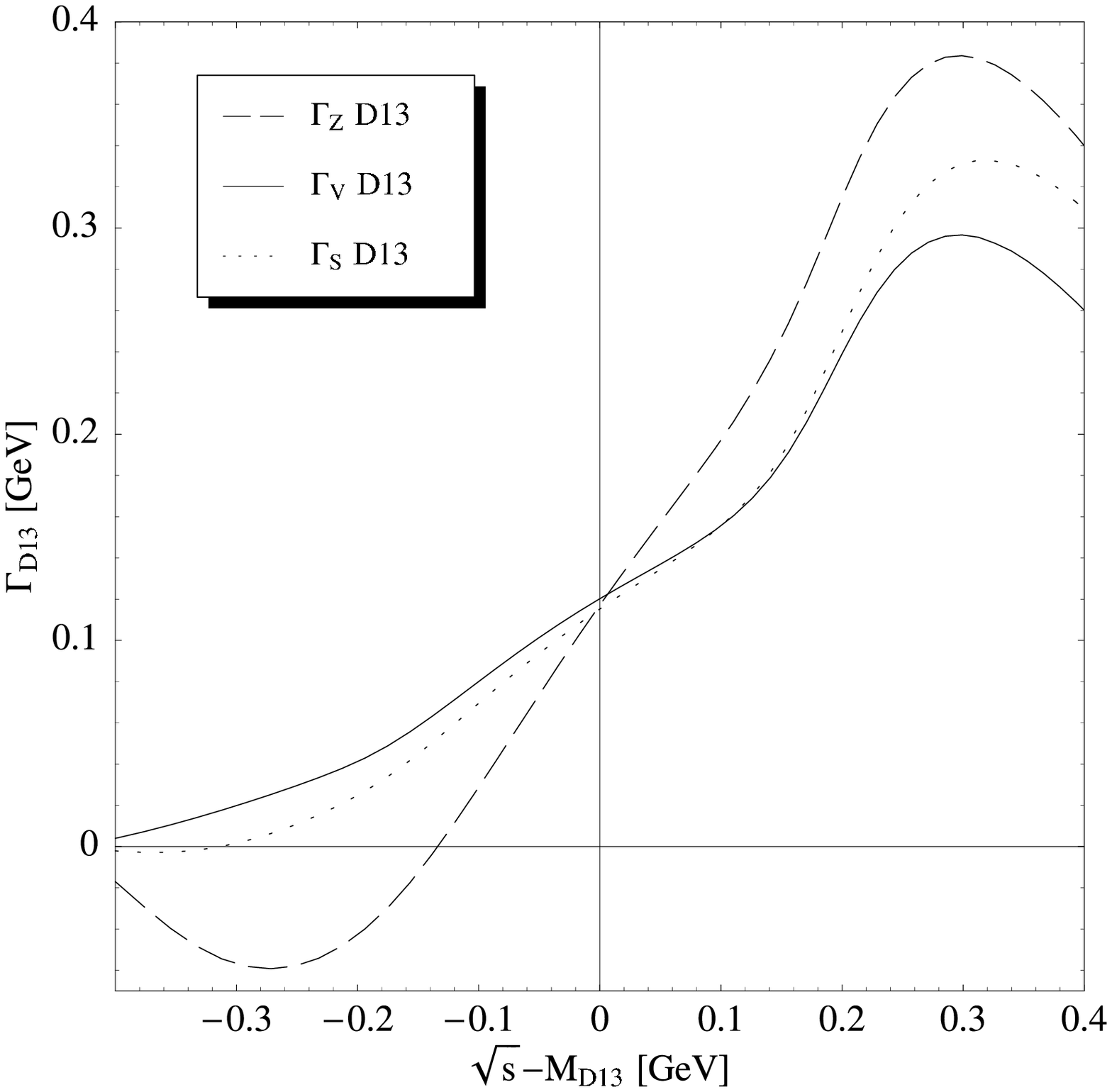,width=7.4cm} \hfill
\epsfig{file=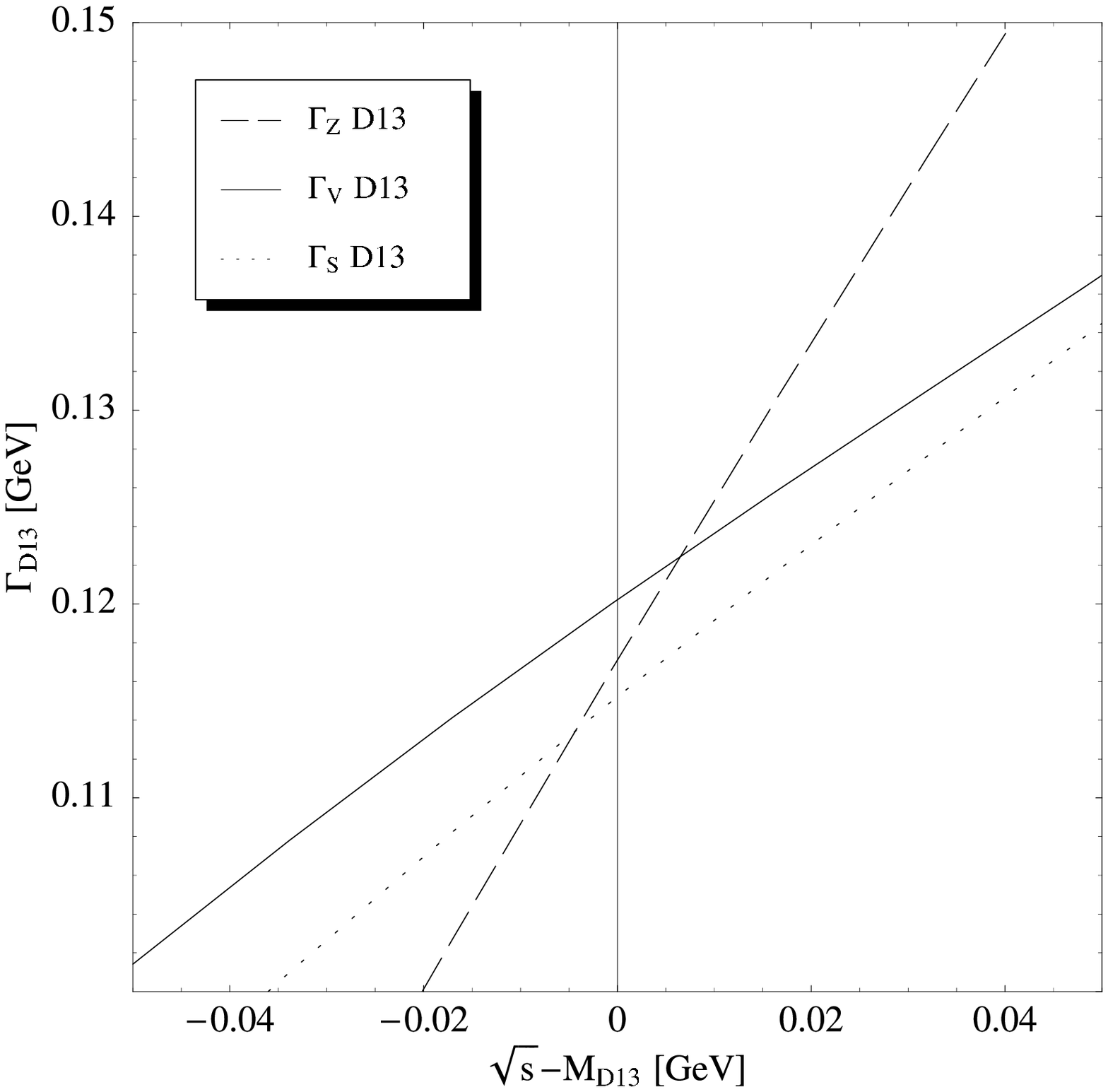,width=7.4cm}
\caption{Left: Different definitions for the width $\Gamma_Z$, $\Gamma_V$ and $\Gamma_S$, Right: $\Gamma_Z$, $\Gamma_V$ and $\Gamma_S$ zoomed at on-shell energy. \label{fig:altwidht}}
\end{figure}

The width of the \Deinsdrei is depicted in figure \ref{fig:widthD13}. On the left hand side the width and the partial widths are depicted for a partial width of the $N \rho$ channel $\Gamma_{N \rho}=10$ MeV and on the right for $\Gamma_{N \rho}=26$ MeV. One sees that the total widths in both cases are highly energy dependent and have their maximum above the mass shell region. Inspecting the partial widths we see that for high energies the coupling to $N \rho$ dominates. This can be understood because the nominal threshold region of this channel lies at $M_N + m_{\rho} \approx 1.7$ GeV, so only the tail of the mass spectrum of the $\rho$-meson contributes at the \Deinsdrei mass shell. Even though the partial width of this channel is comparably small, due to the subthreshold nature, the coupling must be large to gain such a width. This can be seen when the phase space opens up and the $N \rho$ channel dominates the total width.

For the larger value of the partial width of the $N \rho$ channel the energy dependence is even stronger leading to a three times higher maximum value of the width. In the spectral function a transfer of spectral strength to higher energies is expected.

The alternative widths are depicted in figure \ref{fig:altwidht}. They clearly differ as expected. For small $\sqrt{s}$ the quantities $\Gamma_{Z}$ and $\Gamma_{S}$ become negative making them a bad choice for a width. When the real parts of the selfenergies are small, which is the case on the mass shell, all quantities are about the same. Because for the cut-off parameter chosen the real parts are small it is not surprising that all three widths are nearly equal on the mass shell as shown in figure \ref{fig:altwidht}.

Because the real parts are small on the mass shell there is no ambiguity choosing a width in this energy region. Going off-shell it is preferable to take $\Gamma_V$ because it is always a positive definite quantity as shown in section \ref{sec:fermprop}.

In the case of the \Deinsdrei an easy comparison to a non-relativistic width is not possible due to the unstable character of the decay products. To calculate the momentum of the stable particles, a complicated analysis of three-body kinematics would be needed. Even taking the decay products of the \Deinsdrei resonance as stable particles does not solve the problem because the threshold energy of the $N \rho$ channel would be above the on-shell mass region of the \Deinsdrei making it impossible to fit the non-relativistic partial width.

\subsection{The Spectral Function}

Examining the properties of the fermionic spectral representation in section \ref{sec:fermprop} one sees that the resonance is not described by one but two spectral functions. But only one of them $\rho_v$ is normalized and positive definite making it the proper choice for the spectral information of a particle. In the following only the results for $\rho_v$ will be discussed and depicted.

\begin{figure}
\epsfig{file=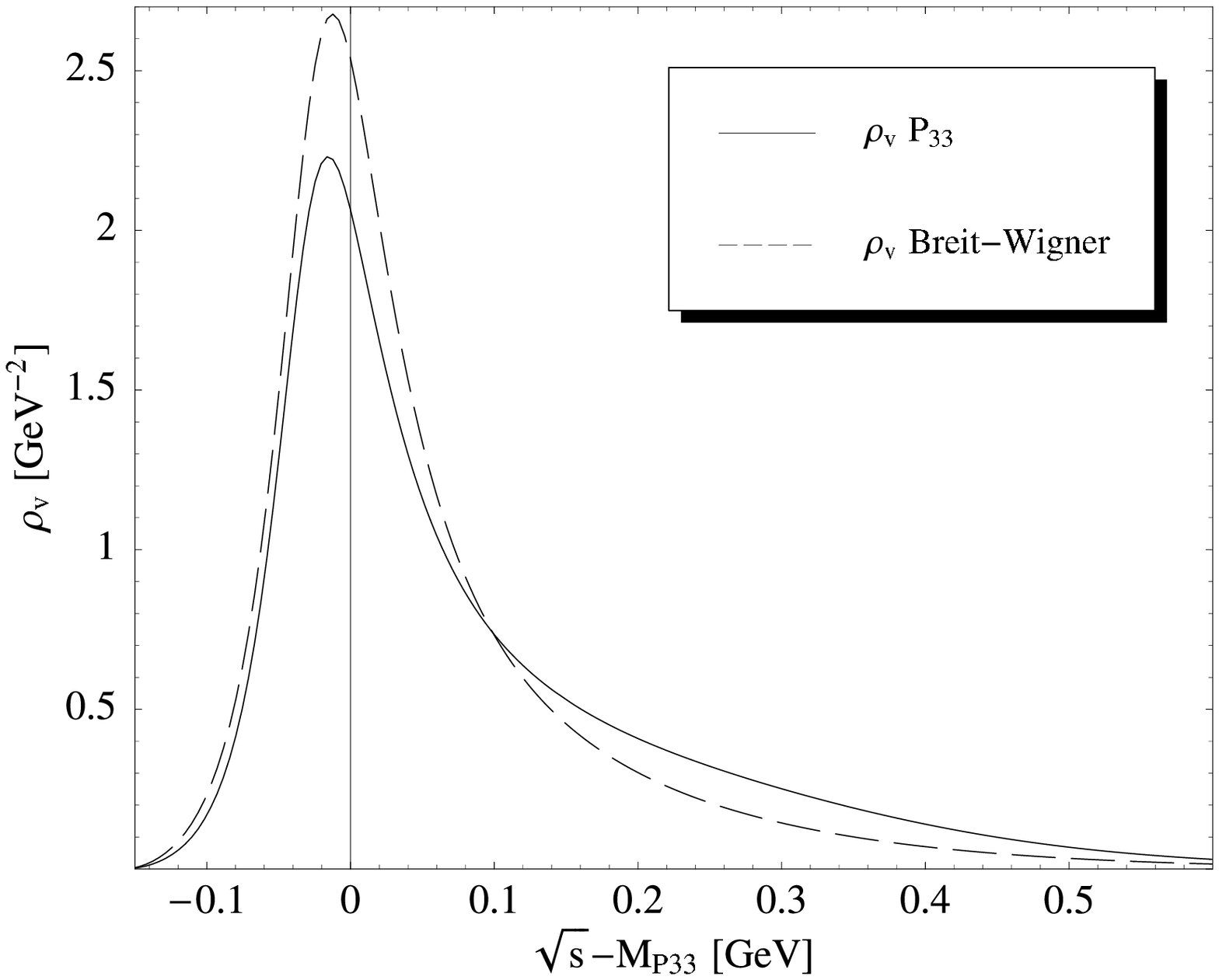,width=7.4cm} \hfill
\epsfig{file=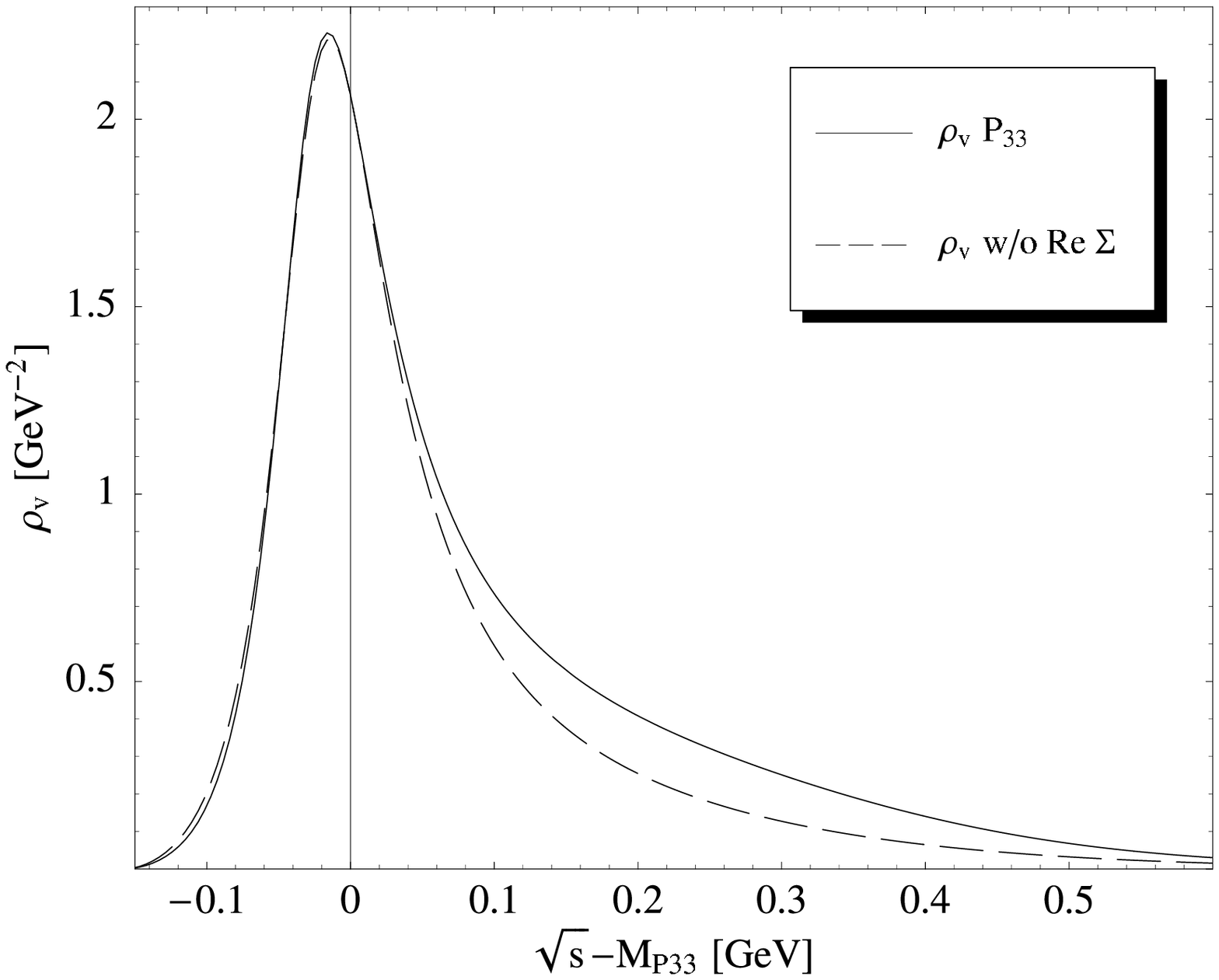,width=7.4cm} 
\caption{Left: Spectral function of the \Pdreidrei resonance compared with a simple Breit-Wigner approximation. Right: Spectral function of the \Pdreidrei resonance with and without real parts.}
\label{fig:SpeksBWForm}
\end{figure}

The spectral function of the \Pdreidrei shown in figure \ref{fig:SpeksBWForm} has the expected asymmetric form. It rises quickly and decreases with a rather large tail. The large tail is induced because the width of the \Pdreidrei has its maximum in this region. The form of the spectral function is typical for Breit-Wigner type quantities. Comparing the spectral function with a "pseudo-relativistic" spectral function, deduced from a Breit-Wigner form, as discussed in section \ref{sec:bosonspec} and especially equation (\ref{eq:BWFormRhoPseudo}), makes its similarity clear. For the width in (\ref{eq:BWFormRhoPseudo}) equation (\ref{eq:non-rel}) is used and fitted to the correct width of 120 MeV on the mass shell. $N$ of equation (\ref{eq:BWFormRhoPseudo}) is deduced by demanding that the "pseudo-relativistic" spectral function is normalized to one. Comparing the full spectral function of the \Pdreidrei with this very simple Breit-Wigner approximation, depicted in figure \ref{fig:SpeksBWForm}, shows that only for small $\sqrt{s}$ both quantities agree. The "pseudo-relativistic" spectral function shifts too much spectral strength to lower energies. This is the case because at higher $\sqrt{s}$ relativistic effects take place which are not accounted for in the approximation. This means that such a simple approximation already fails for the simplest spin 3/2 resonance.

As already discussed in section \ref{sec:fermprop} the mass of the resonance cannot be read off as the energy at the peak of the spectral function. This is possible when the width is a constant then
\begin{equation*}
\left. \frac{d}{ds} \rho_v(s) \right|_{s=M_R^2} = 0.
\end{equation*}
Taking the width as an $s$ dependent function will shift the peak away from the mass shell. This is already true for the simple Breit-Wigner form depicted at the left hand side of figure \ref{fig:SpeksBWForm}. The mass of the resonance is defined as the solution of equation (\ref{eq:defmass}).

\begin{figure}
\begin{center}
\epsfig{file=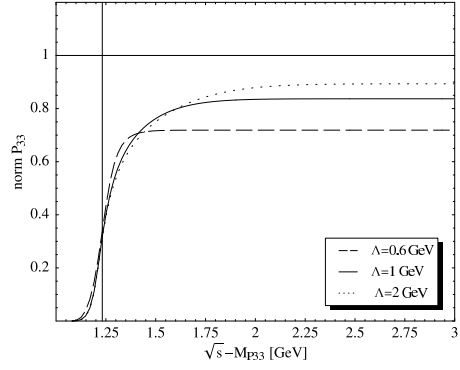,width=7.4cm}
\end{center}
\caption{Right: Normalization of the \Pdreidrei for different cut-off parameters, without real part of the selfenergy.}
\label{fig:normP33}
\end{figure}

Beside the Breit-Wigner form it is possible to approximate the spectral function by neglecting the real parts. Because the real parts are calculated through a dispersion relation their calculation includes some numerical effort. The real parts are small indicating that the changes will not be dominant. Neglecting the real parts gives a better approximation than the "pseudo-relativistic" spectral function. It underpredicts the values for large $\sqrt{s}$ but around the mass shell both agree well. The real parts play a dominant role in rising the tail as can be seen in figure \ref{fig:SpeksBWForm}, right hand side. This is needed because by neglecting the real part of the selfenergies the normalization of the spectral function is violated as depicted in figure \ref{fig:normP33}. There a normalization function corresponding to equation (\ref{eq:normferm}) is depicted. It is defined as 
\begin{equation}
\label{eq:normfermion}
norm(s) = \int_{(M_N + m_{\pi})^2}^s ds^{\prime} \rho_v(s^{\prime}) .
\end{equation}
In principle $norm(s)$ should reach 1 for large $s$. The violation depends on the cut-off parameter and is approx. 20\% in the case of $\Lambda=1$~GeV. 

\begin{figure}
\epsfig{file=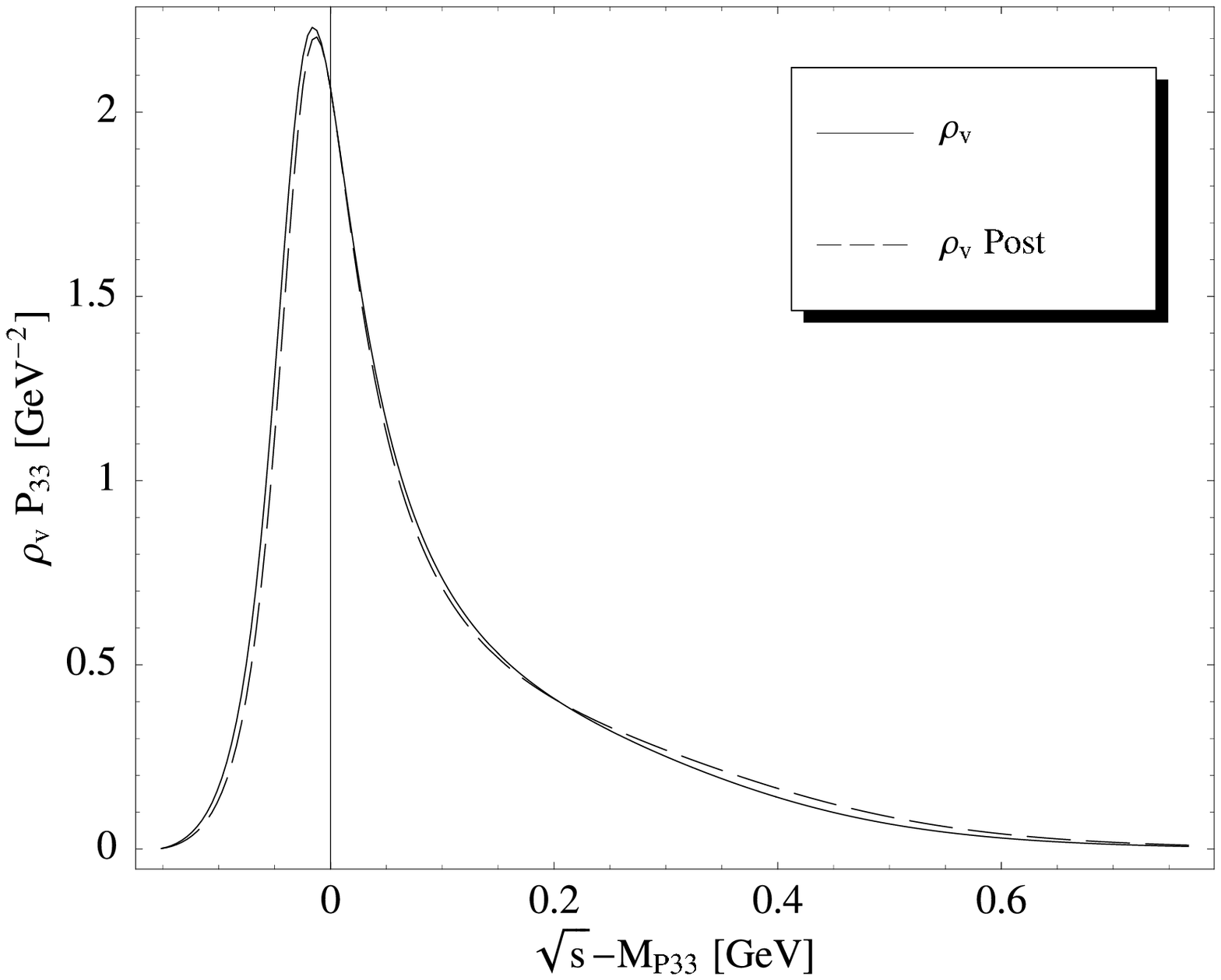,width=7.4cm} \hfill
\epsfig{file=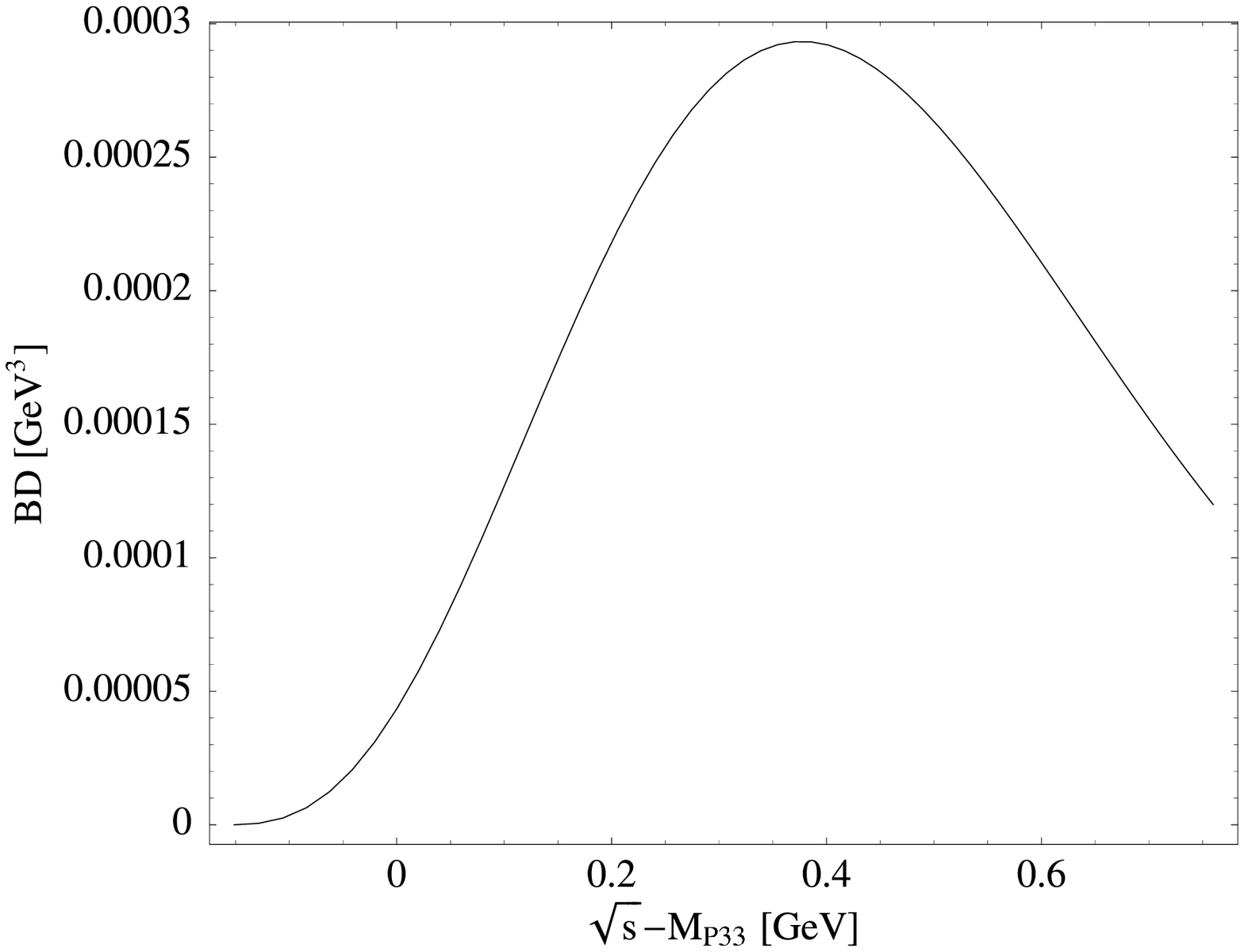,width=7.4cm}
\caption{Left: Comparison of the full spectral function of the \Pdreidrei resonance with simplified version proposed by Post et al.~\cite{Post:2000qi}. Right: Bjorken-Drell function (\ref{eq:BDfunc}) for the \Pdreidrei resonance. \label{fig:P33Post}}
\end{figure}

A third approximation was proposed by Post et al.~in \cite{Post:2000qi} using a simplified propagator. The full relativistic structure of the fermionic propagator as introduced in section \ref{sec:fermprop} is quite involved. The complication arises due to the Dirac structure of the self\-energy appearing in the denominator of the dressed propagator. When taking an averaged scalar selfenergy motivated as an averaging over the spins
\begin{equation*}
\langle \Sigma(p) \rangle = \frac{1}{2} \sum_s \bar{u}_s(p) \Sigma(p) u_s(p) = \frac{1}{2} \Tr \left[ \slashp + \sqrt{k^2} \Sigma(k) \right]
\end{equation*}
one can define a simplified propagator:
\begin{equation}
\mathcal{G}(p) = \frac{\slashp + M}{p^2 - M^2 - \langle \Sigma \rangle} \quad . \label{eq:simpprop}
\end{equation}
But as shown in \cite{Post:2000qi} inconsistencies with the Bjorken-Drell relation (\ref{eq:rho1rho2ende}) arise. In turn, this even leads to negative cross sections, see \cite{Post:2000qi} for details. Therefore it is an important consistency requirement to insist on the validity of (\ref{eq:rho1rho2ende}). When computing $\rho_s$ and $\rho_v$ from (\ref{eq:simpprop}) one finds
\begin{equation*}
M \rho_v(p) - \rho_s(p) = 0 .
\end{equation*}
For $\sqrt{p^2} > M$ equation (\ref{eq:rho1rho2ende}) is violated. To solve this problem it was suggested in \cite{Post:2000qi} to change $M \to \sqrt{p^2}$ in the numerator of (\ref{eq:simpprop}). This leads to an equation for the $\rho$'s of the form
\begin{equation*}
\sqrt{p^2} \rho_v(p) - \rho_s(p) = 0
\end{equation*}
which is in agreement with the Bjorken-Drell relation (\ref{eq:rho1rho2ende}). When comparing this to the full results one sees that in the full calculation the Bjorken-Drell relation is fulfilled but the Bjorken-Drell function (\ref{eq:BDfunc}) is not necessarily zero. Deviations in the spectral functions are expected when for given $p$ the Bjorken-Drell function (\ref{eq:BDfunc}) is larger than zero. This can be seen for the spectral function of the \Pdreidrei(1232) resonance depicted on the left hand side of figure \ref{fig:P33Post} where the deviations are very small. This can be understood because the Bjorken-Drell function is always small for all $p$, depicted on the right hand side of figure \ref{fig:P33Post}.

To summarize, the \Pdreidrei resonance has only one major decay channel giving it a less involved structure than the \Deinsdrei. Even for such a simple spectral function a "pseudo-relativistic" approach will lead to bad agreements already in the mass shell region. Neglecting the real parts of the selfenergy gives a good approximation from threshold energy to the mass shell region but spoils the normalization condition. A very good approximation is given by the simplified propagator of Post et al.~\cite{Post:2000qi} leading to good agreement for all $\sqrt{s}$ without spoiling the normalization.

\begin{figure}
\epsfig{file=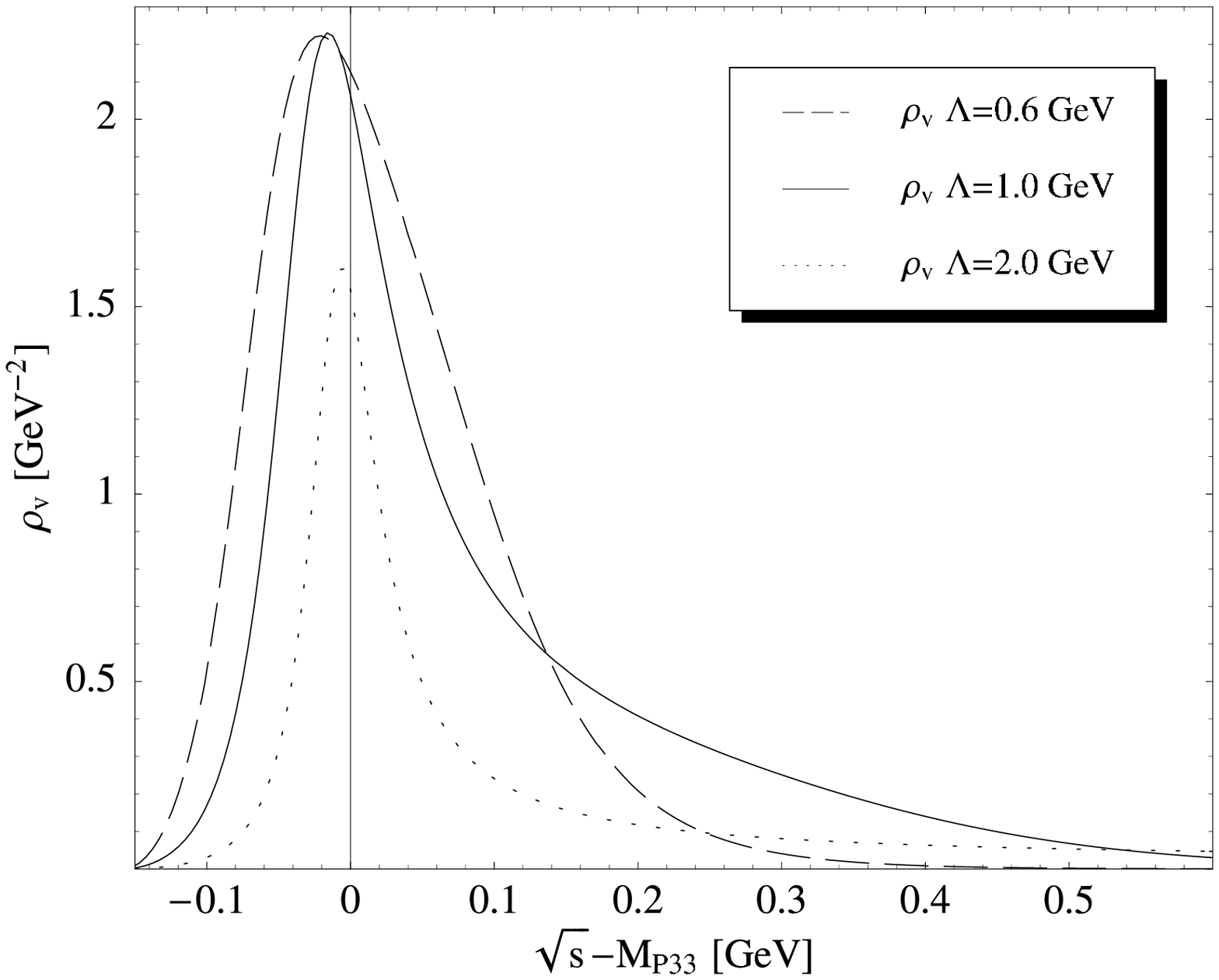,width=7.4cm} \hfill
\epsfig{file=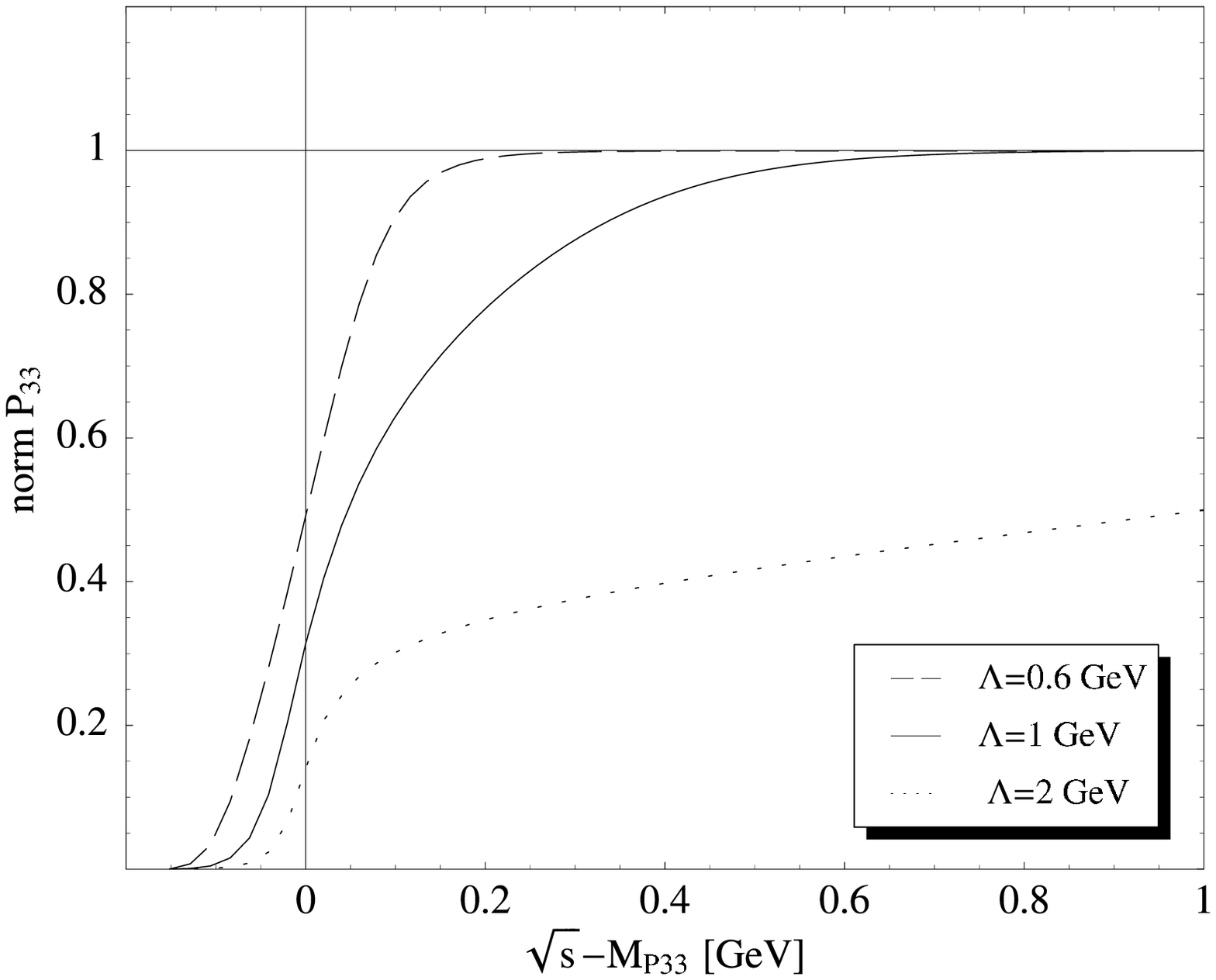,width=7.4cm} 
\caption{Left: Spectral function of the \Pdreidrei resonance with different values of the cut-off parameter $\Lambda$. Right: Normalization function for \Pdreidrei for different cut-off parameters $\Lambda$.}
\label{fig:P33cutoffnorm}
\end{figure}

As discussed above, the cut-off parameter is not fitted to experimental data. The spectral function of the \Pdreidrei resonance depends largely on this parameter as can be seen in figure \ref{fig:P33cutoffnorm}. The shape of all three functions is different. Small values of $\Lambda$ lead to a broadening of the spectral function around the peak by decreasing the tail. This can be understood when looking at the normalization functions for different $\Lambda$ as depicted in figure \ref{fig:P33cutoffnorm}. For small $\Lambda$ unity is reached earlier. For large $\Lambda$ the normalization function reaches unity at $\sqrt{s}-M_{P_{33}}\approx 4$ GeV leading to a strongly compressed spectral function with a long tail. Because the energy region where unity is reached is far beyond the physical energy region modeled here such large $\Lambda$'s should be excluded. From the discussion in the beginning also small $\Lambda$'s are physically unreasonable. From this consideration it can be concluded that reasonable cut-off parameters will be found in the region of 1 GeV.

\begin{figure}
\epsfig{file=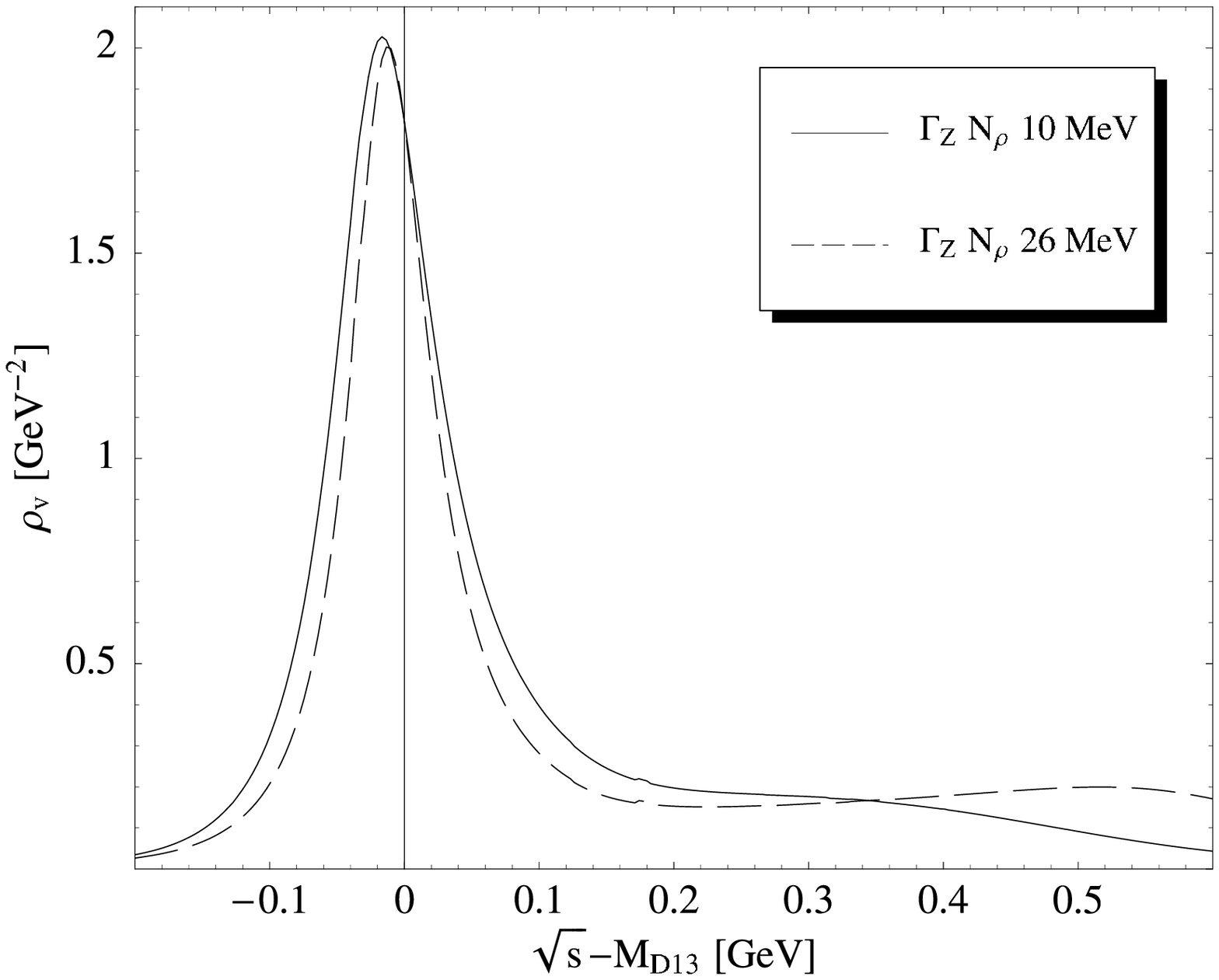,width=7.4cm}
\epsfig{file=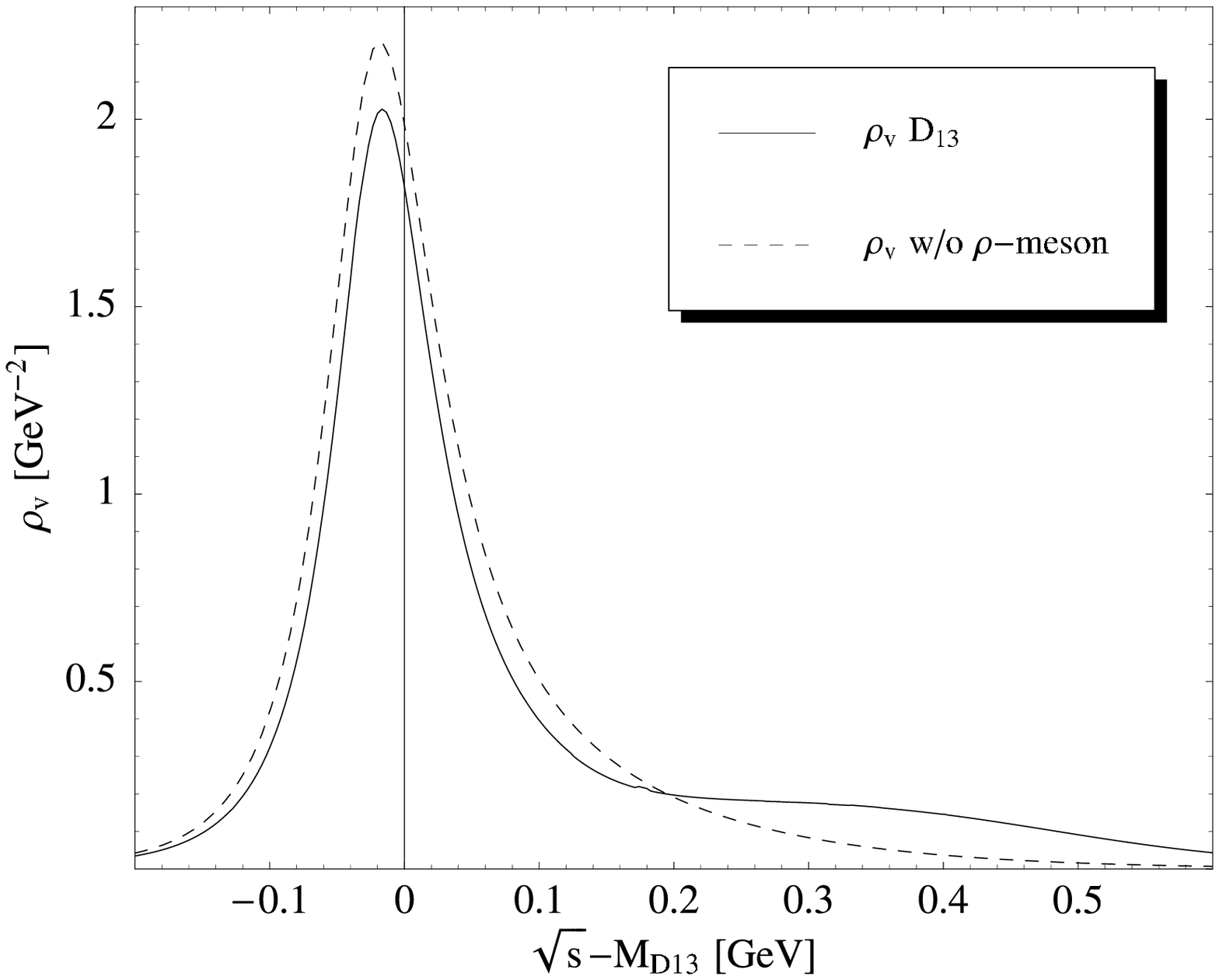,width=7.4cm}
\caption{Left: Spectral function of the \Deinsdrei resonance with both widths of the $N \rho$ channel. Right: Comparison with the spectral function of the \Deinsdrei resonance when the decay into the \rhomeson is ignored.}
\label{fig:Speks}
\end{figure}

The spectral function of the \Deinsdrei compared to the case of the \Pdreidrei is more symmetric around the peak with a shoulder arising in the region of the invariant mass of the $N \rho$ channel as can be seen in figure \ref{fig:Speks}. This shoulder is larger when taking a larger partial width for the $N \rho$ channel. When using this value for the width of the $N \rho$ channel structures arise in the region of the $N \rho$ threshold due to the opening phase space for this reaction. In the last section it was shown that the width becomes heavily larger in this energy region when taking the higher partial width for the $N \rho$ channel. This means that spectral strength is transfered to higher energies leading to the higher shoulder.

The shoulder vanishes when the $N \rho$ channel is ignored as depicted in figure \ref{fig:Speks}. Spectral strength is transfered to lower energies leading to a broadening of the spectral function. These changes indicate the importance of the \rhomeson in this energy region.

\begin{figure}
\epsfig{file=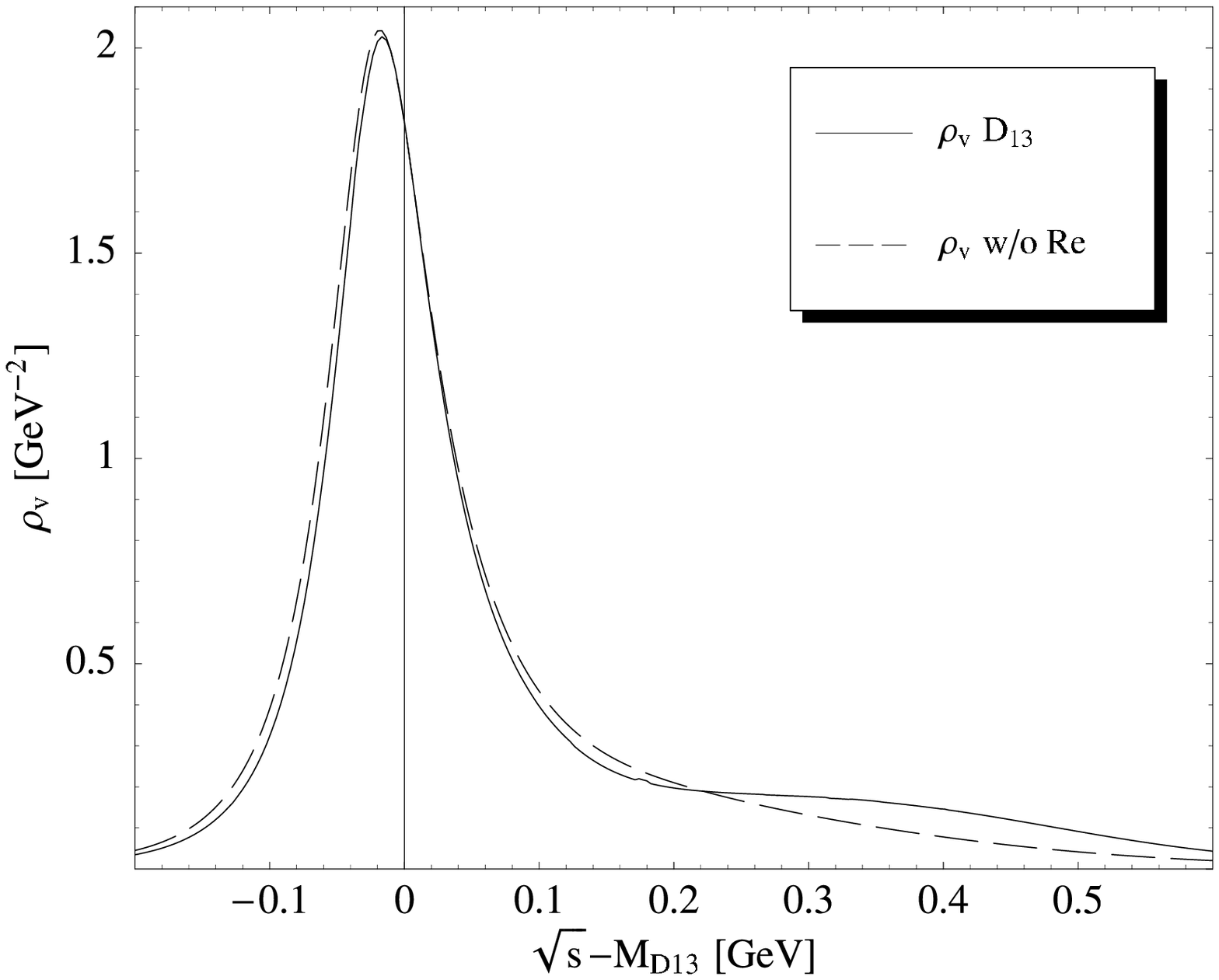,width=7.4cm} \hfill
\epsfig{file=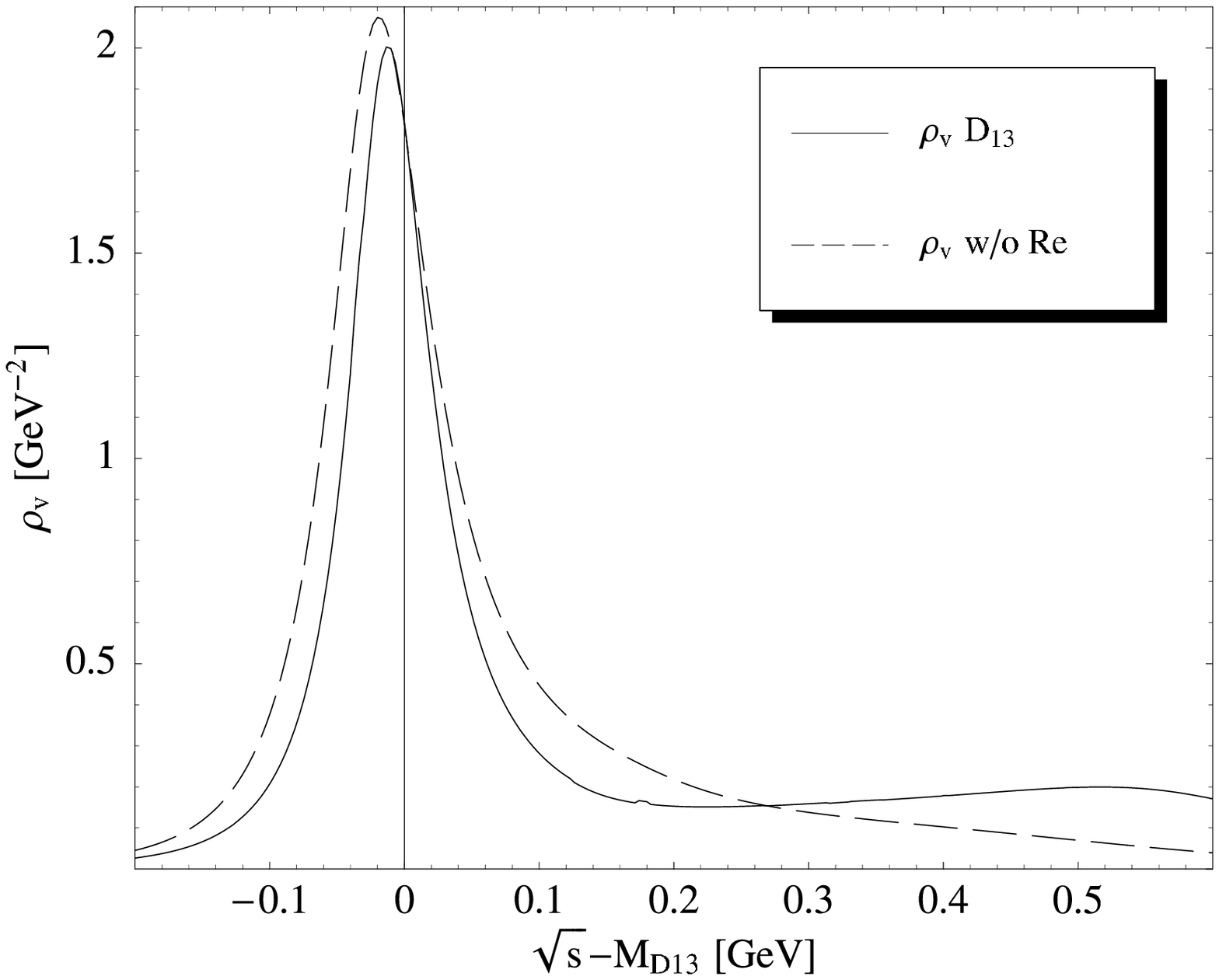,width=7.4cm}
\caption{Spectral function of the \Deinsdrei resonance with and without real parts. Left: Width of the $N \rho$ channel is 10 MeV. Right: Width of the $N \rho$ channel is 26 MeV.}
\label{fig:SpeksOhneRe}
\end{figure}

\begin{figure}
\epsfig{file=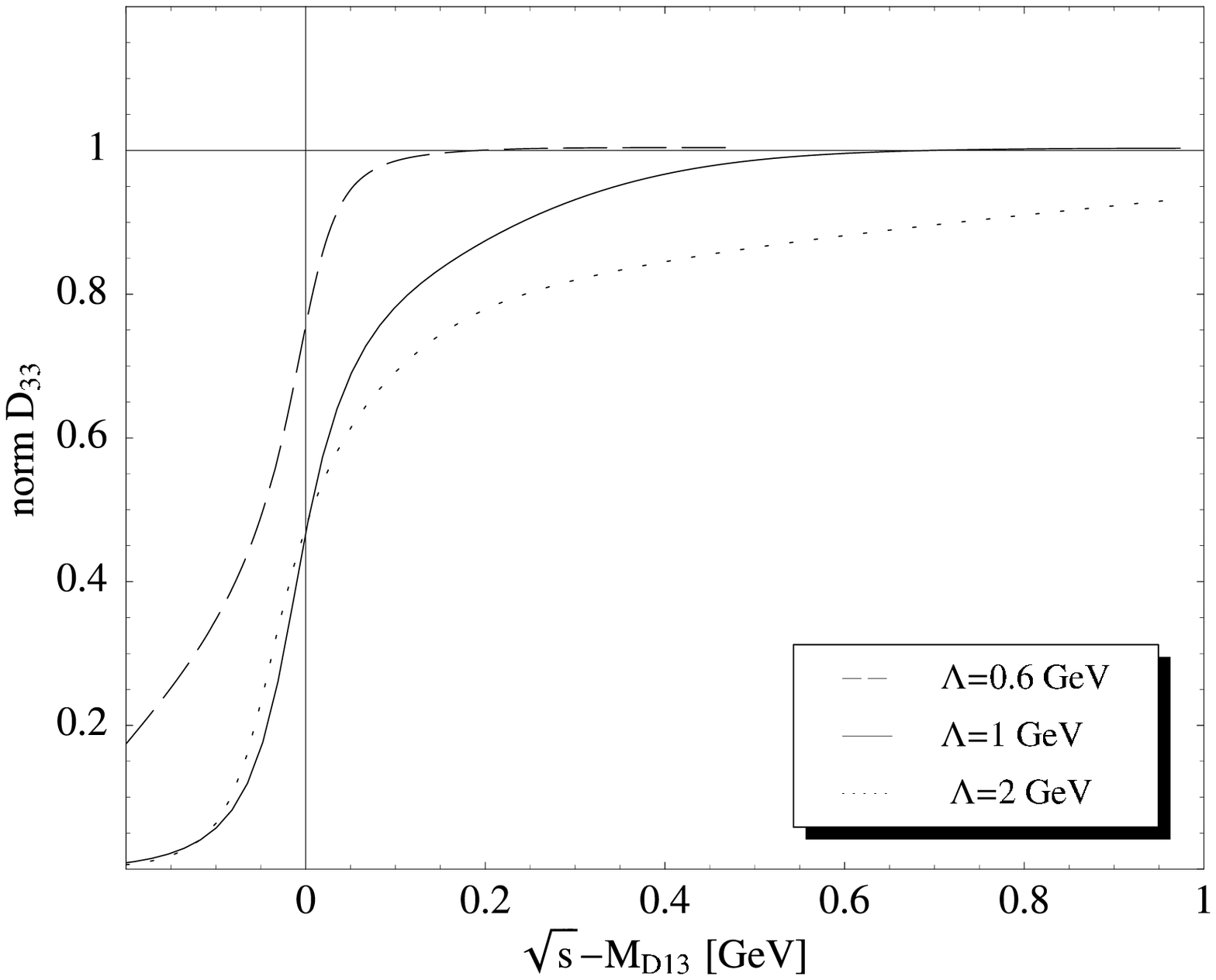,width=7.4cm} \hfill
\epsfig{file=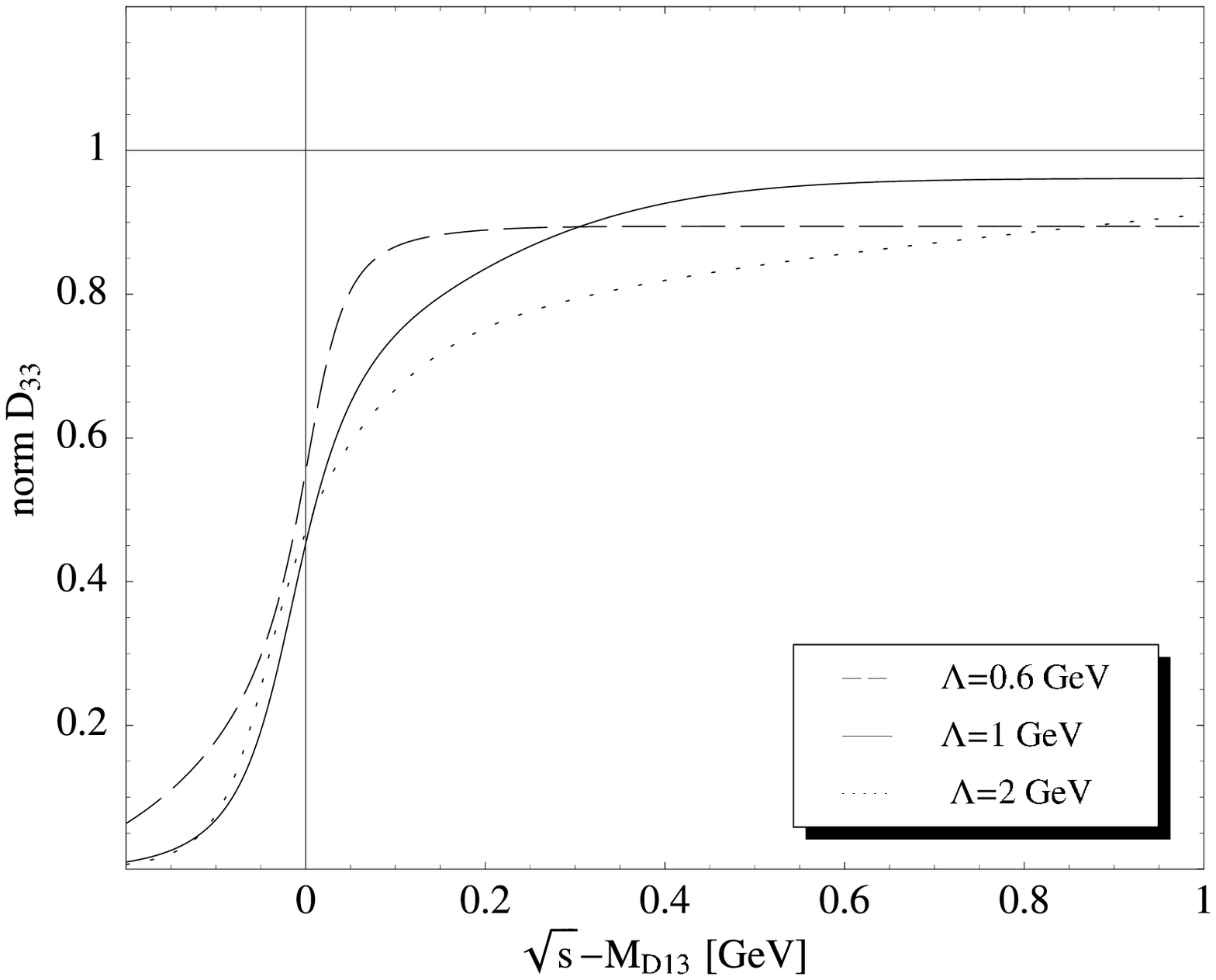,width=7.4cm}
\caption{Normalization function for \Deinsdrei. Left: For different cut-off parameter. Right: For different cut-off parameters without the real part of the selfenergy.}
\label{fig:normD13}
\end{figure}

In the discussion of the selfenergy the strong energy dependence of the real parts indicated a shoulder for the spectral function. This claim can be approved when neglecting the real parts. This is depicted in figure \ref{fig:SpeksOhneRe}. There one sees that the shoulder decreases when the real parts are switched off, showing that the shoulder is induced by the real parts of the selfenergy. This is needed to fulfill the normalization condition given in section \ref{sec:fermprop}. Without the real parts of the selfenergy the normalization is spoiled as shown in figure \ref{fig:normD13}.

\begin{figure}
\epsfig{file=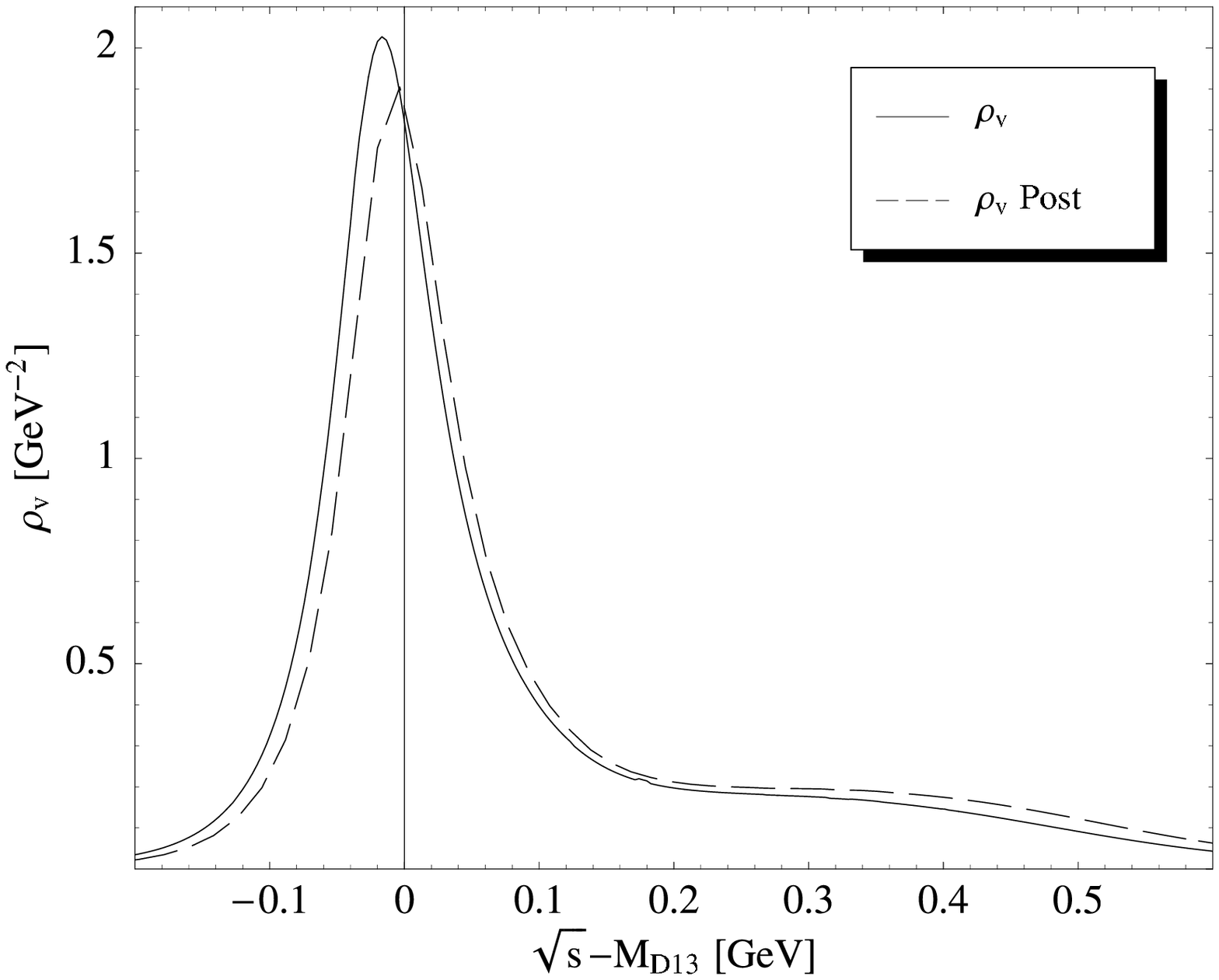,width=7.4cm} \hfill
\epsfig{file=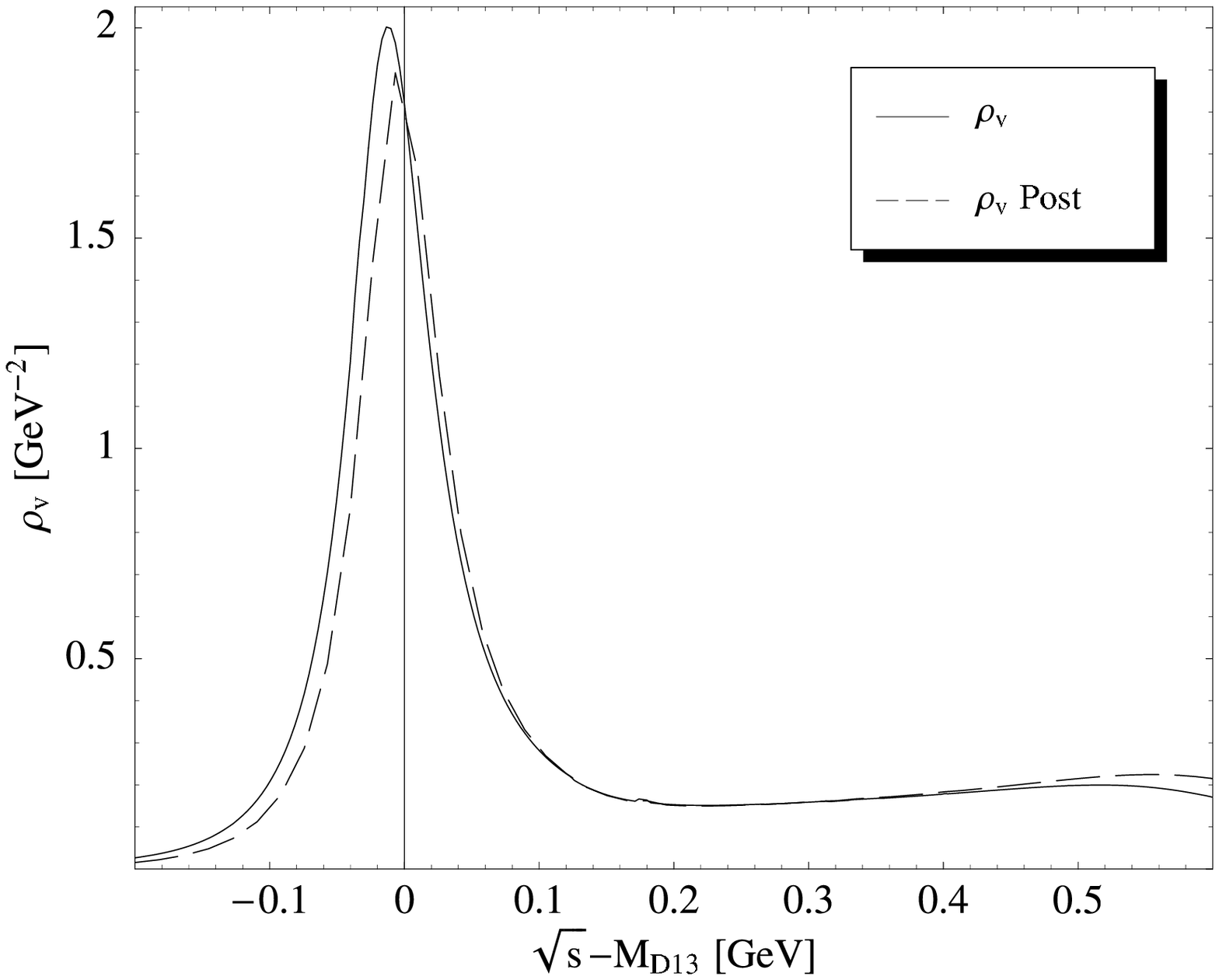,width=7.4cm}
\caption{Spectral function of the full calculation for the \Deinsdrei resonance compared with the simplified calculation of \cite{Post:2000qi}. Left: Width of the $N \rho$ channel is 10 MeV. Right: Width of the $N \rho$ channel is 26 MeV.}
\label{fig:FullAndPost}
\end{figure}

\begin{figure}
\begin{center}
\epsfig{file=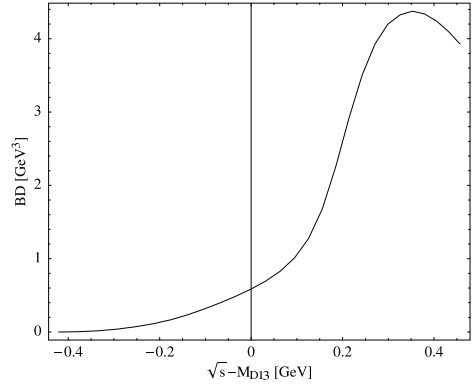,width=7.4cm}
\end{center}
\caption{Same as figure \ref{fig:P33Post}, r.h.s., but for \Deinsdrei.}
\label{fig:BDRelation}
\end{figure}

Because the normalization condition is spoiled when neglecting the real parts of the selfenergy this is not a good approximation for the spectral function. A "pseudo-relativistic" approximation was already unsatisfactory in the simple case of the \Pdreidrei. In addition it is difficult to construct non-relativistic widths due to the sub-threshold behavior of the $N \rho$ channel. This means that a comparison with a "pseudo-relativistic" spectral function is not possible. But the spectral function of the \Deinsdrei can be simplified using the method of Post et al. as discussed in the case of the \Pdreidrei. The results are depicted in figure \ref{fig:FullAndPost}. Spectral strength is transfered to higher energies by the approximation and the peak is shifted towards the mass shell region. This shift takes place because the denominator in the approximation is much simpler than in the full case. But still the results are in good agreement for both widths of the $N \rho$ channel. The deviations for the \Deinsdrei case are larger than for the \Pdreidrei case because also the Bjorken-Drell function defined in equation (\ref{eq:BDfunc}) has larger values than in the \Pdreidrei case as depicted in figure \ref{fig:BDRelation}. In the simplified version the Bjorken-Drell function is always zero. The Bjorken-Drell function (\ref{eq:BDfunc}) can be seen as a measure of the quality for such an approximation for a given particle.

\begin{figure}
\epsfig{file=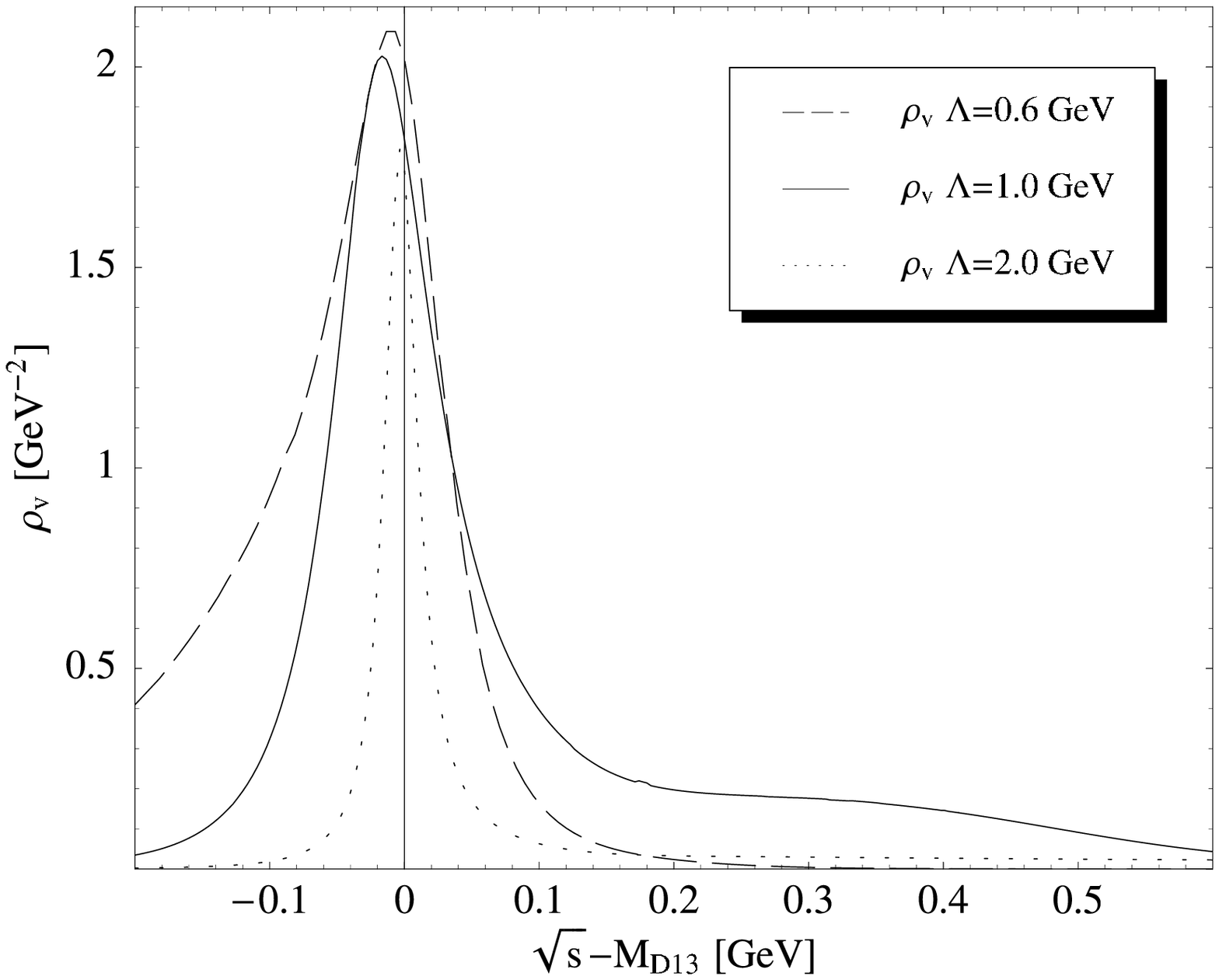,width=7.4cm} \hfill
\epsfig{file=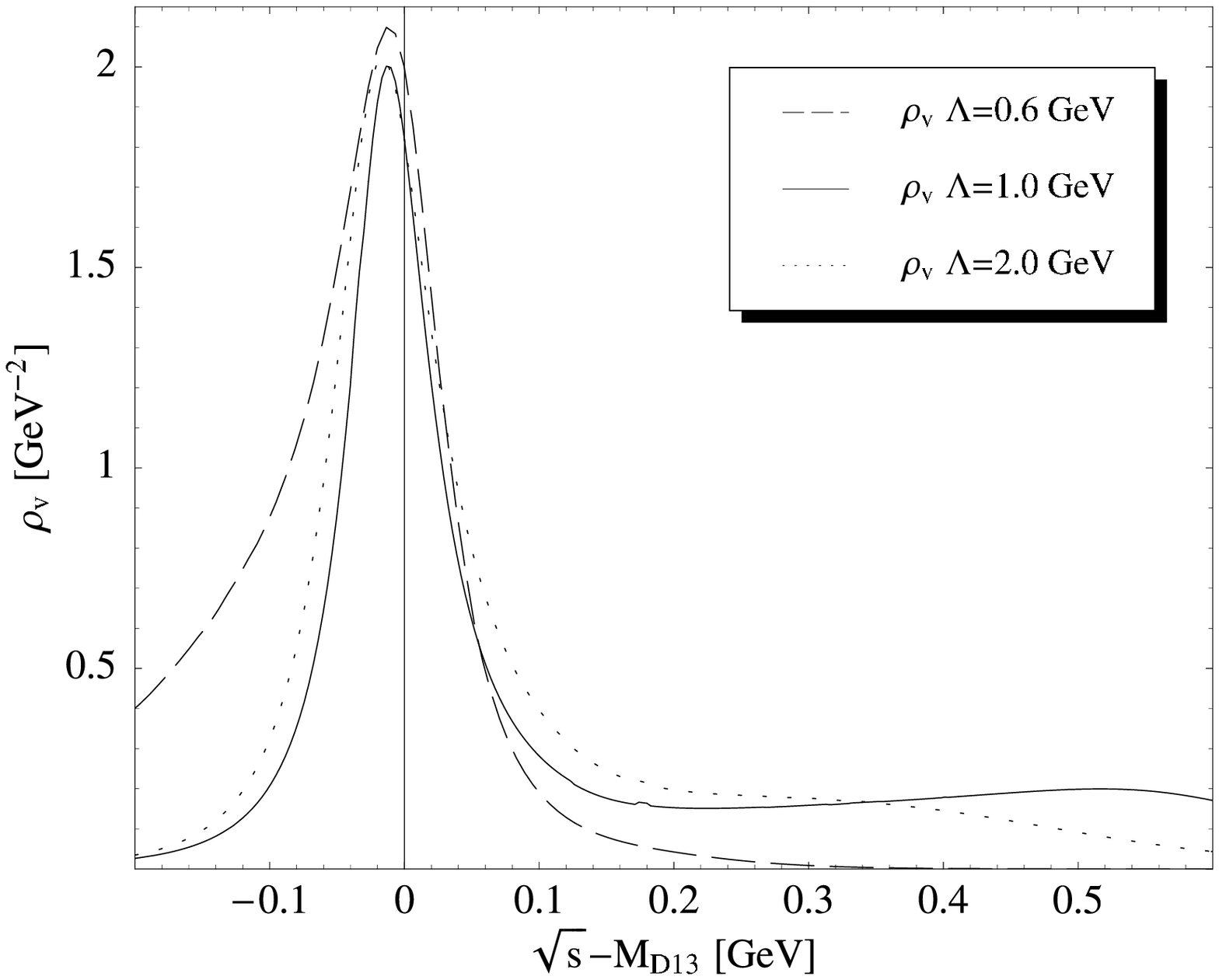,width=7.4cm}
\caption{Spectral function of the \Deinsdrei resonance with different values of the cut-off parameter $\Lambda$. Left: Partial width $\Gamma_{N \rho} = 10$ MeV. Right: Partial width $\Gamma_{N \rho} = 26$ MeV.}
\label{fig:cutoffvar}
\end{figure}

The spectral function of the \Deinsdrei depends strongly on the cut-off parameter $\Lambda$ as depicted in figure \ref{fig:cutoffvar}. Similar to the \Pdreidrei case this comes due too different distributions of the spectral strength over different energies. The distribution can be seen when comparing the normalization functions for different cut-off parameters, depicted in figure \ref{fig:normD13}. Larger cut-off parameters distribute the strength in a larger region and into the high energy parts of the spectrum. As in the case of the \Pdreidrei such a broad distribution of spectral strength outside the physical meaningful energy region is disturbing. This excludes large values of $\Lambda$ as a reasonable choice.

A smaller cut-off parameter suppresses strongly already in the physical meaningful region and large portions of spectral strength can be found at energies much below the mass shell leading to a large tail for small energies. This tail is induced by the large energy dependence of the real parts in this energy region. Already in the beginning of this section we have argued that small values of $\Lambda$ are unreasonable when looking at the  variation of the coupling in the physically meaningful region. Taking all arguments together supplies strong indications that small cut-off parameters will not give a correct physical description of the resonance. We conclude that also for the \Deinsdrei resonance a physically meaningful value for the cut-off parameter will be around $\Lambda=1$ GeV.

\subsection{Influence of Unstable Particles}

One of the major difficulties when calculating the selfenergy and the spectral function of the \Deinsdrei resonance compared to the \Pdreidrei resonance is the fact that the \Deinsdrei de\-cays into unstable particles. The complications arise because when going from a stable to an unstable particle a $\delta$-function in the selfenergy has to be exchanged by a spectral function (see section \ref{sec:fermprop}). This leads to a higher numerical effort because the integrations cannot be solved analytically anymore. If the unstable particle is a baryon as in the case of the $\Delta \pi$ channel the spectral function will even have a more complicated structure leading again to much higher numerical effort.

\begin{figure}
\epsfig{file=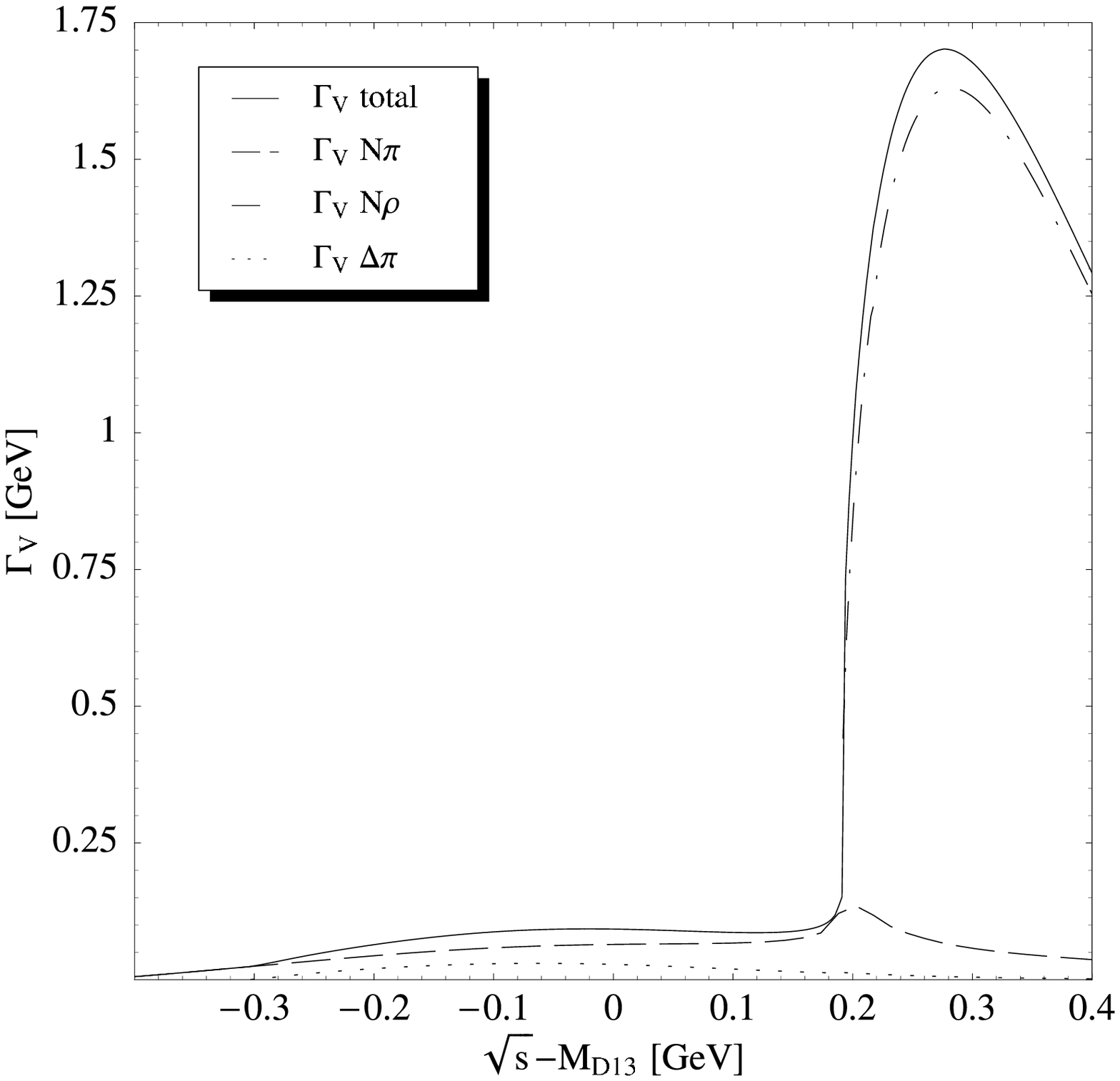,width=7.4cm} \hfill
\epsfig{file=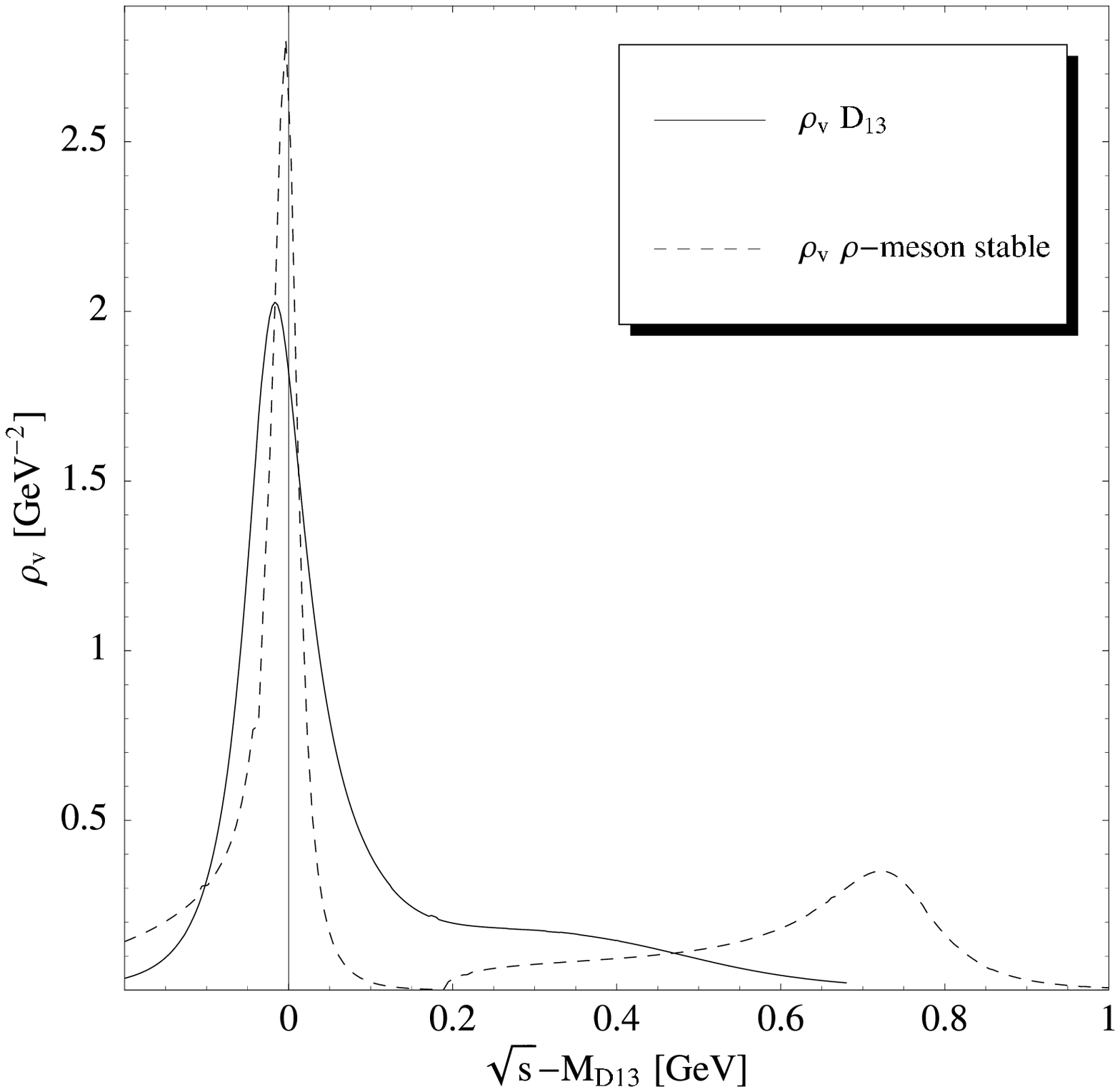,width=7.4cm}
\caption{Left: Partial and total width of \Deinsdrei when considering the \rhomeson as a stable particle. Right: Spectral function of \Deinsdrei for the case of an unstable or stable $\rho$-meson.}
\label{fig:rhostable}
\end{figure}

Assuming the particles to be stable would be a large simplification especially for the $\Delta \pi$ channel. Because in this work the full propagator is available it is interesting to compare the results of the full spectral function to a simplified one where one particle is assumed to be stable. 

When inserting the \rhomeson as a stable particle its threshold value will be higher than the mass shell of the \Deinsdrei. So it is not possible anymore to fit the coupling constant via the partial width. In figure \ref{fig:rhostable} the partial widths and the spectral function are depicted taking the same input as for the unstable case. The partial width of the \rhomeson opens up dramatically reaching nearly seven times the value than in the unstable case. Due to this dramatic increase of the width around the threshold energy of $N \rho$, large amount of spectral strength is transfered above this threshold creating a new peak. The total shape of the spectral function changes leading to the conclusion that this simplification is not a good approximation.

\begin{figure}
\epsfig{file=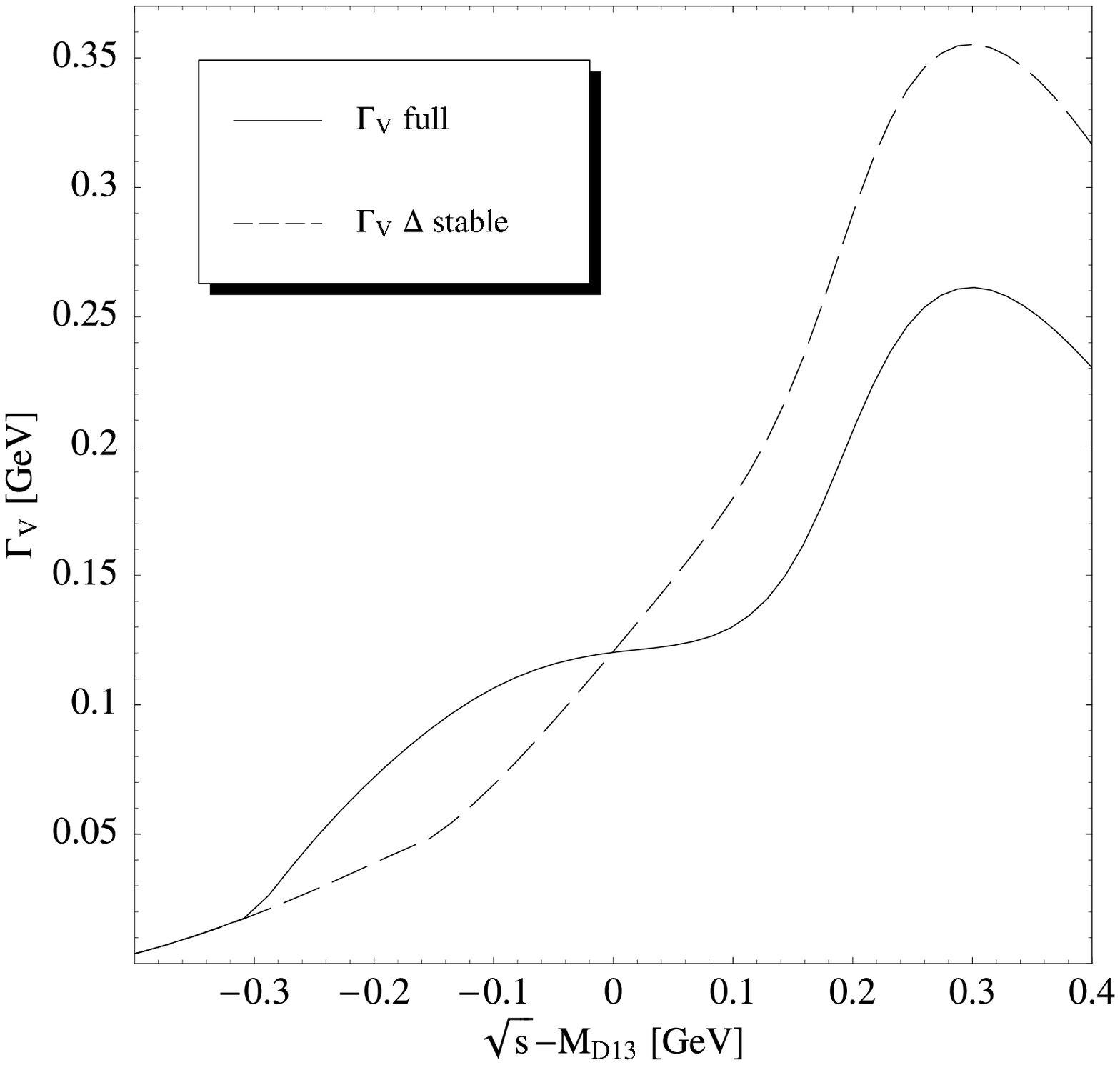,width=7.4cm} \hfill
\epsfig{file=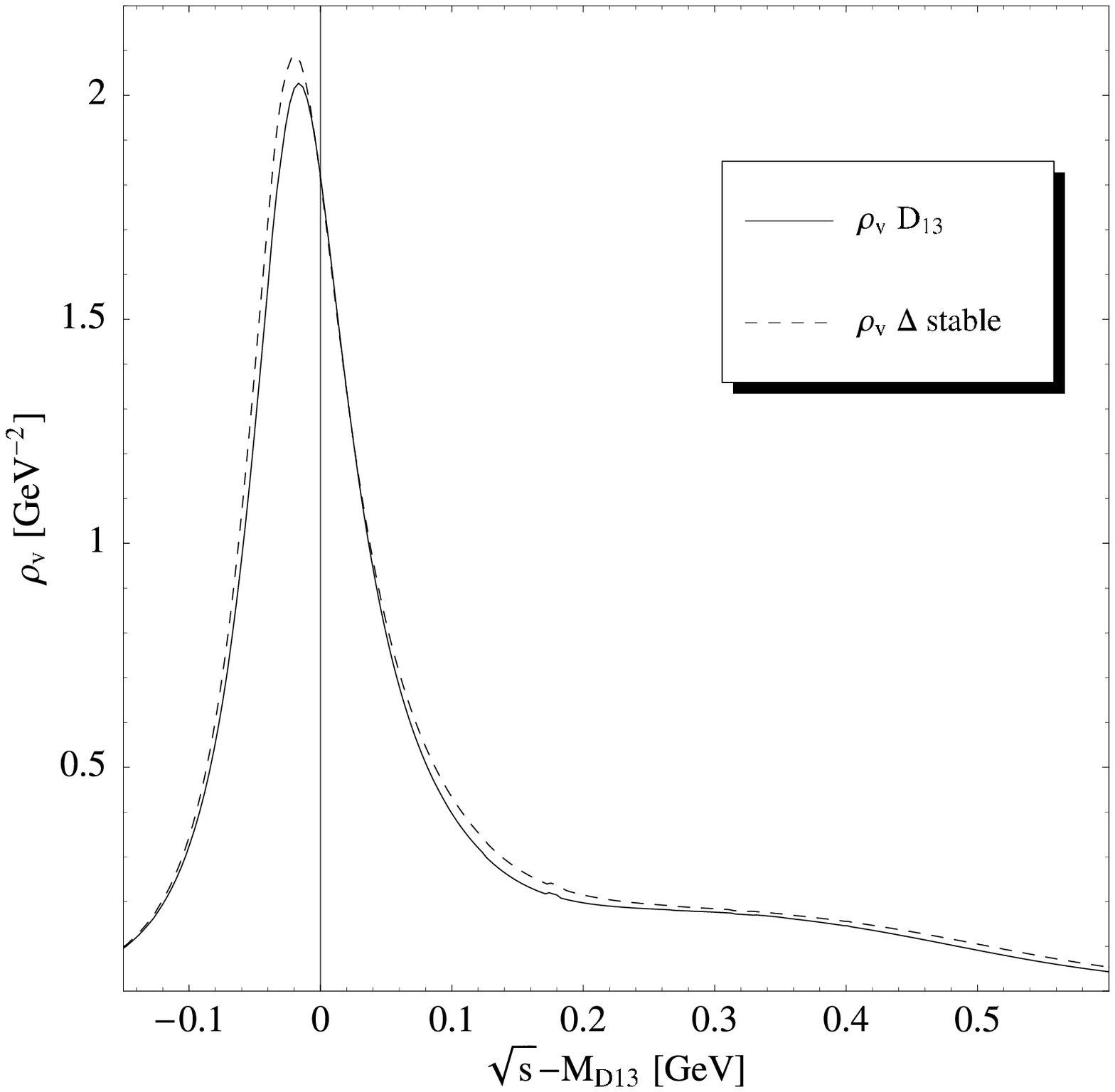,width=7.4cm}
\caption{Left: Total width of \Deinsdrei when considering the $\Delta$ resonance as an unstable (full line) and stable (dashed line)  particle. Right: Spectral function of \Deinsdrei for the case of an unstable or stable $\Delta$.}
\label{fig:deltastable}
\end{figure}

When considering the $\Delta$ as a stable particle it is possible to refit the coupling constants of the $\Delta \pi$ channel. 
The coupling strength basically remains the same.

The differences between a calculation using a stable or an unstable $\Delta$ will be mainly induced by the shift of the threshold for the width of this channel. Because the spectral function is normalized spectral strength has to be transfered to higher energies leading to an increased width for energies higher than the threshold region of $\Delta \pi$. This can be seen in the left plot in figure \ref{fig:deltastable} where the width of the \Deinsdrei is plotted for the full case and taking the $\Delta$ as a stable particle. On the right hand side of figure \ref{fig:deltastable} the spectral function is compared with the result of the full calculation. One sees the shift of spectral strength to higher energies leading to a lowered start, higher peak and a slightly larger shoulder. The effect can be seen but is still considerably small.

Taking the $\Delta$ as a stable particle is a huge simplification for the calculation due to the complicated structure of its spectral function. Comparing the effect of this simplification on the spectral function of the \Deinsdrei and the possible simplification of the calculation leads to the conclusion that it is a good approximation when computing power is restricted.

It was possible to make the simplification where the $\Delta$ is taken as a stable particle because the mass shell of the \Deinsdrei is above the energy threshold to create a $\Delta$. The effect by taking the $\Delta$ as a stable particle only shifts some spectral strength due to the shifted threshold of the partial width which is comparably small. Additionally the coupling of the \Deinsdrei to the $\Delta$ is not too strong so only a small amount of spectral strength is transfered to higher energies. This leads to rather small changes in the spectral function making it a considerable simplification.

In the case of the $\rho$ such a simplification changes the whole structure of the spectral function. This can be understood because the on-shell energy of the \Deinsdrei is less than the threshold energy to create $N \rho$. Due to the strong coupling of the \rhomeson to the \Deinsdrei a considerably large partial width has to be taken into account even at these energies. Additionally the mass of the \rhomeson is quite high leading to a large energy gap between the threshold energy of the unstable \rhomeson and the stable one. Taking the \rhomeson as a stable particle pushes all this spectral strength above threshold where it has a predominant impact on the shape of the spectral function. 

\section{Summary and Outlook}
\label{sec:summary}

In this work hadronic interactions of spin 3/2 resonances are investigated. Such interactions are not trivial and lead to many problems concerning the consistencies of the couplings. It was shown that it is feasible to calculate fully relativistic propagators for spin 3/2 resonances in the Pascalutsa framework. These considerations can serve as a prelude to more complicated treatments for in-medium calculations of hadronic properties.

Further specific calculations were performed for the case of the \Pdreidrei and \Deinsdrei res\-o\-nances which are needed to overcome complications that occurred in simplifications. Such complications (violation of the Bjorken-Drell relation (\ref{eq:BDfunc}) accompanied by negative cross sections) vanish when the full relativistic structure is incorporated.

In section \ref{sec:fermprop} we calculated the full relativistic structure of the dressed propagator for spin 3/2 resonances in an analytical form. With this starting point it was possible to calculate the analytical form of the spectral function for spin 3/2 resonances. It could be shown that in the Pascalutsa framework the spectral function for spin 3/2 resonances has basically the same structure as spin 1/2 states. This is a major simplification and means that it is not only feasible to calculate spin 3/2 resonances in the Pascalutsa framework but also easier than in the conventional approach.

The selfenergy and the spectral function describes fully a resonance. But because they are not measurable quantities the width and the mass of a resonance were introduced. They are closer to theory than experimental data and closer to experiment than spectral functions. But as pure numbers they cannot represent the whole spectral shape. This leaves an ambiguity how to define them. In this work the width and the mass are defined by comparing the spectral function to a relativistic Breit-Wigner form. In the case of a spin 3/2 resonance there are three possible candidates for a width from which only one is a generally positive definite quantity.

To achieve the goal to calculate the spectral function of the \Deinsdrei resonance the selfenergies of the major decay channels are needed. These are $N \pi$, $N \rho$ and $\Delta \pi$ and calculated in section \ref{D13channels}. For $N \pi$ Pascalutsa proposed a consistent coupling which was used. For the $N \rho$ and $\Delta \pi$ channel no couplings were previously available. Using the method proposed by Pascalutsa they were derived. To check the results they are expanded for the non-relativistic limit by assuming stable particles, leading to the correct phase space behavior of the width. The $N \pi$ channel leads to a P-wave and D-wave for a particle with positive and negative parity, respectively. The $N \rho$ and $\Delta \pi$ channels lead to P-wave (S-wave) for particles with positive (negative) parity. The selfenergies were calculated in a general way making it possible to use them for all spin 3/2 resonances decaying into these three channels. Because the \Pdreidrei resonance decays into $N \pi$ it was possible to calculate the properties of this resonance. It was also needed as a part of the \Deinsdrei selfenergy in the $\Delta \pi$ channel because the unstable character of the $\Delta$ must be taken into account. The selfenergy of this channel contains an integration over the spectral function of the $\Delta$ making it a complicated and numerically involved quantity.

The results for the \Pdreidrei and \Deinsdrei resonances were discussed in section \ref{sec:results}. Our spectral functions automatically fulfill all Bjorken-Drell conditions summarized in section \ref{sec:fermprop}. In some simplifications as proposed by Post et al.~\cite{Post:2000qi} the propagator had to be changed by hand to fulfill these conditions. It could be shown in this work, that by taking the full propagator such problems do not arise.

The spectral function of the \Pdreidrei resonance has a typical Breit-Wigner form. It is asymmetric with a quick rise and a long tail. The \Deinsdrei resonance on the other hand has a long shoulder for high energies due to the strong coupling of the $\rho$-meson. This shoulder vanishes when the \rhomeson is neglected and more structure arise when the partial width is increased from $\Gamma_{N \rho} = 10$ MeV to $\Gamma_{N \rho} = 26$ MeV. These changes indicate the importance of the $N \rho$ channel for the whole structure of the \Deinsdrei spectral function. 

For the widths of the resonances it could be shown that for the \Pdreidrei all three candidates for a width are proper choices. All three are positive definite and are similar over the whole energy range. In the case of the \Deinsdrei the three candidates for a width differ largely and only one is positive definite making it the only choice for an off-shell width.

To investigate the dependence of the spectral function on the cut-off parameter, the latter was varied. The shape of the spectral function changes dramatically for both resonances. Because the parameters were only fitted to two quantities it was not possible to fit $\Lambda$ precisely leaving an ambiguity which value is a proper one. Although it was not possible to pin down the number exactly it could be shown that only values of around 1 GeV give physically meaningful results. This resolves the ambiguity and gives some constraints for further fits. To pin down the input parameters more precisely it is desirable to fit them to experimental data.

The full spectral function was compared with various simplifications to rate the quality of these simplifications. First, we compared the results of the \Pdreidrei spectral function with a simple Breit-Wigner approach where the width is taken in a non-relativistic form. It could be shown that such a simple approach does not give a reasonable approximation for energies higher than the mass shell. It is not an unexpected result because such a "pseudo-relativistic" approach will not work for higher energies where relativistic effects are not negligible. 

Next it could be shown that neglecting the real parts spoils the normalization conditions for $\rho_v$. It has mainly an effect on the tails and shoulders of the spectral functions which are reduced.

Comparison of the full calculation with the approximation proposed by Post et al.~\cite{Post:2000qi} shows that this is a reasonable approximation. The difference is mainly a slightly shifted peak and some spectral strength is transfered to higher energies. This effect is much smaller for the \Pdreidrei than for the \Deinsdrei case. Looking at the Bjorken-Drell function (\ref{eq:BDfunc}) one sees that for the \Pdreidrei case it is nearly zero. Because in the \text{Post} et al.~approach it is always zero, deviations from zero in the Bjorken-Drell function will also lead to deviations in the spectral functions. For the \Deinsdrei the deviation from zero are larger leading to larger but not dramatic modifications of the spectral function.

Taking the $\Delta$ and the $\rho$ as stable particles leads to different conclusions. Taking the $\rho$ as a stable particle has a major impact on the structure of the spectral function. Such an approximation is not reasonable. On the other hand, assuming the $\Delta$ as a stable particle has only marginal influence on the structure of the spectral function. Due to the large numerical effort needed when implementing the $\Delta$ as an unstable particle such a simplification is reasonable.

Using the results for the selfenergies it is possible to calculate the widths and spectral functions of all spin 3/2 resonances decaying into $N \pi$, $N \rho$, $\Delta \pi$. As an outlook we point out that with the explicit results for the \Deinsdrei resonance it is possible to calculate reactions going over the \Deinsdrei into dileptons or two pions as an end product, e.g.
\begin{align*}
N \pi &\rightarrow \Deinsdrei \rightarrow N \rho \rightarrow N e^+ e^-,\\
N \pi &\rightarrow \Deinsdrei \rightarrow N \rho \rightarrow N \pi \pi,\\
N \pi &\rightarrow \Deinsdrei \rightarrow N \Delta \rightarrow N \pi \pi. 
\end{align*}
Furthermore it is possible to rate the quality of approximations of spectral functions by comparing them to the results of the full calculation as was done in this work for some approximations.

\acknowledgments The authors thank U.~Mosel for discussions and continuous support.


\bibliographystyle{apsrev}
\bibliography{PaperPaper2}

\end{document}